%% file: paper.tex
\documentclass[aps,prx,twocolumn,superscriptaddress,longbibliography]{revtex4-2}
\usepackage[section]{placeins}

\include{header}
\include{acronyms}

\begin{document}
\def\thetitle{Stable Bipolarons in Open Quantum Systems}

\title{\thetitle}
\author{M.~Moroder}
\affiliation{Department of Physics, Arnold Sommerfeld Center for Theoretical Physics (ASC), Munich Center for Quantum Science and Technology (MCQST), Ludwig-Maximilians-Universit\"{a}t M\"{u}nchen, 80333 M\"{u}nchen, Germany.}
\author{M.~Grundner}
\affiliation{Department of Physics, Arnold Sommerfeld Center for Theoretical Physics (ASC), Munich Center for Quantum Science and Technology (MCQST), Ludwig-Maximilians-Universit\"{a}t M\"{u}nchen, 80333 M\"{u}nchen, Germany.}
\author{F.~Damanet}
\affiliation{Institut de Physique Nucléaire, Atomique et de Spectroscopie, CESAM, University of Liège, B-4000 Liège, Belgium}
\author{U.~Schollw\"ock}
\affiliation{Department of Physics, Arnold Sommerfeld Center for Theoretical Physics (ASC), Munich Center for Quantum Science and Technology (MCQST), Ludwig-Maximilians-Universit\"{a}t M\"{u}nchen, 80333 M\"{u}nchen, Germany.}
\author{S.~Mardazad}
\affiliation{Department of Physics, Arnold Sommerfeld Center for Theoretical Physics (ASC), Munich Center for Quantum Science and Technology (MCQST), Ludwig-Maximilians-Universit\"{a}t M\"{u}nchen, 80333 M\"{u}nchen, Germany.}
\author{S.~Flannigan}
\affiliation{Department of Physics $\&$ SUPA, University of Strathclyde, Glasgow G4 0NG, United Kingdom}
\author{T.~K\"ohler}
\affiliation{Department of Physics and Astronomy, Uppsala University, Box 516, S-751 20, Uppsala, Sweden}
\author{S.~Paeckel}
\affiliation{Department of Physics, Arnold Sommerfeld Center for Theoretical Physics (ASC), Munich Center for Quantum Science and Technology (MCQST), Ludwig-Maximilians-Universit\"{a}t M\"{u}nchen, 80333 M\"{u}nchen, Germany.}
\begin{abstract}

	Recent advances in numerical methods significantly pushed forward the understanding of electrons coupled to quantized lattice vibrations.
	At this stage, it becomes increasingly important to also account for the effects of physically inevitable environments.
	\discussive{Here, we combine state-of-the-art tensor\hyp network and quantum trajectories methods in order to study the impact of dissipation on realistic condensed matter models including highly\hyp excited phononic modes.}
	In particular, we study the transport properties of the Hubbard\hyp Holstein Hamiltonian that models a large class of materials characterized by strong electron-phonon coupling, in contact with a dissipative environment.
	%
	%
	%
	We combine the non\hyp Markovian \acrlong{HOPS} method and the Markovian \acrlong{QJ} method with the newly introduced \acrlong{PP-DMRG_long}, creating powerful tensor network methods for dissipative quantum many\hyp body systems.
	Investigating their numerical properties, we find a significant speedup up to a factor $\sim 30$ compared to conventional tensor\hyp network techniques. 
	We apply these methods to study \discussive{dissipative} quenches, aiming for an in\hyp depth understanding of the formation, stability, and quasi\hyp particle properties of bipolarons.
	Surprisingly, our results show that in the metallic phase, dissipation localizes the bipolarons \discussive{which is reminiscent of an indirect quantum Zeno effect.}
	However, the bipolaronic binding energy remains mainly unaffected, even in the presence of strong dissipation, exhibiting remarkable bipolaron stability.
	These findings shed new light on the problem of designing real materials exhibiting phonon\hyp mediated high\hyp $T_\mathrm{C}$ superconductivity.
\end{abstract}
\maketitle
\section{Introduction}\label{sec:introduction}
Spectrally structured environments are omnipresent in any realistic setup~\cite{DeVegaReview,weissbook}, and it is crucial to understand their effects on quantum many-body systems.
This becomes even more relevant given the remarkable development of experimental platforms such as ultracold quantum gases~\cite{Anderson198,Jaksch1998,Bloch2008,Lim2008,Escobar2009}, high\hyp quality electromagnetic cavities~\cite{Mekhov2007,Murch2008,Groeblacher2009,Verdu2009,Purdy2010,Kollath2016}, time\hyp resolved pump\hyp probe experiments on photosynthetic complexes~\cite{erling_fmo}, and large arrays of superconducting qubits~\cite{Zhang2014,Krantz2019,Arute2019,Yost2020,Zhou2020,Fedele2021,McEwen2022}.
These platforms make it possible to study the effects of structured environments in cleaner setups but also to investigate the possibility to exploit them as a resource to engineering new phenomena in \gls{OQS}~\cite{non_mark_mpo,mark_nonmark_mps_oqs,bofin_heom,hops_stuart_francois,Gao_2022, diss_pt_rabi, signature_diss_pt_rabi}.
The past decades have also seen a rapid development of highly efficient numerical tools, enabling simulations of a large number of quantum mechanical degrees of freedom.
In particular, the \gls{DMRG} in its \gls{MPS} formulation~\cite{White92,White93,schollwoeck_dmrg,schollwoeck_mps_rev} provides a well\hyp established framework in today's efforts with application ranging from (near\hyp) equilibrium studies of low\hyp dimensional lattice systems~\cite{pseudo_site_98,Benthien2004,Heidrich-Meisner2006,Yan2011,Holzner2011,Depenbrock2012,Shirakawa2017,Zheng2017}, out\hyp of\hyp equilibrium simulations following global quenches~\cite{Jeckelmann2002,Laeuchli2008,Langer2009,Manmana2009,Karrasch2013,Essler2014,Sorg2014,Schwarz2018,time_ev_methods,Paeckel2020,Tang2020}, impurity solvers for quantum embedding techniques~\cite{Garcia2004,Guettge2013,Aoki2014,Wolf2014,Wolf2015,Bramberger2021} or as solver in coupled\hyp cluster techniques to study large molecules~\cite{White1999,Marti2010,Lin2011,Wouters2014,Xie2019,singlet_fission}.
Despite its large success on isolated quantum systems, effective numerical schemes to simulate \glspl{OQS} using \gls{MPS} are typically applicable only in the Markovian regime~\cite{Lesanovsky2013,Vega2015,Schroeder2016,Wolff2020,Nueseler2020,Reh2021,Kordas_2015,Atomic_Three_Body_Loss,Kantian_2009,Barmettler,Superoperator_QJUMPS,lindblad_stst,Pichler_2010,Schachenmayer_Pollet}.
\begin{figure*}[!t]
	\centering
	\includegraphics[width=\textwidth]{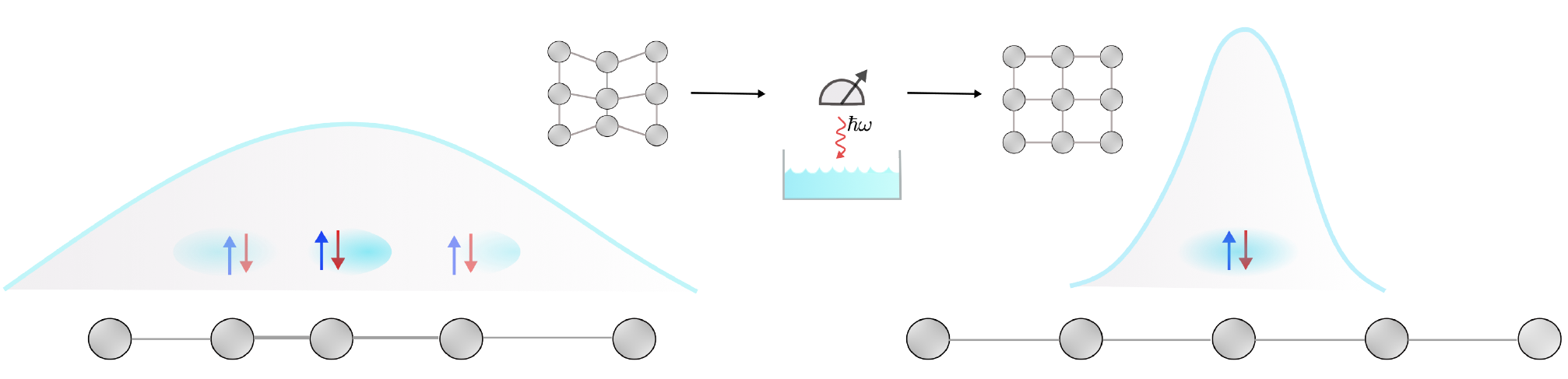}
	\caption
	{
		\label{fig:first_figure} 
		Summary of our main finding: Dissipation tends to localize the bipolarons through effective, non\hyp projective measurements.
		However, in the metallic regime, the bipolaronic binding energy remains mainly unaffected, i.e., bipolarons are stable even for strong dissipation.
	}
\end{figure*}
\begin{figure}[!t]
	\centering
	\ifthenelse{\boolean{buildtikzpics}}%
	{%
		\tikzsetnextfilename{markovian_non_markovian_test}%
		\begin{tikzpicture}%
			\begin{scope} [node distance = 0.25]%
				\node[inner sep = 0pt] (graphic) { \includegraphics[width=0.5\textwidth-9pt, clip, trim = 0 0 963 0]{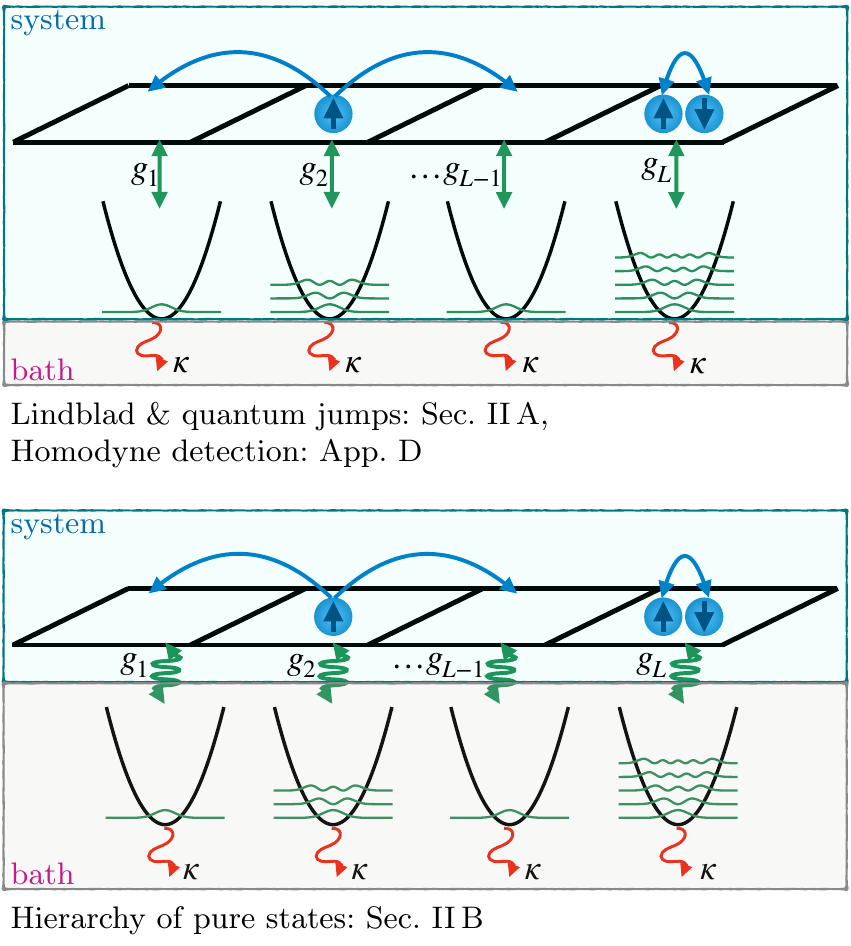} };%
				\node[anchor = north west] at (graphic.north west) {\textcolor{cbColorE}{system}};
				\node[anchor = south west] at (graphic.south west) {\textcolor{cbColorD}{bath}};
				\node [text width = 0.4545\textwidth, below = 0em of graphic.south west, anchor = north west] (text) 
				{
					Lindblad \& \acrlong{QJ}:~\cref{sec:q-jumps}, Homodyne detection:~\cref{app:sec:homodyne-detection}
				};%
				\node[inner sep = 0pt, below = of text.south west, anchor = north west] (graphic) { \includegraphics[width=0.5\textwidth-9pt, clip, trim = 963 0 0 0]{figures/markovian_non_markovian_test} };%
				\node[anchor = north west] at (graphic.north west) {\textcolor{cbColorE}{system}};
				\node[anchor = south west] at (graphic.south west) {\textcolor{cbColorD}{bath}};
				\node [text width = 0.4545\textwidth, below = 0em of graphic.south west, anchor = north west] {\Acrlong{HOPS}:~\cref{sec:hops}};%
			\end{scope}%
		\end{tikzpicture}%
	}%
	{%
		\includegraphics{markovian_non_markovian_test}%
	}%
	\caption
	{
		\label{fig:markovian_non_markovian_test}
		Cartoon of two possible system-bath partitioning of electron-phonon systems.
		Markovian system (left): when considering the electrons and the phonons as the system, the dissipative terms acting on the phonons can be modelled as Markovian.
		Non\hyp Markovian system (right): if only the electrons are treated as the system, the damped phonon modes constitute a non-Markovian bath.
		Below both images, a short list of the Markovian and non-Markovian methods analyzed here is given, together with the corresponding section.
		For the non\hyp Markvian \gls{HEOM} method, we refer to \cite{heom1,heom2}.
	}
\end{figure}
\discussive{This work aims to close the gap between the necessity of unbiased descriptions of \gls{OQS} on the one hand, and numerically efficient lattice representations, operating on the required large local Hilbert spaces on the other hand.
For that purpose, we build upon a recently introduced efficient representation of bosonic Hilbert spaces~\cite{PP,proj_pur_2} as well as both Markovian~\cite{Daley_QJUMPS} and non\hyp Markovian methods for \gls{OQS}~\cite{hops_original,hops_stuart_francois,Gao_2022}.
Based on \gls{MPS} representations, we combined both approaches, yielding a powerful numerical tool to study the impact of dissipation on realistic, phononic, condensed matter models, which were previously out of reach.}
We test and benchmark the obtained methods at the example of the dissipative Hubbard\hyp Holstein model in a large parameter space and explore their applicabilities as a function of the electron\hyp phonon coupling and dissipation strength.
Having these tools at hand, we are able to study the effect of realistic phonon\hyp anharmonicities on electron\hyp phonon quasi-particles (polarons, bipolarons), originating from the dissipative character of the phonons.
Here, our main focus is to answer the question of whether or not dissipation enhances the metallic behaviour of (tightly) bound bipolarons.
We conduct a systematic analysis of their binding energy and effective mass, whose ratio serves as a measure of their metallicity.
Surprisingly, in the strong coupling regime, we find a significant suppression of the metallicity compared to the non\hyp dissipative case, which, however, does not affect the quasiparticles' stability.
We complement these findings by studying the bipolaron delocalization in real space ~\cite{Souza2000}, which has a straightforward generalization to higher dimensions and is directly related to the experimentally\hyp measurable optical conductivity.
The phenomenon of dissipation\hyp induced bipolaron localization is summarized in \cref{fig:first_figure}.
We interpret it as an instance of an indirect, dissipation\hyp induced quantum Zeno effect \cite{zeno_2008,indirec_quantum_zeno, comment_indirec_quantum_zeno}.
In a more general frame, our findings indicate that even though the environment is coupled to the electronic degrees of freedom via an indirect path, there is still significant decoherence.
However, the decoherence is induced only on the level of the quasi\hyp particles of the isolated system, i.e., quasi\hyp particles are not destroyed but rather localized.
The decoherence itself is generated from averaging over the various phonon configurations, a mechanism which is generic to electron\hyp phonon systems.
We, therefore, believe that our findings are relevant to the general situation of mixtures containing phonons, which are coupled to an environment.
%
%

The article is structured as follows. 
In \cref{sec:methods} we briefly review the Markovian \gls{QJ} method \cite{Daley_QJUMPS} and the non\hyp Markovian \gls{HOPS} method \cite{hops_original}, and introduce their efficient \gls{MPS} realization, using the recently developed \gls{PP} mapping. 
Then, in \cref{sec:results} we apply \gls{HOPS} and \gls{QJ} to study the effect of dissipation on the bipolarons in the Hubbard\hyp Holstein model, and in \cref{sec:conclusion} we summarize our findings. 
In \cref{sec:benchmark}, a systematic comparison between \gls{QJ} and \gls{HOPS} can be found.

\section{Methods}\label{sec:methods}
Dissipative electron\hyp phonon systems can be described in two different ways, depending on how they are decomposed into a ``system" and an ``environment". 
Thus, in this section, we present both a Markovian (system = electrons + phonons) and a non-Markovian (system = electrons only) open system method.
These methods can be combined with \gls{MPS} techniques in order to be able to treat many\hyp body systems.
The electron\hyp phonon Hamiltonian we considered takes the form:
\begin{align}
	 \hat{H}^{\nodagger}_{\text{\tiny{tot}}} &= \overbrace{ \underbrace{\hat{H}^{\nodagger}_{\text{\tiny{f}}}}_{\text{non-Markovian sys.}} + \: \hat{H}^{\nodagger}_{\text{\tiny{b}}} \: + \: \hat{H}^{\nodagger}_{\text{\tiny{int}}} }^{\text{Markovian sys.}} \notag\\
	 &= 
	 \hat{H}^{\nodagger}_{\text{\tiny{f}}} +  \sum_{j}  \omega_j^{\nodagger} \hat{a}_{j}^\dagger  \hat{a}_{j}^\nodagger +  \sum_{j} g_j^{\nodagger} \big (\hat{L}_{j}^\nodagger \hat{a}_{j}^\dagger + \hat{L}_{j}^\dagger \hat{a}_{j}^\nodagger \big ) \;,
	 \label{eq:hops:ham-appl}
\end{align}
where $\hat{H}_{\text{\tiny{f}}}^{\nodagger}$ is an arbitrary Hamiltonian acting on the fermionic degrees of freedom, $\hat{H}_b^{\nodagger}$ describes a collection of harmonic oscillators representing the phonons and $\hat{L}$ is an operator acting on the fermions. 
The index $j$ labels the lattice sites and the parameters $\omega_j^{\nodagger}$ and $g_j^{\nodagger}$ are the vibration frequencies of the harmonic oscillators and the electron\hyp phonon coupling constants, respectively. 
In addition to the unitary dynamics described by \cref{eq:hops:ham-appl}, we consider dissipation of the form of phonon losses so that the time evolution of the ``electron+phonon" density matrix is described by the Lindblad master equation \cite{Pearle_lindblad}:
\begin{equation}
 \partial_t \hat{\rho} = -i [\hat{H}^{\nodagger}_{\text{\tiny{tot}}}, \hat{\rho}] +\sum_j \hat{D}^{\nodagger}_j\hat{\rho} \hat{D}^{\dagger}_j -\frac{1}{2} \{ \hat{D}^{\dagger}_j\hat{D}^{\nodagger}_j, \hat{\rho} \} \;,
 \label{eq:Lindblad}
\end{equation}
where $\hat{D}_j = \sqrt{\kappa} \hat{a}_j$ are the corresponding Lindblad operators acting on each phononic lattice site.
\subsection{\Acrlong{QJ}}\label{sec:q-jumps}
In the left panel of \cref{fig:markovian_non_markovian_test} we show a system decomposition where both the electrons and the phonons are part of the physical system, and dissipation acts on the phonons only.
This representation can be modelled as Markovian via the master equation \cref{eq:Lindblad}, which can be rewritten as an evolution for pure states with a stochastic process so that averaging over its samples gives the correct expectation values for the observables.
From a numerical point of view, this is highly beneficial since for each random process one only has to store the $\mathcal{O}(\sqrt{N_{\rho}})$ complex coefficients, with $N_{\rho}$ being the number of entries of the density matrix of the electron\hyp phonon system.
A typical so\hyp called pure state unravelling of the Lindblad equation \cref{eq:Lindblad} is given by the \gls{QJ} method (we discuss a different unravelling, the homodyne detection method in \cref{app:sec:homodyne-detection}).
Working with pure states, a stochastic process $\mathcal Q$ is introduced so that the density matrix, time\hyp evolved by the Lindblad equation, is obtained from averaging over many realizations of the stochastic process:
\begin{equation}
	\mathcal{E} [\ket{\Psi(t)}_{\text{\tiny{q}}} \bra{\Psi(t)}_{\text{\tiny{q}}} ] = \hat{\rho}(t) \;,
 \label{eq:av_over_proj}
\end{equation}
where $q \in \mathcal Q$ is a collection of pseudo\hyp random numbers identifying a so\hyp called trajectory.
Here, every single step $q$ in a trajectory $\mathcal Q$ is specified by (i) deciding if a dissipative event (quantum jump) has to occur and (ii) choosing the lattice site where the jump happens.
Thereby, instead of constructing the density matrix one computes the expectation values of an observable $\hat{O}$ for every trajectory and averages them according to:
\begin{equation}
 \langle \hat{O} \rangle (t) = \mathcal{E} [ \bra{\Psi(t)}_{\text{\tiny{q}}} \hat{O} \ket{\Psi(t)}_{\text{\tiny{q}}} ] \;.
 \label{eq:av_over_observables}
\end{equation}
In \cref{app:sec:quantum-jumps} we provide a detailed derivation of the \gls{QJ} method together with a sketch of the algorithm.
\subsection{\Acrlong{HOPS}}\label{sec:hops}
Another bipartition of \cref{eq:hops:ham-appl} is possible by treating only the electrons as system, wrapping the phononic system into a non\hyp Markovian bath, as shown in the right panel of~\cref{fig:markovian_non_markovian_test}.
Tracing out the phonons in \cref{eq:Lindblad} makes it possible to derive a non-Markovian stochastic Schrödinger equation \cite{Di_si_1997_qsd} for the fermionic degrees of freedom only $\ket{\psi(t)}$:
\begin{align}
 \partial_t \ket{\psi(t)} =& -i \hat{H}_{\text{\tiny{f}}} \ket{\psi(t)} + g \sum_j \hat{L}_j z^*_j(t) \ket{\psi(t)} \notag \\
 &-g \sum_j \hat{L}_j^{\dagger} \int_0^t \mathrm{d} s \,\alpha_j^*(t-s) \frac{\delta  \ket{\psi(t)}}{\delta  z_j^*(s)}.
 \label{eq:non-mark_qsd}
\end{align}
Here $\alpha_j(t)$ represents the environment correlation function, which on site $j$ and at zero temperature is given by the Fourier\hyp transform of the spectral density $J_j(\omega)$.
Furthermore, $z_j(t)$ denotes a colored noise that satisfies $\mathcal{E} \big[ z_j(t) z_{j'}^*(t') \big] = \alpha(t-t') \delta_{j,j'}$, while the term $\delta/\delta z_j^*(s)$ represents the functional derivative with respect to $z^*$.
The observables for the electronic system are then obtained by averaging the dynamics of \cref{eq:non-mark_qsd} over many trajectories. 
In practical calculations, solving \cref{eq:non-mark_qsd} is exceptionally challenging because of the last term of the right\hyp hand side, which is non\hyp local in time \cite{hops_hartmann_strunz}.
This problem can be solved efficiently by the \gls{HOPS} method~\cite{hops_original,hops_stuart_francois}, where one defines:
\begin{equation}
 \ket{\psi^{(1,j)}(t) } = D_j(t) \ket{\psi(t)} \equiv \int_0^t \mathrm{d}s \, \alpha_j^*(t-s) \frac{\delta  \ket{\psi(t)}}{\delta  z_j^*(s)}
\end{equation}
which is labeled \textit{first auxiliary state} relative to site $j$.
One then introduces the \textit{k\hyp th auxiliary state} in a recursive manner:
\begin{equation}
 \ket{\psi^{(k,j)}(t) } = [D_j(t)]^k \ket{\psi(t)} \; ,
\end{equation}
and defines a state on the combined fermionic and bosonic Hilbert space as:
\begin{equation}
	\ket{\Psi(t)} = \sum_{\mathbf{k}} C_{\mathbf{k}}(t) \ket{\psi^{(\mathbf{k})}(t)} \otimes \ket{\mathbf{k}}^{\text{bos}},
\end{equation}
where $\ket{\mathbf{k}}^{\text{bos}} \equiv \otimes_{j} \ket{ k }_j^{\text{bos}}$ labels an effective bosonic mode corresponding to the k\hyp th auxiliary state and $C_{\mathbf{k}}(t)$ is a time\hyp dependent coefficient.
The hierarchy then takes the form of a simple Schr\"odinger equation for the state on the combined fermionic and bosonic Hilbert space (see~\cref{app:sec:hops} for a detailed description and the full representation of the effective Hamiltonian and a sketch of the \gls{HOPS} algorithm).
Being a pure state method, \gls{HOPS} \cite{Suess2015} is more suited for many\hyp body systems than its density matrix formulation, the so\hyp called \gls{HEOM} method \cite{heom1,heom2}. 
Moreover, time\hyp evolving density matrices with \gls{MPS} methods is non\hyp trivial since one needs to guarantee the positivity of $\rho$ at all times \cite{rho_MPS_Cirac}.
In the next section, we present how the open systems methods described above can be hybridized with many\hyp body approaches to tackle the non\hyp Markovian dynamics of many\hyp body systems.
\subsection{\Acrlongpl{MPS} and \Acrlong{PP}}\label{sec:PP}
\Acrlongpl{MPS} \cite{PhysRevB.55.2164,PhysRevLett.99.220405,PhysRevLett.93.040502}, also known as tensor trains, provide well\hyp established numerical representations for 1D quantum many\hyp body systems.
There are efficient \gls{MPS} algorithms available for both ground state \cite{schollwoeck_dmrg,schollwoeck_mps_rev,intro_to_TNs} and time-dependent \cite{time_ev_methods} problems.
Here, we provide a very short introduction to \gls{MPS} and \gls{PP-DMRG}~\cite{PP,proj_pur_2}, focussing on the relevant technical aspects to combine them with \gls{QJ}~\cite{Daley_QJUMPS} and \gls{HOPS}~\cite{hops_original,hops_hartmann_strunz}. We note that the combination of \gls{HOPS} with \gls{MPS} was originally presented in \cite{hops_stuart_francois}.
Importantly, exploiting the \gls{PP} mapping is required to treat the large local bosonic Hilbert spaces efficiently and thus rendering the discussed \gls{OQS}\hyp techniques suitable for \gls{MPS} algorithms.
For any pure state with $L$ sites and a finite number of local degrees of freedom $\sigma_1,\sigma_2, ...,\sigma_L$ ($\sigma_i = 1,2,...,d_i$ with local dimensions $d_i$) the coefficient tensor $c_{\sigma_1,\sigma_2, ...,\sigma_L}$ can be reshaped as
\begin{align}
 \ket{\Psi} &= \sum_{\sigma_1,...,\sigma_L} c_{\sigma_1\cdots\sigma_L} \ket{\sigma_1\cdots\sigma_L} \xrightarrow{} \ket{\Psi}_{\text{\tiny{MPS}}} \notag\\
 &= \sum_{\substack{ \sigma_1,...,\sigma_L \\ m_0,...,m_L }} M^{\sigma_1}_{1;m_0,m_1} \cdots M^{\sigma_L}_{L;m_{L-1},m_L} \ket{\sigma_1\cdots\sigma_L} \;,
 \label{eq:general_mps}
\end{align}
where $\{ M^{\sigma_i}_{i;m_{i-1},m_i} \}$ are $m_{i-1} \times m_i$ rectangular matrices. 
This representation has two main advantages: it allows for optimal and physically motivated compression of the state via \glspl{SVD} and decomposes the coefficient tensor into local objects, which, moreover, can be related to the system\hyp environment picture of the original \gls{DMRG}\cite{White92,White93}.
\Glspl{MPS} and \glspl{MPO}, which follow the same structure, are often represented graphically in terms of tensor network diagrams.
Therein, geometric shapes represent the rank\hyp 3 or rank\hyp 4 tensors.
It is essential to note that the dimensions of the \gls{MPS} tensors on some site $j$, called bond dimensions $m_j$, typically grow exponentially with the entanglement when partitioning the system at the sites $j-1,j$. 
When it comes to time\hyp evolution methods, \gls{TDVP}~\cite{Haegeman2011,Haegeman2016} is a well\hyp established technique, which is based on the Dirac\hyp Frenkel variational principle and consists of subsequently updating a small number (typically one or two) of site\hyp tensors~\cite{time_ev_methods}. 
One must bear in mind, however, that in its original formulation, this method is particularly prone to cause significant errors when used for time\hyp evolving a product state with a large local Hilbert space dimension~\cite{time_ev_methods}.
Clearly, ~\gls{MPO}\hyp based techniques, such as the \gls{TEBD}~\cite{PhysRevLett.93.040502} or the $W^{\mathrm{I,II}}$~\cite{PhysRevB.91.165112}, can overcome this limitation, but are also suffering from systematic Trotter errors~\cite{Suzuki1993}.
However, we found it to be sufficient to time\hyp evolve the state with the slower but more accurate global Krylov method~\cite{time_ev_methods} up to the point where the bond dimension is as large as the local Hilbert space dimension and then to switch to \gls{TDVP}.

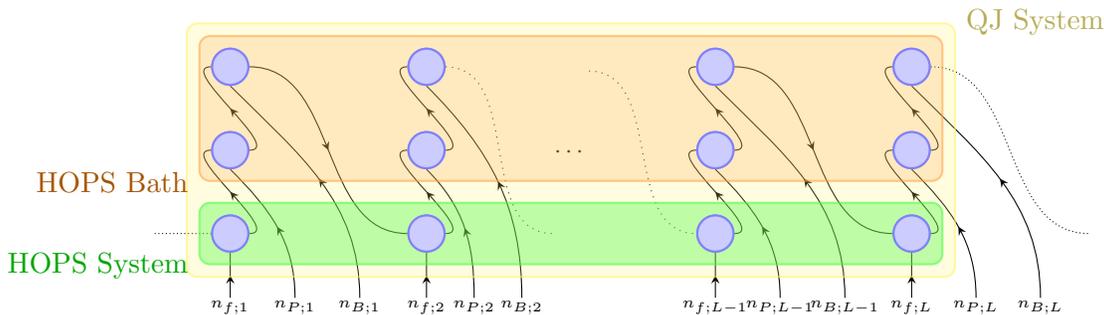
\begin{figure*}[!t]
	\centering
	\ifthenelse{\boolean{buildtikzpics}}%
	{%
 		\tikzsetnextfilename{mps2tripleQCvsHOPS}%
		\begin{tikzpicture}%
			\begin{scope}[node distance = 0.6 and 2.1]%
				\node[ghost] (site0a) {};%
				\node[site] (site1a) [right=0.75 of site0a] {};%
				\node[site] (site1b) [above=of site1a] {};%
				\node[site] (site1c) [above=of site1b] {};%
				\node[site] (site2a) [right=of site1a] {};%
				\node[site] (site2b) [above=of site2a] {};%
				\node[site] (site2c) [above=of site2b] {};%
				\node[unsite] (site4a) [right=1.425 of site2a] {};%
				\node[unsite] (site4b) [above=of site4a] {};%
				\node[unsite] (site4c) [above=of site4b] {};%
				\node[ghost] at ($(site4a)!0.5!(site4c)$) {$\cdots$};%
				\node[site] (site5a) [right=1.425 of site4a] {};%
				\node[site] (site5b) [above=of site5a] {};%
				\node[site] (site5c) [above=of site5b] {};%
				\node[site] (site6a) [right=of site5a] {};%
				\node[site] (site6b) [above=of site6a] {};%
				\node[site] (site6c) [above=of site6b] {};%
				\node[unsite] (site7a) [right=of site6a] {};%
				\node[ld] (sigma1a) [below=of site1a] {\tiny$\ingoing{n}_{f;1}$};%
				\node[ld] (sigma2a) [below=of site2a] {\tiny$\ingoing{n}_{f;2}$};%
				\node[ld] (sigma4a) at (site4a|-sigma2a) {};%
				\node[ld] (sigma5a) at (site5a|-sigma2a) {\tiny$\ingoing{n}_{f;L-1}$};%
				\node[ld] (sigma6a) at (site6a|-sigma2a) {\tiny$\ingoing{n}_{f;L}$};%
				\node[ld] (sigma7a) at (site7a|-sigma2a) {};%
				\node[ld] (sigma1b) at ($(sigma1a)!0.33!(sigma2a)$) {\tiny$\ingoing{n}_{P;1}$};%
				\node[ld] (sigma2b) at ($(sigma2a)!0.33!(sigma4a)$) {\tiny$\ingoing{n}_{P;2}$};%
				\node[ld] (sigma5b) at ($(sigma5a)!0.33!(sigma6a)$) {\tiny$\ingoing{n}_{P;L-1}$};%
				\node[ld] (sigma6b) at ($(sigma6a)!0.33!(sigma7a)$) {\tiny$\ingoing{n}_{P;L}$};%
				\node[ld] (sigma1c) at ($(sigma1a)!0.66!(sigma2a)$) {\tiny$\ingoing{n}_{B;1}$};%
				\node[ld] (sigma2c) at ($(sigma2a)!0.66!(sigma4a)$) {\tiny$\ingoing{n}_{B;2}$};%
				\node[ld] (sigma5c) at ($(sigma5a)!0.66!(sigma6a)$) {\tiny$\ingoing{n}_{B;L-1}$};%
				\node[ld] (sigma6c) at ($(sigma6a)!0.66!(sigma7a)$) {\tiny$\ingoing{n}_{B;L}$};%
				\draw[densely dotted] (site0a) -- (site1a);%
				\draw[->-,out=0,in=180] (site1a.east) to (site1b.west);%
				\draw[->-,out=0,in=180] (site2a.east) to (site2b.west);%
				\draw[->-,out=0,in=180] (site5a.east) to (site5b.west);%
				\draw[->-,out=0,in=180] (site6a.east) to (site6b.west);%
				\draw[->-,out=0,in=180] (site1b.east) to (site1c.west);%
				\draw[->-,out=0,in=180] (site2b.east) to (site2c.west);%
				\draw[->-,out=0,in=180] (site5b.east) to (site5c.west);%
				\draw[->-,out=0,in=180] (site6b.east) to (site6c.west);%
				\draw[->-,out=0,in=180] (site1c.east) to (site2a.west);%
				\draw[->-,out=0,in=180] (site5c.east) to (site6a.west);%
				\draw[out=0,in=180,densely dotted] (site6c.east) to (site7a.west);%
				\draw[out=0,in=180,dotted] (site2c.east) to (site4a.west);%
				\draw[out=0,in=180,dotted] (site4c.east) to (site5a.west);%
				\draw[->-] (sigma1a) to (site1a);%
				\draw[->-] (sigma2a) to (site2a);%
				\draw[->-] (sigma5a) to (site5a);%
				\draw[->-] (sigma6a) to (site6a);%
				\draw[->-,out=90,in=-45] (sigma1b.north) to (site1b.south);%
				\draw[->-,out=90,in=-45] (sigma2b.north) to (site2b.south);%
				\draw[->-,out=90,in=-45] (sigma5b.north) to (site5b.south);%
				\draw[->-,out=90,in=-45] (sigma6b.north) to (site6b.south);%
				\draw[->-,out=90,in=-45] (sigma1c.north) to (site1c.south);%
				\draw[->-,out=90,in=-45] (sigma2c.north) to (site2c.south);%
				\draw[->-,out=90,in=-45] (sigma5c.north) to (site5c.south);%
				\draw[->-,out=90,in=-45] (sigma6c.north) to (site6c.south);%
				\node[draw,,fill=green!75,draw=green!75,fill opacity=.2,draw opacity=0,thick,rounded corners,fit=(site1a) (site6a),inner sep=0.4em] (ferm) {};%
				\node[draw,,fill=orange!75,draw=orange!75,fill opacity=0,draw opacity=0,thick,rounded corners,fit=(site1b) (site6c),inner sep=0.4em] (bos) {};%
				\node[draw,,fill=yellow!75,draw=yellow!75,fill opacity=0.2,draw opacity=.6,thick,rounded corners,fit=(ferm) (bos),inner sep=0.4em] (full) {};%
				\node[draw,,fill=green!75,draw=green!75,fill opacity=.2,draw opacity=.6,thick,rounded corners,fit=(site1a) (site6a),inner sep=0.4em] (ferm) {};%
				\node[draw,,fill=orange!75,draw=orange!75,fill opacity=.2,draw opacity=.6,thick,rounded corners,fit=(site1b) (site6c),inner sep=0.4em] (bos) {};%
				\node[site] (site1a) at (site1a) {};%
				\node[site] (site1b) at (site1b) {};%
				\node[site] (site1c) at (site1c) {};%
				\node[site] (site2a) at (site2a) {};%
				\node[site] (site2b) at (site2b) {};%
				\node[site] (site2c) at (site2c) {};%
				\node[site] (site5a) at (site5a) {};%
				\node[site] (site5b) at (site5b) {};%
				\node[site] (site5c) at (site5c) {};%
				\node[site] (site6a) at (site6a) {};%
				\node[site] (site6b) at (site6b) {};%
				\node[site] (site6c) at (site6c) {};%
				\node[anchor=east, green!65!black] at (ferm.south west) {\gls{HOPS} System};%
				\node[anchor=east, orange!65!black] at (bos.south west) {\gls{HOPS} Bath};%
				\node[anchor=west, yellow!65!black] at (full.north east) {\gls{QJ} System};%
			\end{scope}%
		\end{tikzpicture}%
	}%
	{%
		\includegraphics{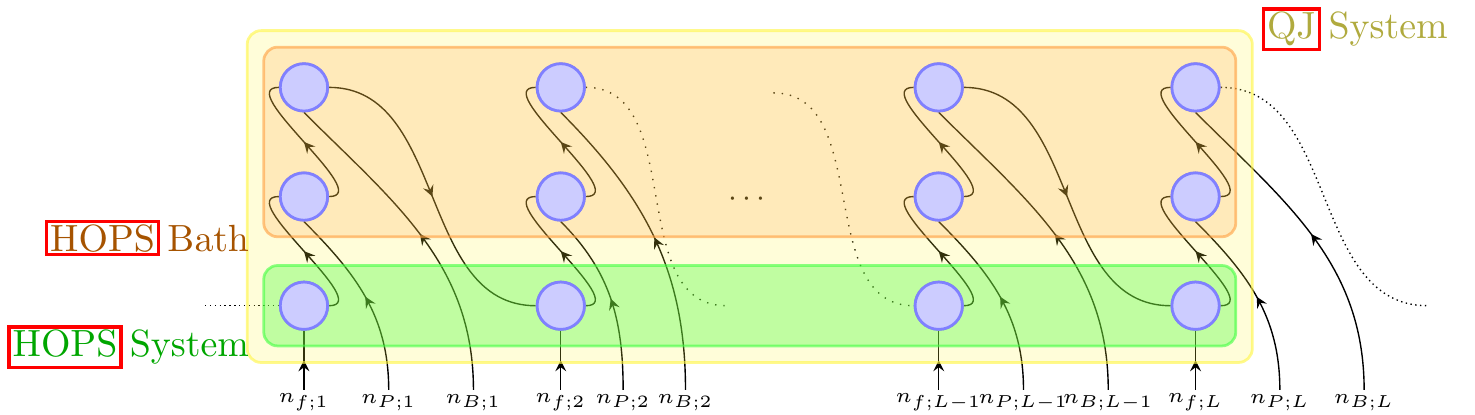}%
	}%
	\caption
	{
		\label{fig:mpstensornetwork}\label{fig:dummy_mps_hops_qjumps} 
		\Gls{MPS} representation in an enlarged Hilbert space with each physical site consisting of a physical fermionic, a physical bosonic, and a bosonic bath site.
		The circles labelled by $n_f$ (bottom row) represent the fermions, while the ones labelled by $n_P$ and $n_B$ represent the physical bosons and the projected\hyp purified bath bosons, respectively (middle and upper top).
		The background colors indicate that, as explained in the main text, for \gls{QJ} the bosonic sites correspond to actual physical phonons while for \gls{HOPS} they are related to the auxiliary states.
		Adapted from \cite{PP}.
	}
\end{figure*}
The description of bosonic degrees of freedom has posed substantial challenges to \gls{MPS} methods because of their infinite\hyp dimensional Hilbert spaces.
Much work has been devoted to an accurate and efficient truncation of bosonic Hilbert spaces, resulting in successful techniques such as the \gls{PS} method \cite{pseudo_site_98} and the \gls{LBO} method \cite{lbo_fhm_2015,lbo_bursill,lbo_friedmann,lbo_wong,Jansen2022}. 
In this context, a newly\hyp introduced \gls{MPS} method is the so\hyp called \acrlong{PP}\cite{PP}. 
For the class of Hamiltonians described by \cref{eq:hops:ham-appl}, the electron\hyp phonon interaction term $\hat{H}_{\text{\tiny{int}}}$ does not conserve the number of phonons. 
The breaking of the associated $U(1)$ symmetry prevents the site tensors of the \gls{MPS} from having a block\hyp diagonal structure, resulting in a significant slow\hyp down of matrix operations \cite{TN_global_symm}. 
For a thorough presentation of the method, we refer to Refs.~\cite{PP}. 
The main idea of the \gls{PP} method is to restore the $U(1)$ symmetry artificially by doubling the bosonic Hilbert space, precisely as one does for the thermal purification method \cite{rho_MPS_Cirac} (see \cref{fig:mpstensornetwork}), and to modify the bosonic creation and annihilation operators as follows:
\begin{equation}
 \begin{array}{l}
 \hat{a}^{\dagger}_j \xrightarrow{} \hat{a}^{\dagger}_{P;j} \otimes \hat{b}^{\nodagger}_{B;j} \\
 \hat{a}_j \xrightarrow{} \hat{a}^{\nodagger}_{P;j} \otimes \hat{b}^{\dagger}_{B;j}\;, 
 \end{array} 
 \label{eq:balancing_ops}
\end{equation}
where $\hat{b}^{\nodagger}_{B;j}$, $\hat{b}^{\dagger}_{B;j}$ are the bare operators defined in \cref{eq:bare_ops} of \cref{app:sec:hops}. 
Accompanied by this transformation, a local gauge condition on the allowed states is imposed, i.e., on each pair of physical and bath sites, the sum of the number of physical particles $n_P$ and bath particles $n_B$ has to be conserved $n_P + n_B = n_\text{\tiny{ph,max}}-1$, where $n_\text{\tiny{ph,max}}$ is the maximal phononic local Hilbert space dimension.
The second key ingredient of the \gls{PP} method consists in adopting a truncation method for the local Hilbert space dimension of the phononic sites that is analogous to the one exploited by \gls{MPS} algorithms for truncating the bond dimension. 
Thereby, imposing a discarded weight $\delta$, defined as the maximally allowed leakage of spectral weight for density matrices belonging to any lattice bipartition, determines a truncation in both the physical dimensions and the bond dimensions.
Thus, if the diagonal elements of the phononic reduced density matrices decay fast enough, truncations can reduce the actually used local dimensions: $d_{\mathrm{max}} \le n_\mathrm{ph, max}$.

\begin{figure}[!t]
	\centering
	\ifthenelse{\boolean{buildtikzpics}}%
	{%
		\input{hops_qj_pp_speedup}
	}%
	{%
		\includegraphics{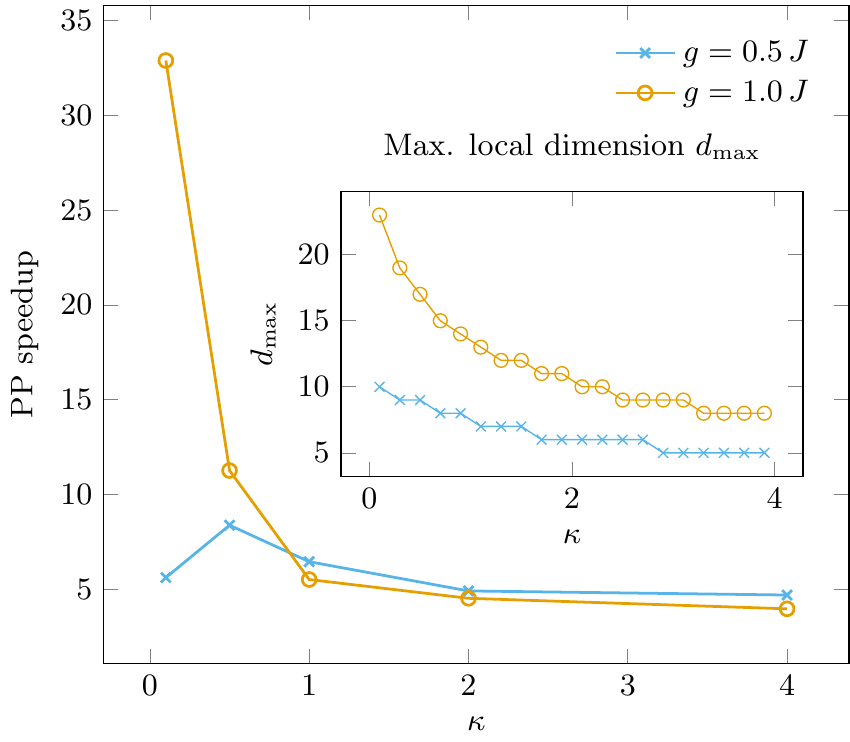}%
	}%
	\caption%
	{%
		\label{fig:PP_speedup}%
		\Gls{PP} speedup for intermediate (blue) and strong (orange) electron-phonon coupling as a function of the dissipation strengths.
		The speedup factor is significant for large electron-phonon couplings and small dissipation strength, corresponding to a large bosonic local Hilbert space dimension as indicated in the inset.
		The time evolutions were performed with the \gls{QJ} method for systems with $L=20$ sites for a single trajectory.
		All other parameters were the same as those described in \cref{fig:k_max_surface}.
	}%
\end{figure}
From a more general point of view, it is the decay of the \gls{1RDM} diagonal elements $\rho_{\sigma^\noprime_j, \sigma^\noprime_j}$ that controls the possible speed up generated by the \gls{PP} mapping.
Therefore, while large local dimensions are doable within \gls{PP-DMRG}, in practice, one has to check for converged diagonal elements of the \gls{1RDM} and, if required, increase the maximally allowed local dimension to keep the truncation error $\delta$ at an acceptable level.
In~\cref{sec:benchmark} we present a detailed benchmark and convergence analysis of the \gls{PP}\hyp enhanced \gls{HOPS} and \gls{QJ} methods for the dissipative Hubbard\hyp Holstein model.
We want to emphasize that in certain parameter regimes, both a rescaling of the auxiliary states for \gls{HOPS} (\cref{eq:new_auxiliary_states} and \cref{fig:new_aux_states}) and a very large phononic Hilbert space dimension (made manageable by \gls{PP}) are critical for reliably computing fermionic observables (\cref{fig:4_points_dim_conv}).
Most importantly, both methods are numerically stable and well\hyp controlled in different physical situations, rendering a combination of both an ideal toolset for studying \gls{OQS} dynamics.
While \gls{QJ} allows for an efficient simulation of weak\hyp and intermediate\hyp dissipation, \gls{HOPS} reveals its strengths when considering the limit of intermediate to strong dissipation.
However, both methods benefit significantly when combined with \gls{PP}.
In \cref{fig:PP_speedup} we illustrate the speedup provided by adopting the \gls{PP} mapping for a system of $L=20$ lattice sites and one trajectory (note that larger system sizes were out of reach for the reference calculations).
We find that the runtime is significantly reduced when using the \gls{PP} mapping for all analyzed parameters.
In particular, we observe a substantial speed\hyp up of a factor of $\sim 30$ for small dissipation and strong electron\hyp phonon coupling, i.e., for large local Hilbert space dimensions $>20$, while it is less significant for medium and strong dissipation (factor $\sim 5$).
Note that the reduced speedup in the strongly dissipating regime is not severe for the overall runtime.
This can be attributed to the fact that strong dissipation naturally reduces the correlations in the system and thus the bond dimension, too (c.f., \cref{fig:bond_dim} in \cref{sec:benchmark}).
Therefore, combining \gls{QJ} with \gls{PP} allows for the numerically efficient application of \gls{QJ} in exactly that parameter regime, where \gls{QJ} was also found to be the method of choice.
\section{Impact of dissipation on bipolaronic quasi\hyp particles}\label{sec:results}
The Hubbard\hyp Holstein Hamiltonian describes spinful fermions coupled to Einstein phonons \cite{Holstein1959}.
We consider the one\hyp dimensional case of the form of \cref{eq:hops:ham-appl} that reads: 
\begin{align}
	\hat{ H}_{\text{\tiny{HH}}}
	=&
	-J\sum_{j=1}^L \sum_{\sigma= \uparrow, \downarrow} \left( \hat c^\dagger_{j,\sigma} \hat c^\nodagger_{j+1,\sigma} + \text{h.c.} \right)
	+
	U \sum_{j=1}^L \hat n^\nodagger_{j,\uparrow} \hat n^\nodagger_{j,\downarrow}
	\notag\\
	&+
	\omega \sum_{j=1}^L \hat a^\dagger_j \hat a^\nodagger_j
	+
	g\sum_{j=1}^L \left( \hat a^\nodagger_j + \hat a^\dagger_j \right) \hat n^\nodagger_j \;.
	\label{eq:hubbard-holstein-hamiltonian}
\end{align}
Here, $U$ denotes the onsite Hubbard\hyp interaction while $g$ measures the electron\hyp phonon coupling, and the phonon frequency is given by $\omega$.
In the following, we fix $J$ as the unit of energy and $J^{-1}$ as the unit of time.
Despite its conceptional simplicity, \cref{eq:hubbard-holstein-hamiltonian} provides a minimal model for the complex interplay between lattice vibration and electronic degrees of freedom in the strong coupling regime.
Such a physical situation occurs, for instance, in Alkali\hyp doped $\mathrm{C}_{60}$ fullerene molecules~\cite{Hebard1991,Tanigaki1991}, a class of unconventional superconductors that recently has been investigated for optically induced superconductivity~\cite{Mitrano2016,Nava2018,Budden2021}.
However, understanding, in particular, the regime of competing (spinless) fermion\hyp phonon and onsite Hubbard\hyp interaction remains a challenging numerical task even in equilibrium, with lots of numerical effort conducted in the past decade~\cite{Tezuka,Clay_Hubb_Hol,Hubb_hol_spin_charge,lbo_fhm_2015,Weber2015,Lavanya2017,Reinhard2019,Stolpp2020,Jansen2021,Brink2022}.
We aim to push the limit towards complete microscopic modelling of the out\hyp of equilibrium dynamics, incorporating the effect of dissipation on a strongly\hyp correlated quantum many\hyp body system with up to $L=40$ lattice sites.
We note that the dissipative Hubbard\hyp Holstein model considered here can be derived from a more general perspective, where the electronic degrees of freedom are coupled to a global bosonic environment (see \cref{app:sec:phys_motiv}).
We emphasize that in contrast to previous works, we made no strong assumptions about the phonons to render it more tractable~\cite{Lavanya2017,Debnath2021}. 
\begin{figure}[!t]
	\centering
	\subfloat[\label{fig:hh-phase-diag}]{%
		\ifthenelse{\boolean{buildtikzpics}}%
		{%
			\tikzsetnextfilename{tezuka-phaseboundary}%
			\begin{tikzpicture}%
				\begin{axis}%
				[%
					width = 0.55\textwidth-17.1pt,%
					height = 0.3\textheight,%
					xmin = 0.,%
					xmax = 1.5,%
					ymin = 0.,%
					ymax = 4.5,%
					xlabel = {$U\, [J]$},%
					ylabel = {$\nicefrac{2g^2}{\omega}\, [J]$},%
				]%
					\addplot [draw = none, name path = BottomLine] {0};%
					\addplot [draw = none, name path = TopLine] {4.5};%
					\addplot%
					[%
						colour = cbColorB!30!cbColorD,
						thick,%
						dashed,%
						name path = SdwToMetallic,%
					]%
						{x};%
					\addplot%
					[%
						colour = cbColorD!50!cbColorE,
						thick,%
						dotted,%
						name path = MetallicToCdw,%
					]%
						table%
						[%
							x expr = \thisrowno{0},%
							y expr = \thisrowno{1},%
						]%
							{data/tezuka-cdw-phase-boundary.dat};%
					\addplot [fill=cbColorB, opacity = 0.1] fill between[of=BottomLine and SdwToMetallic];%
					\addplot [fill=cbColorD, opacity = 0.1] fill between[of=SdwToMetallic and MetallicToCdw];%
					\addplot [fill=cbColorE, opacity = 0.1] fill between[of=MetallicToCdw and TopLine];%
					\node [text = cbColorB,] at (axis cs: 1.3,0.5) {SDW};%
					\node [text = cbColorD,] at (axis cs: 0.7,2.25) {metallic};%
					\node [text = cbColorE,] at (axis cs: 0.4,3.8) {CDW};%
					\node [circle, fill=cbColorB!20, draw=cbColorB!50, thick, inner sep=1.5pt] at (axis cs: 1.0,0.5) {};%
					\node [circle, fill=cbColorD!20, draw=cbColorD!50, thick, inner sep=1.5pt] at (axis cs: 1.0,1.5) {};%
					\node [circle, fill=cbColorE!20, draw=cbColorE!50, thick, inner sep=1.5pt] at (axis cs: 1.0,4.0) {};%
					\draw [->, thick, cbColorB, bend right] (axis cs: 1.0,0.0) to (1.0,0.5);%
					\draw [->, thick, cbColorD, bend right] (axis cs: 1.0,0.0) to (1.0,1.5);%
					\draw [->, thick, cbColorE, bend right] (axis cs: 1.0,0.0) to (1.0,4.0);%
					\coordinate (coSDW) at (axis cs: 1.0, 0.55);%
					\coordinate (coMET) at (axis cs: 1.0, 1.5);%
					\coordinate (coCDW) at (axis cs: 1.0, 4.5);%
				\end{axis}%
			\end{tikzpicture}%
		}%
		{%
			\includegraphics{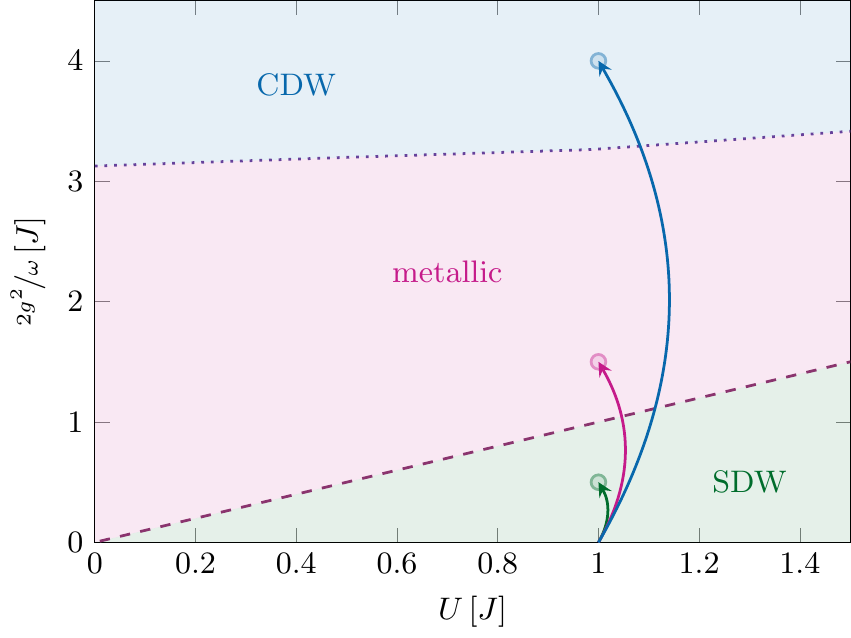}%
		}%
    }%
    \hfill%
    \subfloat[\label{fig:hops:quench:hubbard-gs:double-occupations}]{
        \ifthenelse{\boolean{buildtikzpics}}%
        {%
        	\input{hops_quench_time_dependent_double_occupancies}%
        }%
	    {%
			\includegraphics{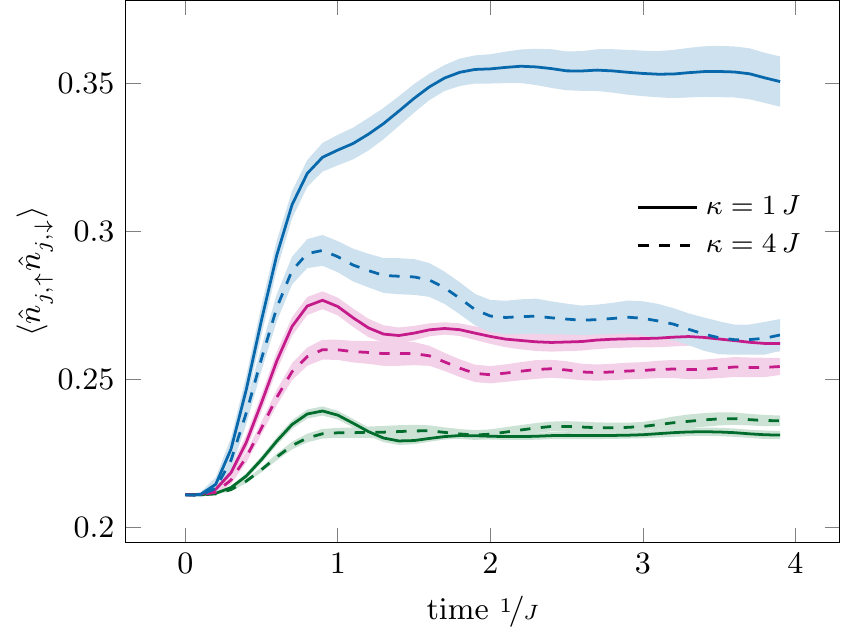}%
		}%
    }
    \caption
    {
        \protect\subref{fig:hh-phase-diag} phase diagram of the Hubbard\hyp Holstein model at a constant phonon oscillation frequency $\omega=2 \, J$ adapted from Ref.~\protect\cite{Tezuka}.
        The coloured arrows indicate quenches from the Hubbard ground state at $U=J$ into the \acrshort{SDW}, the metallic and the \acrshort{CDW} phase.
        \protect\subref{fig:hops:quench:hubbard-gs:double-occupations} double occupancy dynamics after global quenches from the Hubbard ground state.
        They were performed with \gls{HOPS} at intermediate dissipation $\kappa=J$ (solid line) and strong dissipation $\kappa=4 \, J$ (dashed line). 
        The double occupancy dynamics in the electron systems depend more strongly on the phonon loss rate for large electron-phonon coupling $g$.
%
    }
\end{figure}
The phase diagram of the Hubbard\hyp Holstein model at half\hyp filling sketched in~\cref{fig:hh-phase-diag} has been investigated comprehensively, and in the regime of large phonon frequencies, the picture of three different phases has been established~\cite{Takada2003,Clay_Hubb_Hol,Koller2005,Hardikar2007,Tezuka,Greitemann2015,sam_thesis}.
In the limit of vanishing electron\hyp phonon coupling $g/U \rightarrow 0$, a correlated \gls{SDW} phase exists, reminiscent of a Hubbard Mott phase.
In the opposite limit $g/U \rightarrow \infty$, strong phonon fluctuations drive the system into a Peierls state, usually referred to as \gls{CDW} phase.
This limit is understood most easily when transforming the Hubbard\hyp Holstein model into a polaronic description through a Lang\hyp Firsov transformation~\cite{Lang1963}.
Then, the Hubbard on\hyp site interaction is renormalized by the phonons as $U \rightarrow U - \frac{2g^2}{\omega}$ and for sufficiently large electron\hyp phonon couplings, a dominant attractive interaction between the polarons features a spontaneous breaking of the system's translational symmetry.
For intermediate couplings $U \sim \frac{2g^2}{\omega}$, the competition between attractive phonon\hyp mediated polaron\hyp polaron and repulsive electron\hyp electron interactions drive the system towards a metallic Luther\hyp Emery phase~\cite{Greitemann2015}.
There has been a vivid debate about whether this metallic regime may also realize superconductivity, with today's assessment being that superconducting correlations are always subdominant, compared to charge\hyp correlations~\cite{Tezuka, Greitemann2015}.
However, when incorporating gaussian or quartic anharmonicities in the phonon potentials, a strengthening of the metallic behaviour has been observed and the question of whether anharmonic phonons may even drive the Hubbard\hyp Holstein model into a superconducting state arises~\cite{hubb_hol_anharmonicity, Frick1991HightemperatureSI,Lavanya2017}.
Here, we study the effect of a realistic source of anharmonicities, namely a dissipative coupling of the phonons to an environment.
\paragraph{Dissipation and double occupancy.}
Previously, the effect of dissipation in the Hubbard\hyp Holstein model has been investigated using \gls{HOPS}, reporting an enhancement of superconducting correlations following a quench from a Neel state \cite{hops_stuart_francois}. 
We connect to these findings and evaluate the dynamics of the double occupancy $\braket{\hat n^\nodagger_{j,\uparrow} \hat n^\nodagger_{j,\downarrow}}$.
As the initial state, we choose the ground state of the Hubbard model ($g=0$, $\kappa=0$) at $U=J$, and perform a quench to a point in the \gls{SDW} phase ($\nicefrac{2g^2}{\omega}=0.5 \,J$), one in the metallic phase ($\nicefrac{2g^2}{\omega}=1.5\,J$), and one in the \gls{CDW} phase ($\nicefrac{2g^2}{\omega}=4 \,J$).
As a method, we use \gls{HOPS}, which is particularly tailored for quenching in both $\kappa$ and $g$.
In \cref{fig:hops:quench:hubbard-gs:double-occupations}, we show the dynamics of the double occupancy on the central site of a $20$\hyp electron system for intermediate ($\kappa=J$) and strong dissipation ($\kappa=4 \,J$).
Quenching into the \gls{SDW} regime of the Hubbard\hyp Holstein phase diagram (green curve), we find only a weak dependency on the dissipation strength.
This is consistent with dominant spin\hyp spin correlations in the \gls{SDW} phase, which are relatively insensitive to the phonon occupations.
On the other hand, quenching into the \gls{CDW} regime of the Hubbard\hyp Holstein phase diagram (blue curve), there is a strong dependency on the dissipation.
This can be understood by noting that strong phonon fluctuations drive charge correlations and the formation of double occupations in the Peierls phase.
However, increasing the dissipation strength allows the phonons to escape the system, weakening charge correlations.
Surprisingly, the quenches into the metallic regime (purple curve) resemble the behaviour found in the \gls{SDW} quenches.
The weak dependency on the dissipation strength indicates a strong suppression of charge correlations, already for moderate dissipation, an observation that counteracts the reported observation of enhanced metallicity driven by gaussian or quartic phonon anharmonicities~\cite{hubb_hol_anharmonicity, Frick1991HightemperatureSI,Lavanya2017}.
On the other hand, these findings are still consistent with enhanced superconducting correlations~\cite{hops_stuart_francois}.
\begin{figure*}[!t]
	\centering
	\ifthenelse{\boolean{buildtikzpics}}%
	{%
		\input{qj_quench_time_dependent_single_particle_energies}%
	}%
	{%
		\includegraphics{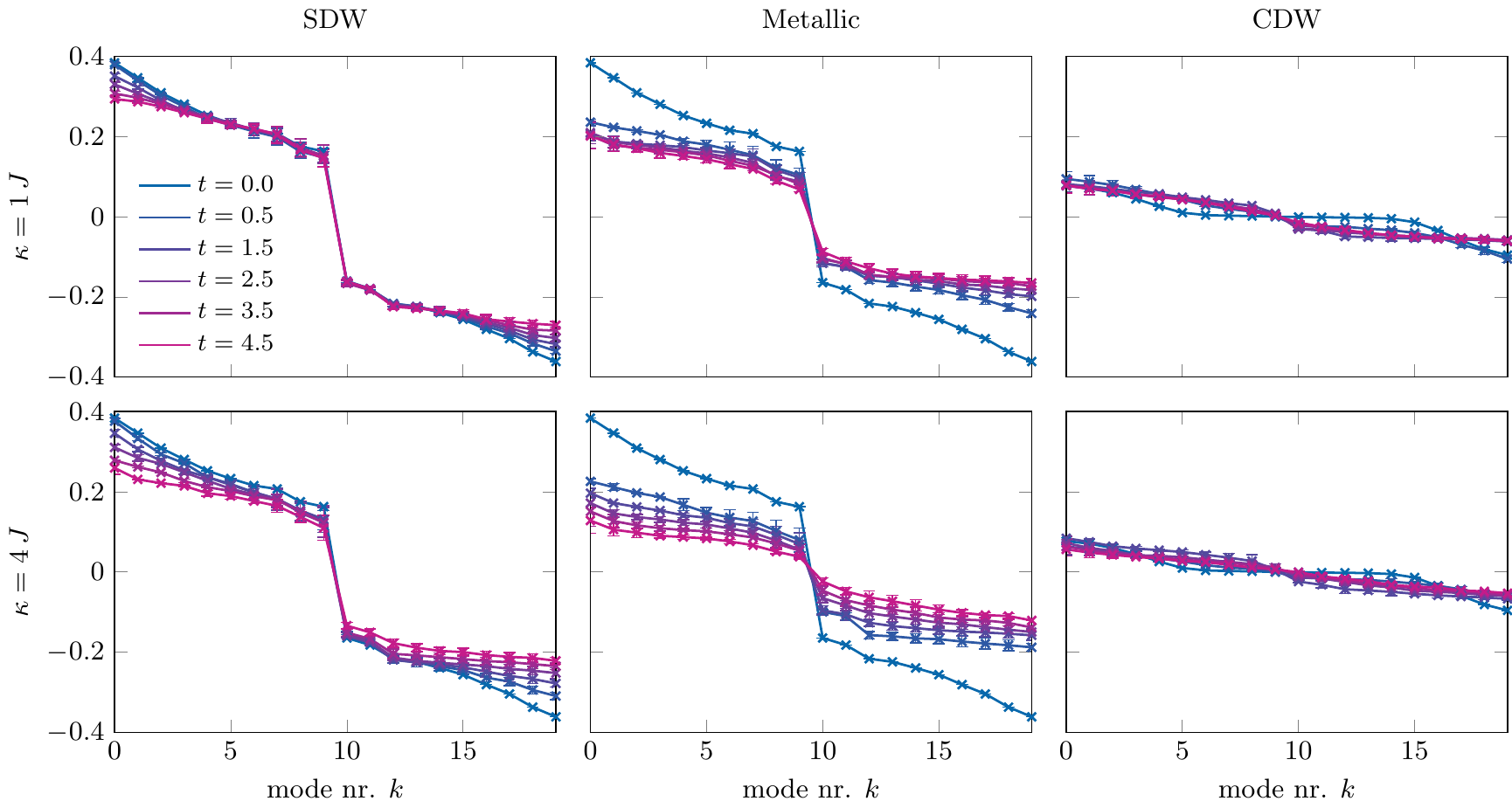}%
	}%
	\caption
	{
		\label{fig:qj:quench:hubbard-holstein-gs:bipolarons:sp-ekin}
		Eigenvalues $t_k$ of bipolaronic hopping matrix $\hat T^{bp}_{i,j,\sigma}$ as a function of time, after turning on dissipation to $\kappa=1\,J$ and $\kappa=4\,J$ in the three different regions of the ground\hyp state phase diagram (indicated by the three circles in \cref{fig:hh-phase-diag} ).
		In the \gls{SDW} phase, a large single\hyp particle gap indicates a small bipolaronic effective mass, while the flat band in the \gls{CDW} phase represents heavy bipolarons. 
		Both phases are basically insensitive to dissipation. 
		In the metallic phase, the gap closing shows that strong dissipation $\kappa=4$ significantly increases the bipolarons' effective mass.
		We simulated a system with $L=20$ sites with \gls{QJ}, with timestep $dt=0.01 \, J^{-1}$ and computed $\lvert \mathcal Q \rvert = 200$ trajectories, using $k_{\text{max}}=40$ local basis states, a max.~bond dimension of $m = 2000$ fixing the discarded weight to $\delta=10^{-10}$.
	}
\end{figure*}
\begin{figure}[h]
	\centering
	 \ifthenelse{\boolean{buildtikzpics}}%
	 {%
	\input{qj_quench_time_dependent_mobilities}%
	 }%
	 {%
	 	\includegraphics{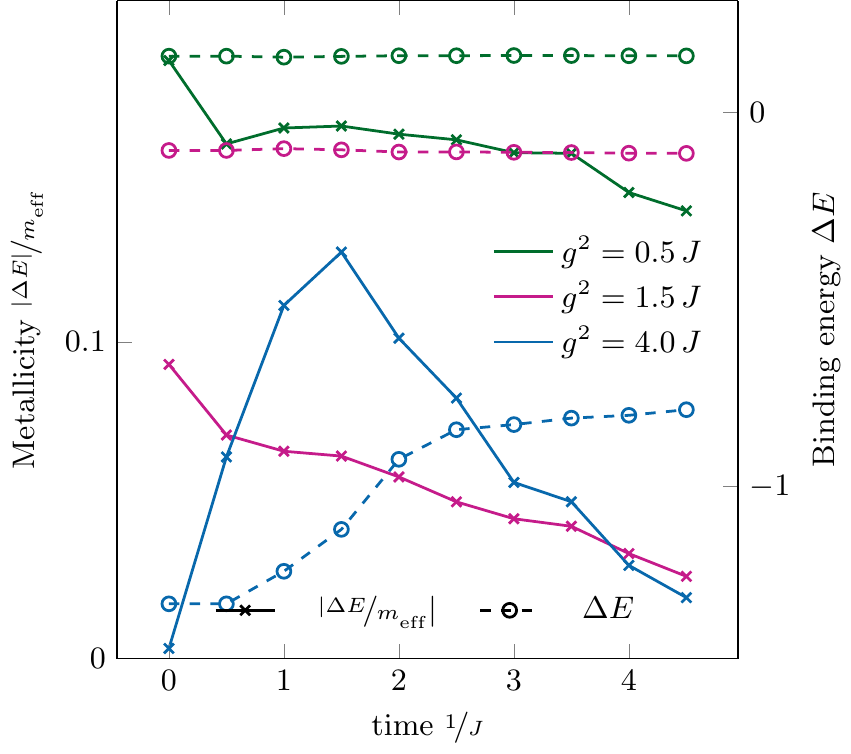}%
	 }%
	\caption%
	{%
		\label{fig:qj:quench:hubbard-holstein-gs:mobilities}%
		Binding energy (circles) and metallicity (crosses), after the dissipative quenches at $\kappa=4 \, J$ from the three points in the Hubbard\hyp Holstein phase diagram considered in \cref{fig:qj:quench:hubbard-holstein-gs:bipolarons:sp-ekin}. 
		Analyzing the sign of the binding energy $\Delta E$, we observe the formation of stable bipolarons in the metallic and in the \gls{CDW} phase, but not in the \gls{SDW} phase. 
		Most interestingly, in the metallic phase, strong dissipation localizes the bipolarons (the metallicity decreases) without disrupting their stability ($\Delta E$ is constant). 
	}%
\end{figure}%
\begin{figure}[h]
	\centering
 	\ifthenelse{\boolean{buildtikzpics}}%
	{%
		\input{time_dependent_localization_lengths}
	}%
	{%
 		\includegraphics{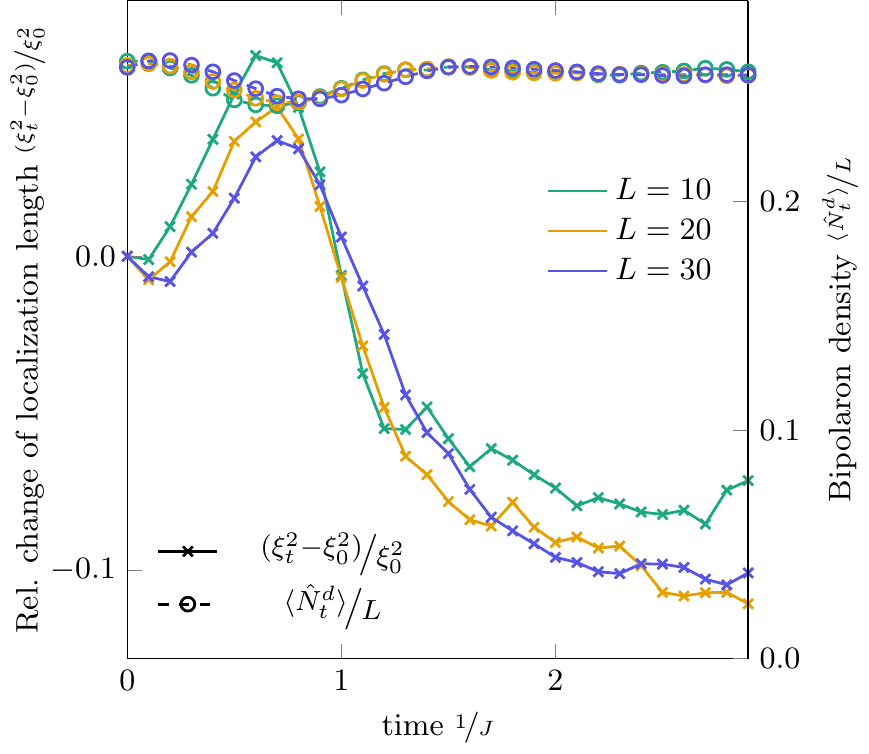}%
	}%
	\caption
	{
		\label{fig:quench:localization-lengths}
		Relative change of localization length $\xi^2_t$ after the dissipative quenches for $\kappa=4 \, J$ from the metallic region in the Hubbard\hyp Holstein phase diagram considered in \cref{fig:hh-phase-diag} for various system sizes.
		Note the decay of the localization length when turning on strong dissipation.
		For convenience, the dashed lines (right $y$-axis) display the time\hyp dependent bipolaron densities, which are constant after an initial dip, consistent with the constant binding energy.
		Thus, the reduced localization length is not an artefact of the bipolaron number $N^d_t$.
	}
\end{figure}
\paragraph{Polarons and bipolarons.}
To disentangle the roles of $g$ and $\kappa$ and study the impact of dissipation on quasi\hyp particle formation and their metallicity, we investigate further global quenches from the ground state of the Hubbard\hyp Holstein Hamiltonian at finite $g$, switching on dissipation.
For that purpose, we decompose the electronic annihilation (creation) operators into strictly single- and two\hyp particle operators
\begin{equation}
	\hat{c}^{\nodagger}_{j,\sigma} = \hat{s}^{\nodagger}_{j,\sigma} +\text{sgn}(\sigma) \hat{s}^{\dagger}_{j,\bar{\sigma}} \hat{d}^{\nodagger}_j,
	\label{eq:spilt_electron_annihilator}
\end{equation}
where $\hat{s}^{\nodagger}_{j,\sigma} = \hat{c}^{\nodagger}_{j,\sigma}(1-\hat{n}^{\nodagger}_{j,\bar{\sigma}} )$ and $\hat{d}_{j}^{\nodagger}=\hat{c}^{\nodagger}_{j,\downarrow} \hat{c}^{\nodagger}_{j,\uparrow}$.
Upon applying a Lang\hyp Firsov transformation~\cite{Lang1963}, the Hubbard\hyp Holstein Hamiltonian acquires the form
\begin{align}
	\hat{H}_\mathrm{LF}
	&=
	-J\sum_{j,\sigma} 
	\left(
		\hat{D}_j^{\dagger}
		\left(\frac{g}{\omega}\right)
			\hat{c}^{\dagger}_{j,\sigma}\hat{c}^{\nodagger}_{j+1,\sigma}
		\hat{D}_{j+1}^{\nodagger}\left(\frac{g}{\omega}\right)
		 + \mathrm{h.c.}
	\right)
	\notag \\
	&\phantom{=}+
	U_\mathrm{eff} \sum_j \hat n^\nodagger_{j,\uparrow} \hat n^\nodagger_{j,\downarrow}
	+
	\omega \sum_j \hat a^\dagger_j \hat a^\nodagger_j \notag \\
	&=
	-J\sum_j
	\left(
		\hat T^{bp}_{j,j+1}
		+
		\sum_\sigma \hat T^p_{j,j+1,\sigma }
		+
		\hat U^{bp}_j
	\right)
	\notag\\
	&\phantom{=}+
	\omega \sum_j \hat a^\dagger_j \hat a^\nodagger_j \, ,
	\label{eq:hub-hol:lf-transformed}
\end{align}
where $U_\mathrm{eff} = \left(U - \frac{2 g^2}{\omega} \right)$.
We, furthermore, introduced the bipolaron potential energy $\hat U^{bp}_j = U_\mathrm{eff} \hat d^\dagger_j \hat d^\nodagger_j$, the displacement operator $\hat{D}_j^{\dagger}\left(\frac{g}{\omega}\right) = e^{g/\omega(\hat{a}^\dagger_j - \hat{a}^\nodagger_j)}$, and the polaronic and bipolaronic hopping operators, $T^p_{i,j,\sigma}$ and $T^{bp}_{i,j}$, respectively:
\begin{align}
	\hat T^p_{i,j,\sigma} &=  \hat{D}_i^{\dagger}\left(\frac{g}{\omega}\right) \hat s^\dagger_{i,\sigma} \hat s^\nodagger_{j,\sigma} \hat{D}_j^{\nodagger}\left(\frac{g}{\omega}\right) + \mathrm{h.c.} \; , \label{eq:polaron:hop-matrix} \\
	\hat T^{bp}_{i,j} &= \hat{D}_i^{\dagger}\left(\frac{g}{\omega}\right) \hat d^\dagger_i \left(\sum_\sigma \hat s^\nodagger_{i,\sigma} \hat s^\dagger_{j,\sigma}\right) \hat d^\nodagger_{j} \hat{D}_j^{\nodagger}\left(\frac{g}{\omega}\right) + \mathrm{h.c.} \; . \label{eq:bipolaron:hop-matrix}
\end{align}
Measuring the full hopping matrix $\hat T^{bp}_{i,j}$, we can study the kinetic energies $t_k$ of bipolaronic quasi\hyp particles from a diagonalization of $\braket{\hat T^{bp}_{i,j}} \equiv t^{bp}_{ij}$ where we label the eigenstates by quasi momenta $k_n\equiv \frac{2\pi}{L} n$ with corresponding eigenvalues $t_k$.
As for the quench from the Hubbard ground state, in the following we consider a system with $L=20$ sites and compute $\lvert \mathcal Q \rvert =200$ trajectories with maximal local dimension $d_{\text{max}}=40$, maximal bond dimension $m = 2000$, discarded weight $\delta=10^{-8}$ and timestep $dt=0.01 \, J^{-1}$.
We also check, by Fourier transforming the hopping matrix, that assigning the ordered eigenvalue numbers $n$ with quasi momenta is reasonable.
From the kinetic energies, we determine the maximal quasi\hyp particle velocity $v_\mathrm{eff}$ by taking the discretized derivative at $k_\mathrm{eff}=\pi/2$.
Then, in the quasi\hyp particle picture, we introduce an estimation for the bipolaronic quasi\hyp particle mass via
\begin{equation}
	m_\mathrm{eff} = \frac{k_\mathrm{eff}}{v_\mathrm{eff}} = k_\mathrm{eff} \left( \left.\frac{\Delta t_k}{\Delta k}\right\rvert_{k_\mathrm{eff}} \right)^{-1} \; .
\end{equation}
If there are stable bipolaronic quasi\hyp particles in the system, then $m_\mathrm{eff}$ yields the smallest quasi\hyp particle mass and thereby provides a measure for their metallicity.
This interpretation immediately becomes clear, when inspecting the \gls{CDW} quenches in \cref{fig:qj:quench:hubbard-holstein-gs:bipolarons:sp-ekin} (most right column).
Here, we observe a nearly flat band over the whole simulation time, indicating the insulating character of the \gls{CDW} phase that stems from localized bipolarons.
In turn, in the \gls{SDW} phase, a single\hyp particle gap is found, indicating a very small bipolaron effective mass.
In the metallic phase, we find the strongest dependency on the dissipation strength.
An initially large metallicity is suppressed upon time\hyp evolving for the case of $\kappa=4\,J$, i.e., the single\hyp particle gap closes, indicating localization of bipolaronic quasi\hyp particles.
\paragraph{Bipolarons' stability, metallicity and localization length}
In order to determine the stability of bipolaronic quasi\hyp particles, we furthermore calculated the averaged, bipolaronic binding energy~\cite{Barisic2012}.
Using \cref{eq:polaron:hop-matrix,eq:bipolaron:hop-matrix} this quantity can be written as the difference between the site\hyp averaged bipolaronic and polaronic energies
\begin{align}
    \Delta E =& \frac{1}{L} \sum_j \left(\braket{ \hat{U}^{bp}_j} + \braket{\hat{T}^{bp}_{j,j+1}} \right. \notag\\ 
    &\phantom{\frac{1}{L} \sum_j \left(\right. }- \left. \braket{\hat{T}^{p}_{j,j+1,\uparrow}} - \braket{\hat{T}^{p}_{j,j+1,\downarrow}}\right) \; ,
\end{align}
where $\Delta E>0$ indicates that bipolarons are unstable and tend to decay into two polarons, whereas $\Delta E<0$ signals the formation of stable bipolaronic quasi\hyp particles.
In \cref{fig:qj:quench:hubbard-holstein-gs:mobilities}, the dashed lines represent the obtained bipolaronic binding energies for the case of strong dissipation.
Turning on dissipation in the \gls{SDW}, $\Delta E$ remains constant and positive, i.e., bipolarons are unstable, which is consistent with the insulating character of the antiferromagnetic Hubbard ground state.
For the quench in the \gls{CDW} phase, we find $\Delta E < 0$, which, however, decreases by roughly a factor of two in the scope of the time\hyp evolution on a time scale which is comparable to the phonon frequency $\omega$.
Nevertheless, the bipolaronic binding energy is comparably large over the whole time\hyp evolution, indicating stable bipolaronic quasi\hyp particles.
In the metallic regime, we also observe $\Delta E < 0$, which surprisingly is nearly time\hyp independent.
Thus, in the metallic phase, even in the presence of strong dissipation, phonons that are bound to a bipolaronic quasi\hyp particle do not escape into the environment.
Note that these results are in perfect agreement with the time\hyp dependent double occupations shown in~\cref{fig:hops:quench:hubbard-gs:double-occupations}.
Indeed, in the metallic regime, the double occupation is nearly independent of the dissipation strength, while the decay of double occupations during the dynamics in the~\gls{CDW} phase at $\kappa=4\,J$ occurs on the same time scale as the reduction of the bipolaronic binding energy in~\cref{fig:qj:quench:hubbard-holstein-gs:mobilities}.
The solid lines in \cref{fig:qj:quench:hubbard-holstein-gs:mobilities} illustrate the ratio between the absolute value of the binding energy and the effective bipolaron mass.
This quantity provides a measure for the bipolaronic metallicity where, for constant binding energies, large ratios correspond to highly mobile bipolarons.
The displayed curves provide a compact overview of our analysis, exhibiting the persistent insulating character of both the \gls{SDW} and \gls{CDW} phase, also in the presence of dissipation.
Moreover, we find a significant decrease in the metallicity in the metallic regime, which is generated by the increased quasi\hyp particle mass of the bipolarons.
We further elaborate on the peculiar behaviour of the metallicity when adding strong dissipation, by studying the bipolaron's localization length $\xi^2$, which can be obtained from the connected correlation functions of the bipolaronic density\hyp density correlation matrix~\cite{Souza2000}.
Defining the time\hyp dependent bipolaronic center of mass operator $\hat X_t = \sum_j \nicefrac{j}{L} \cdot \hat n^d_j(t)$, the localization length at time $t$ can be obtained from
\begin{equation}
	\xi^2_t = \frac{\braket{\hat X^2_t} - \braket{\hat X_t}^2}{N^d_t} \; ,
\end{equation}
with $N^d_t = \braket{\sum_j \hat n^d_j(t)}$.
In~\cref{fig:quench:localization-lengths} the dynamics of the relative change $\nicefrac{(\xi^2_t-\xi^2_0)}{\xi^2_0}$ is shown when switching on dissipation in the metallic phase for different system sizes $L$ (solid lines, left $y$\hyp axis).
Following an initial increase, which is mainly generated by the short-time behaviour of the bipolaron density, we observe a quick decrease in the metallic regime, indicating a localization of the bipolaronic quasiparticles.
We checked that the localization is not an artefact of a reduced bipolaron population by monitoring the bipolaron density $\nicefrac{\braket{\hat N^d_t}}{L}$ (right $y$\hyp axis in~\cref{fig:quench:localization-lengths}).
Two main features are present: first of all, apart from the very short\hyp time dynamics, the bipolaron density is independent of the system size.
Second, there is no significant decay of the bipolaron density at late times, which is in agreement with the observed, constant binding energy (c.f.~\cref{fig:qj:quench:hubbard-holstein-gs:mobilities}).
%
%
%
Therefore, the bipolaronic localization length constitutes an alternative measure, which, in combination with the binding energy and the metallicity, strongly suggests a localizing character of dissipation w.r.t. the bipolaronic quasi\hyp particles.
\discussive{A recent study has shown that couplings to an environment can be modelled by measurements, suppressing transport via the formation of decoupled clusters~\cite{Doggen2022}.
In the limit of very strong dissipation, this is reminiscent of the quantum Zeno effect~\cite{yan_zeno_2013}. 
In particular, the fact that intermediate dissipation has no relevant impact on the dynamics (c.f. \cref{fig:qj:quench:hubbard-holstein-gs:bipolarons:sp-ekin}) while large dissipation induces a strong localization of the bipolarons, suggests the existence of the transition between a volume law entangled phase and a quantum Zeno phase described in \cite{zeno_phase_transition}.
Moreover, we stress that the non-projective measurements are performed only on the phononic system, rendering the observed bipolaron localization an instance of an indirect quantum Zeno effect \cite{indirec_quantum_zeno,comment_indirec_quantum_zeno}.}
\section{Conclusion}\label{sec:conclusion}
Incorporating dissipation into the description of strongly\hyp correlated electron systems coupled to phonons paved the way to intriguing phenomena such as light\hyp enhanced or cavity\hyp induced phonon\hyp mediated superconductivity~\cite{Sentef2016,Curtis2019,Schlawin2019}.
Furthermore, in the prototypical Hubbard\hyp Holstein model, recent (semi\hyp) analytical investigations suggested the enhancement of the metallic regime in the presence of anharmonic phonons, posing the question of enhanced superconducting correlations~\cite{hubb_hol_anharmonicity, Frick1991HightemperatureSI,Lavanya2017}.
In this study, we, therefore, investigated the effect of a realistic source of phonon anharmonicities generated by a dissipative coupling of the phonons to an environment.
In order to be able to perform the required, numerically very challenging, dissipative quantum many\hyp body simulations for large systems we combined both \gls{HOPS} and \gls{QJ}, two established out\hyp of equilibrium methods to describe \gls{OQS}, with the recently introduced \gls{PP-DMRG}.
We tested and benchmarked the obtained numerical tools, demonstrating their feasibility in capturing the complex, dissipative out\hyp of equilibrium dynamics after global quenches.
Interestingly, we found that both methods, being comparably computationally efficient, exhibit complementary regimes of the physical model parameter in which they yield precise and numerically well\hyp controlled time\hyp evolution schemes.
In particular, \gls{HOPS} proved to be the method of choice for the case of intermediate and strong dissipation and large electron\hyp phonon couplings, whereas \gls{QJ} yielded excellent performance for weak dissipation and weak to intermediate electron\hyp phonon couplings.
As a consequence, using the \gls{PP}\hyp mapping, we elevated \gls{OQS} methods to be applicable in an efficient and unbiased way to a broad class of dissipative quantum many\hyp body systems, using tensor network algorithms.
We believe that the discussed, tensor network\hyp based Markovian (\gls{QJ}) and non\hyp Markovian (\gls{HOPS}) methods will be very fruitful tools for addressing relevant problems such as thermalization of quantum systems \cite{rigol_thermalization,Lebreuilly_thermalization_2018,Reichental_thermalization}, cooling of quantum many\hyp body systems \cite{Metcalf_2020_cooling,kantian_2008}, exciton dynamics in light\hyp harvesting complexes \cite{nalbach,Erling_exciton_dynamics}, and quantum transport in two\hyp terminal dissipative setups \cite{Damanet_reservoir_engeneering, stuart_francois_quantum_transport,francois_prl,gianmarchi_driven_transport,ginmarchi_transport,quantized_conductance}.
Moreover, as mentioned in \cref{app:sec:hops}, the methods developed here for systems described by \cref{eq:hops:ham-appl,eq:Lindblad}, can be generalized to multiple phonon modes per site, to phonon modes coupled to baths with arbitrary spectral structures, to different kinds of baths (dephasing, absorption), or to non\hyp local phonons coupled to several sites \cite{longrange_cuprates_wang}. 
This latter generalization could, for instance, make it possible to study dissipative versions of the Hubbard\hyp Fröhlich model \cite{froelich_superconductivity_Alexandrov,froelich_superconductivity_hardy}.

Having established the \gls{PP}\hyp enhanced \gls{HOPS} and \gls{QJ} methods, we turned to the question of whether dissipation enhances metallicity in the Hubbard\hyp Holstein model.
For that purpose, we performed a series of quenches, investigating the formation of bipolarons, i.e., phonon\hyp mediated bound two\hyp electron quasi\hyp particles and their metallicity.
Here, we defined metallicity as the ratio between the bipolaronic binding energy and its effective mass.
In the metallic regime of the Hubbard\hyp Holstein ground\hyp state phase diagram, we found that the time dependence of the bipolaronic binding energy remains mainly unchanged, i.e., the phonons that contribute to bound electron pairs do not tend to escape the system.
Studying the bipolaronic kinetic energy dynamics, we observed melting of the bipolaronic single\hyp particle gap upon increasing dissipation, indicating an increased scattering rate.
Consequently, the effect of dissipation is to enhance the bipolaronic effective mass, yielding an overall reduction of the bipolaronic metallicity.
We complement these findings by calculating the bipolaronic localization length, and explicitly find the localization of bipolarons under the action of dissipation, in the metallic regime.
Since our results contrast previous findings when considering gaussian anharmonicities, we calculated the phononic excitation probabilities for the different sources of anharmonic phonons (see \cref{app:sec:gaussian-anharmonicities}).
%
%
%
%

The picture of a quantum jump description of the dissipative dynamics creates an interesting connection to the \discussive{indirect} quantum Zeno effect~\cite{zeno_2008,yan_zeno_2013}. 
\discussive{Moreover, the absence of bipolaron localization for moderate dissipation is in agreement with the transition from a volume law entanglement phase to a quantum Zeno phase described in \cite{zeno_phase_transition}.}
Nevertheless, we also find that the bipolaronic binding energy is very robust against dissipation in the metallic regime.
This is a remarkable observation, in particular, since the calculated binding energies are of the order of $0.15 \, J$ and thereby much smaller than the studied dissipation strengths $\kappa = 1 J,\,4 J$.
Understanding the origin of this unexpected robustness of formed bipolarons in the metallic regime would be an interesting theoretical question, particularly concerning phonon\hyp mediated superconductivity.
Here, investigating the impact of dissipation on light\hyp bipolarons in Peierls\hyp coupled electron\hyp phonon systems and the reported, enhanced values of $T_\mathrm C$ is extremely important for actual physical realizations~\cite{Sous_bipol_liq,Sous_bipolaronic_sc,Sous_optical_pump}.
%
%

The relevance of phononic degrees of freedom in the description of real materials is demonstrated by the vast recent effort on studying, for instance, anharmonic phonons \cite{hubb_hol_anharmonicity}, long\hyp range electorn\hyp phonon coupling \cite{longrange_hubb_hol} and optically pumped \cite{Sous_optical_pump} phonons. In this article, we not only added a new aspect (the coupling to a dissipative bath) which is relevant to real materials but also provided tools from which we believe that all the aforementioned fields can benefit greatly.  
\section{Acknowledgements}
We thank Adrian Kantian and Alexander Wietek for very fruitful discussions.
TK acknowledges financial support by the ERC Starting Grant from the European Union's Horizon 2020 research and innovation program under grant agreement No. 758935.
MM, MG, US, SM and SP acknowledge support by the Deutsche Forschungsgemeinschaft (DFG, German Research Foundation) under Germany’s Excellence Strategy-426 EXC-2111-390814868.
\bibliography{Literatur}
\clearpage
\appendix%
\onecolumngrid%
\begin{center}%
	\textbf{\large Appendix: \thetitle}%
\end{center}%
\setcounter{equation}{0}%
\setcounter{figure}{0}%
\setcounter{table}{0}%
\setcounter{page}{1}%
\makeatletter%
\renewcommand{\theequation}{S\arabic{equation}}%
\renewcommand{\thefigure}{S\arabic{figure}}%
\renewcommand{\bibnumfmt}[1]{[S#1]}%
\section{Dissipation and Gaussian Anharmonicities} \label{app:sec:gaussian-anharmonicities}

\begin{figure}[!t]
	\centering
	\ifthenelse{\boolean{buildtikzpics}}%
	{%
		\input{excitation_probabilities}
	}%
	{%
		\includegraphics{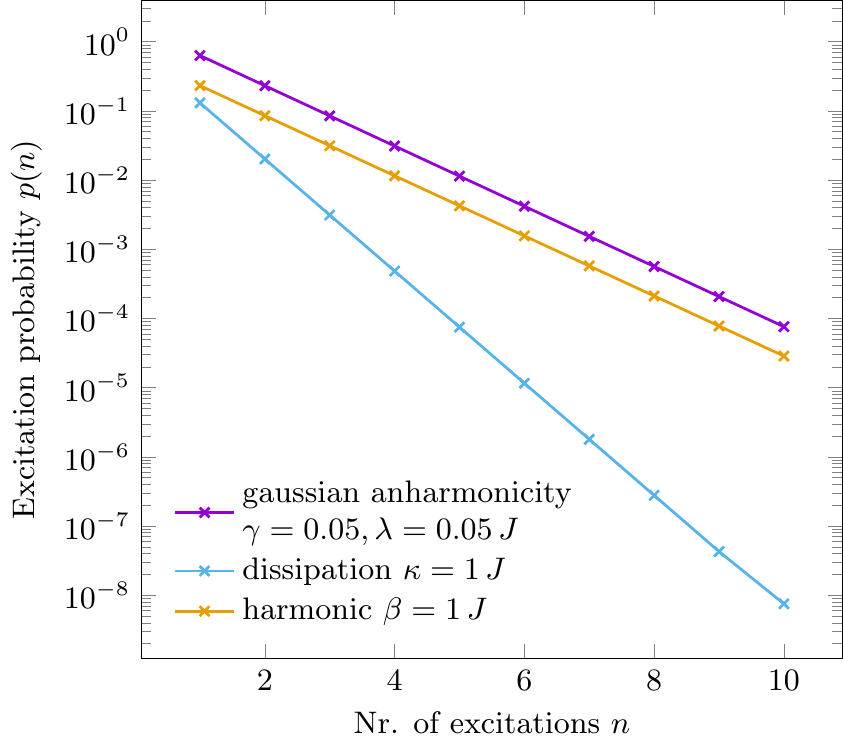}%
	}%
	\caption
	{
		\label{fig:phonon:excitations} 
		Effect of anharmonicities and dissipation on the excitation probability of bosonic modes.
		The yellow curve is obtained from evaluating the excitation probabilities of a harmonic oscillator $\hat H_\mathrm{HO}$ at inverse temperature $\beta=1$.
		We compare these to the probabilities obtained when adding a gaussian, quadratic anharmonicity $\hat H_\mathrm{HO} + \lambda \mathrm{e}^{-\gamma (\hat a^\dagger + \hat a^{\protect\nodagger})^2}$ with $\lambda=0.05\,J, \gamma=0.05$ (purple data) and when incorporating dissipation $\kappa=1\,J$ (blue data).
		While gaussian anharmonicities reduce the spacing between the energy levels and thus increase the probabilities of populating highly excited states, the effect of dissipation is to leak phonons into the environment, increasing the ground state occupation while higher excitations are suppressed.
	}
\end{figure}

Recent theoretical studies considered the effect of anharmonicities on the properties of the metallic phase in the Hubbard\hyp Holstein model, indicating the tendency to stabilize light bipolarons even at larger electron\hyp phonon couplings~\cite{Lavanya2017,Debnath2021}, a crucial requirement for large transition temperatures into a bipolaronic, superconducting state~\cite{Barisic2012,Sous2018,Sous_bipol_liq,Sous_bipolaronic_sc,Sous_optical_pump}.
However, the anharmonic contributions to the phononic oscillator potentials have been incorporated constructively, such that the resulting models can be treated semi\hyp analytically.
This is mainly due to the extremely high numerical costs for simulating phononic degrees of freedom, whose local Hilbert spaces are, in principle, infinite\hyp dimensional.
The computational limitations become even more severe when incorporating more realistic foundations for anharmonicities, such as treating the phononic system as an \gls{OQS} or considering dispersive behaviour \cite{Costa_dispersive_phonons,Jansen2022}.
To connect our results to the reported enhancement of the metallic phase via gaussian and quartic anharmonic modifications of the phononic modes, we compared the effects of dissipation and the anharmonicities investigated in~\cite{Lavanya2017,hubb_hol_anharmonicity} on the excitation probabilities of a single phonon mode.
As a reference distribution, we computed the population of the excited modes by diagonalizing the corresponding Hamiltonian and evaluated the Boltzmann weights at inverse temperature $\beta$ equal to the oscillator frequency $\omega$.
For the dissipative case, it can be shown that the thermal state for a single harmonic oscillator $ \hat{\rho}_{\beta}^{eq} = e^{-\beta \omega \hat{n} } / \mathcal{N} $ is the steady state solution of a Lindblad master equation with Lindblad operators $\hat{D}_1 = e^{-\beta \omega/2} \hat{a}^{\dagger}$,   $\hat{D}_2 = \hat{a}$.
Combining them with the Lindblad operator for dissipation $\hat{D}_3 = \sqrt{\kappa} a$ yields the following equation
\begin{align}
\partial_t \hat{\rho} = &-e^{-\beta \omega} \left( \frac{1}{2} \{\hat{a} \hat{a}^{\dagger}, \hat{\rho} \} - \hat{a}^{\dagger}  \hat{\rho} \hat{a}  \right)  \notag \\
& -(1+\kappa) \left( \frac{1}{2} \{\hat{a}^{\dagger}\hat{a} , \hat{\rho} \} - \hat{a}  \hat{\rho} \hat{a}^{\dagger}  \right)  \; ,
\label{eq:lindblad_thermal_steady_state}
\end{align}
which can be solved numerically.
In \cref{fig:phonon:excitations}, we show the excitation probabilities of a harmonic oscillator $\hat H_\mathrm{HO}=\omega \hat{a}^\dagger \hat{a}^\nodagger $, an anharmonic oscillator with gaussian anharmonicity $\hat H_\mathrm{G}(\lambda,\gamma) = \hat H_\mathrm{HO} + \lambda \mathrm{e}^{-\gamma (\hat a^\dagger + \hat a^{\protect\nodagger})^2}$, and a harmonic oscillator with dissipation \cref{eq:lindblad_thermal_steady_state}.
Here, we illustrate what we believe to be the underlying reason for the seemingly contradicting results: dissipation and gaussian anharmonicities have opposite effects on the population of the excited phonon states.
While the decay of the excitation probability is reduced by gaussian anharmonicities, it is enhanced when considering dissipation.
These observations can be connected to our investigation of the metallicity and the localization length, and by noting that the binding energy is mainly unaffected by dissipation in the metallic phase.
This suggests that the metallicity mainly depends on the mean free path length of the bipolaronic quasi\hyp particles, which appears to be reduced by dissipation.
\section{Quantum Jumps}\label{app:sec:quantum-jumps}
\begin{figure}
	\centering
	 \ifthenelse{\boolean{buildtikzpics}}%
	 {%
	\input{qj_algorithm}
	 }%
	 {%
	 	\includegraphics{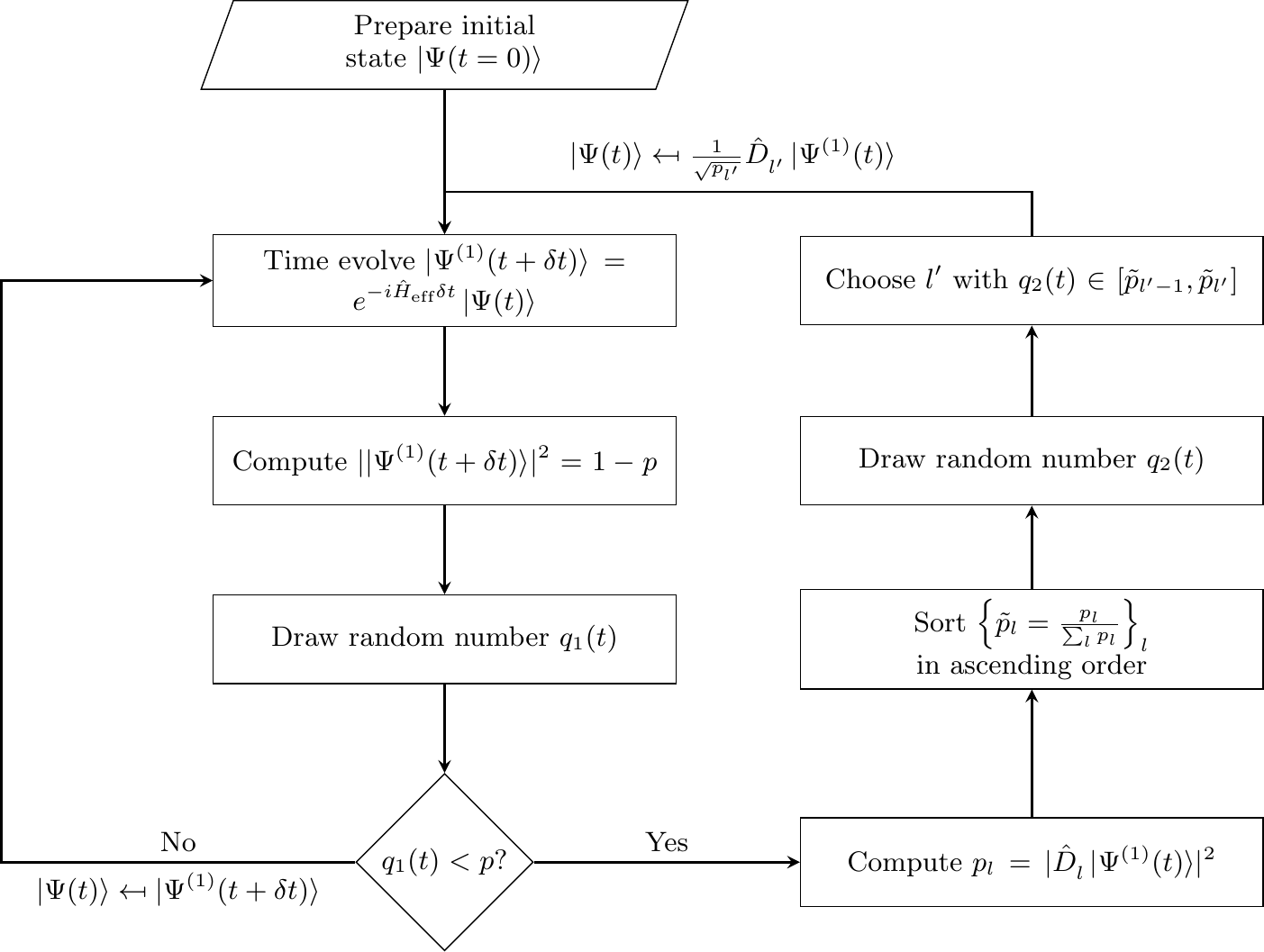}%
	 }%
	\caption
	{
	\label{fig:qj:algorithm}
		Algorithmic sketch of the \acrlong{QJ} method.
	}
\end{figure}

In the following, we sketch the main equations of the method presented in Ref. \cite{Daley_QJUMPS}.
First, it is convenient to define an effective, non-hermitian Hamiltonian: 
\begin{equation}
\hat{H}^{\nodagger}_{\text{eff}} \equiv \hat{H}^{\nodagger}_{\text{\tiny{tot}}} -\frac{i}{2} \sum_l \hat{D}^{\dagger}_l\hat{D}^{\nodagger}_l \; ,
\label{eq:q-jumps_H_eff}
\end{equation}
that allows us to rewrite the Lindblad equation \cref{eq:Lindblad} as: 
\begin{equation}
\partial_t \hat{\rho} = -i ( \hat{H}^{\nodagger}_{\text{eff}} \hat{\rho} - \hat{\rho} \hat{H}^{\dagger}_{\text{eff}}) +\sum_l \hat{D}^{\nodagger}_l\hat{\rho} \hat{D}^{\dagger}_l \;.
\label{eq:linblad_with_h_eff}
\end{equation}
Working with pure states, a stochastic process $\mathcal Q$ is introduced so that the density matrix time-evolved by the Lindblad equation is obtained from averaging over many realizations:
\begin{equation}
\mathcal{E} [\ket{\Psi(t)}_{\text{\tiny{q}}} \bra{\Psi(t)}_{\text{\tiny{q}}} ] = \hat{\rho}(t) \;,
\label{app:eq:av_over_proj}
\end{equation}
where $q \in \mathcal Q$ is a collection of pseudo-random numbers identifying a so\hyp called trajectory.
Thus, instead of constructing the density matrix, one computes observables for every trajectory and averages over them:
\begin{equation}
\langle \hat{O} \rangle (t) = \mathcal{E} [ \bra{\Psi(t)}_{\text{\tiny{q}}} \hat{O} \ket{\Psi(t)}_{\text{\tiny{q}}} ] \;.
\label{app:eq:av_over_observables}
\end{equation}
In \cref{fig:qj:algorithm}, we give a sketch of the described unravelling and the random processes involved.
In practice, typically $\sim 10^2-10^3$ trajectories are needed for getting converged observables.

For a trajectory specified by two uniform random numbers $q = (q_1(t), q_2(t) )$ with $q_i(t)\in [0,1]$, the algorithm to compute the time\hyp evolution of $\ket{\Psi(t)}_{\text{\tiny{q}}}$ is shown in~\cref{fig:qj:algorithm}.
The general idea is to expand the time\hyp evolved state to first order, to decompose the change in its norm
\begin{equation}
\lvert \ket{\Psi^{(1)}(t+\delta t)} \rvert^2 = 1 - p \approx 1 - \delta t \sum_l \bra{\Psi(t)} \hat{D}^{\dagger}_l\hat{D}^{\nodagger}_l \ket{\Psi(t)} \equiv{1-\sum_l p^{\nodagger}_l} \; .
\end{equation}
Then, the random number $q_1(t)$ is picked and compared to the overall norm change $p$ to decide whether a jump has to happen.
If a jump needs to occur, the second random number $q_2(t)$ is picked to choose the actual jump operator, according to the different jump probabilities $\delta t \bra{\Psi(t)} \hat{D}^{\dagger}_l\hat{D}^{\nodagger}_l \ket{\Psi(t)}$.
If the algorithm described above is carried out for each trajectory $q$, averaging over the projectors yields:
\begin{align}
\hat{\rho}(t+\delta t)
&=
\mathcal{E} [\ket{\Psi(t)}_{\text{\tiny{q}}} \bra{\Psi(t)}_{\text{\tiny{q}}} ] \notag \\
&=
(1-p) \frac{ \ket{\Psi^{(1)}(t+\delta t)} } {\sqrt{1-p}} \frac{ \bra{\Psi^{(1)}(t+\delta t)} } {\sqrt{1-p}} + \sum_l \frac{p_l}{p} \frac{\hat{D}^\nodagger_l \ket{\Psi(t)}} { \sqrt{p_l/\delta t} } \frac{ \bra{\Psi(t)}\hat{D}^{\dagger}_l } { \sqrt{p_l/\delta t} } \notag \\
&= \hat{\rho}(t) -i\delta t ( \hat{H}^\nodagger_{\text{eff}} \hat{\rho} - \hat{\rho} \hat{H}^{\dagger}_{\text{eff}}) +\delta t \sum_l \hat{D}^\nodagger_l\hat{\rho} \hat{D}^{\dagger}_l \;,
\label{eq:qj_linblad_equivalence}
\end{align} 
which in the limit $\delta t \to 0$ is precisely the Lindblad equation.
\section{Hops}\label{app:sec:hops}
\begin{figure}[!t]
	\centering
	\ifthenelse{\boolean{buildtikzpics}}%
	{%
		\input{hops_algorithm}
	}%
	{%
		\includegraphics{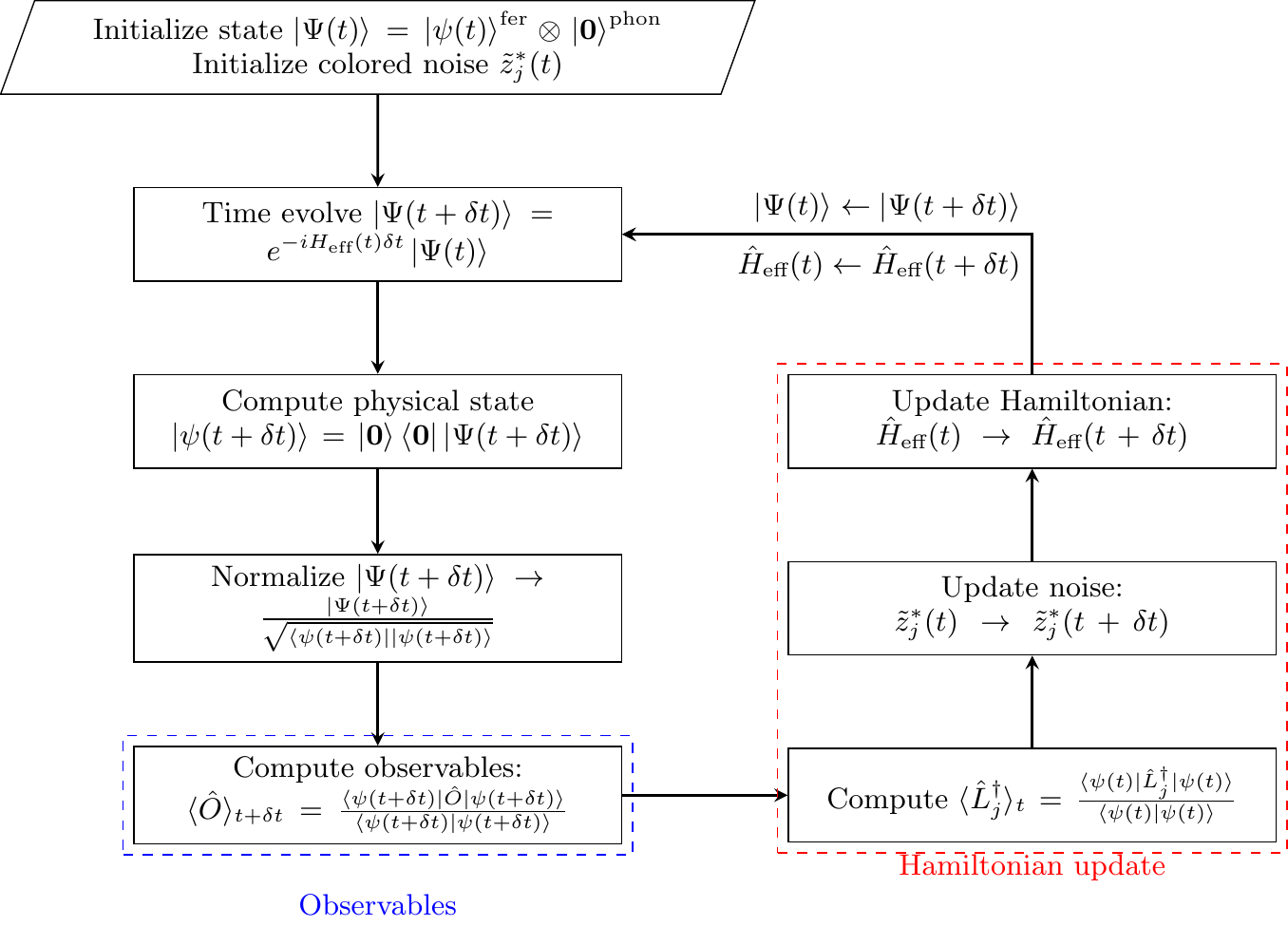}%
	}%
	\caption
	{
		\label{fig:hops_algorithm}
		Algorithmic sketch of the \acrlong{HOPS} method.
	}
\end{figure}
For clarity, we consider a single $g$, $\omega$ and $\kappa$.
Tracing out the phonons transforms the Schr\"odinger equation with the Hamiltonian of \cref{eq:hops:ham-appl} into the non-Markovian quantum state diffusion equation \cite{Di_si_1997_qsd} for the state of the fermionic degrees of freedom $\ket{\psi(t)}$:
\begin{equation}
\partial_t \ket{\psi(t)} = -i \hat{H}_{\text{\tiny{f}}} \ket{\psi(t)} + g \sum_j \hat{L}_j z^*_j(t) \ket{\psi(t)} -g \sum_j \hat{L}_j^{\dagger} \int_0^t \mathrm{d} s \,\alpha_j^*(t-s) \frac{\delta \ket{\psi(t)}}{\delta z_j^*(s)}.
\label{app:eq:non-mark_qsd}
\end{equation}
Here $\alpha_j(t)$ represents the environment correlation function, which on site $j$ and at zero temperature is given by the Fourier\hyp transform of the spectral density $J_j(\omega)$:
\begin{equation}
	\alpha_j(t) \equiv \langle \hat{a}_{j}(t) \hat{a}_{j}^{\dagger}(t') \rangle = \frac{1}{\sqrt{2}} \int_{-\infty}^{+\infty} \mathrm{d}\omega \, J_j(\omega) e^{-i\omega (t-t')} \; ,
	\end{equation}
which can be determined, for instance, from spectroscopic experiments \footnote{At finite temperature the relation between the environment correlation function and the spectral density reads: $\alpha(t) = \frac{1}{\pi} \int_0^{\infty} \mathrm{d} \omega \, J(\omega) \big[ \coth{(\frac{\beta\omega}{2})} \cos{(\omega t)} -i \sin{(\omega t)} \big]$}.
In the following we assume that the environment correlation function is given by a single complex exponential $\alpha(t-t') = e^{-\kappa|t-t'| -i\omega(t-t')}$.
The term $z_j(t)$ in \cref{eq:non-mark_qsd} represents a colored noise that satisfies $\mathcal{E} \big[ z_j(t) z_{j'}^*(t') \big] = \alpha(t-t') \delta_{j,j'}$, which can be generated in practice following e.g. \cite{hops_stuart_francois,DeVega_colored_noise}, while the term $\delta/\delta z_j^*(s)$ represents the functional derivative with respect to $z^*$.
The observables for the electronic system are obtained by averaging the results of \cref{eq:non-mark_qsd} over many trajectories, as explained for \gls{QJ} in \cref{sec:q-jumps}. 

In practical calculations, solving \cref{eq:non-mark_qsd} is exceptionally challenging because of the last term of the right-hand side, which is non\hyp local in time \cite{hops_hartmann_strunz}.
This problem can be solved efficiently by the hierarchy of pure states \gls{HOPS} method~\cite{hops_original,hops_stuart_francois}, where one defines:
\begin{equation}
\ket{\psi^{(1,j)}(t) } = D_j(t) \ket{\psi(t)} \equiv \int_0^t \mathrm{d}s \, \alpha_j^*(t-s) \frac{\delta \ket{\psi(t)}}{\delta z_j^*(s)}
\label{eq:funct_deriv_aux_states}
\end{equation}
which is labeled \textit{first auxiliary state} relative to site $j$. One then defines the \textit{k\hyp th auxiliary state} recursively:
\begin{equation}
\ket{\psi^{(k,j)}(t) } = [D_j(t)]^k \ket{\psi(t)}. 
\end{equation}
%
%
In \cref{fig:hops_algorithm} we sketch the \gls{HOPS} algorithm.
As discussed in \cref{app:sec:aux_states_rescaling}, at least for the model considered in this work, it is highly important to rescale the auxiliary states in the following way inspired by \cite{Gao_2022}: 
\begin{equation}
\ket{\psi^{(k,j)}(t)} \to \frac{1}{\sqrt{\alpha_j(0)^{k} k! } } \ket{\psi^{(k,j)}(t) } \;.
\label{eq:new_auxiliary_states}
\end{equation}
With \cref{eq:funct_deriv_aux_states,eq:new_auxiliary_states}, we can replace \cref{eq:non-mark_qsd} by a hierarchy of equations.
Following \cite{hops_stuart_francois}, it is convenient to define a state on the combined fermionic and bosonic Hilbert space as:
\begin{equation}
\ket{\Psi(t)} = \sum_{\mathbf{k=1}}^{k_{\text{max}}} C_{\mathbf{k}}(t) \ket{\psi^{(\mathbf{k})}(t)} \otimes \ket{\mathbf{k}}^{\text{bos}},
\label{eq:combied_ferm_bos_state}
\end{equation}
where $\ket{\mathbf{k}}^{\text{bos}} \equiv \otimes_{j} \ket{ k }_j^{\text{bos}}$ labels the bosonic mode corresponding to the k\hyp th auxiliary state, $C_{\mathbf{k}}(t)$ is a time\hyp dependent coefficient and $k_{\text{max}}$ is the local bosonic Hilbert space dimension.
The hierarchy then takes the form of a simple Schr\"odinger equation for the state on the combined fermionic and bosonic Hilbert space:
\begin{equation}
\partial_t \ket{\Psi(t)} = -i \hat{H}_{\text{\tiny{eff}}} ^{\text{\tiny{Q}}} \ket{\Psi(t)}\;,
\label{eq:Schro_eq_H_eff_HOPS}
\end{equation}
where the effective, non-hermitian Hamiltonian now reads \cite{hops_original,hops_stuart_francois}:
\begin{equation}
\hat{H}_{\text{\tiny{eff}}} = \hat{H}_{\text{\tiny{s}}} 
+ \sum_j i \Big( \tilde{z}_j^*(t) g \hat{L}^\nodagger_j - ( \kappa +i\omega ) \hat{K}^\nodagger_j 
+g \hat{L}^\nodagger_j \otimes \hat{K}^{1/2}_j \hat{b}_j^{\dagger} -g \big( \hat{L}_j^{\dagger} - \langle \hat{L}_j^{\dagger} \rangle_t \big) \otimes \hat{b}^\nodagger_j \hat{K}^{1/2}_j \Big) \;.
\label{eq:hops_evolution}
\end{equation}
Here, $\hat{K}_j $ is the bosonic number operator acting on site $j$ and $\hat{b}_j^{\dagger}$, $\hat{b}_j$ are the so\hyp called bare creation and annihilation operator, respectively, acting on the bosonic modes as:
\begin{equation}
\begin{array}{l}
\hat{b}^{\dagger}\ket{k} = \ket{k+1} \\
\hat{b}\ket{k} = \ket{k-1}\;.
\end{array}
\label{eq:bare_ops}
\end{equation}
The coloured noise is modified as:
\begin{equation*}
\tilde{z}_j^*(t) = z_j^*(t) + g \int_0^t \mathrm{d} s \, \alpha_j^*(t-s) \langle \hat{L}_j^{\dagger} \rangle_s \;.
\end{equation*}
\cref{eq:Schro_eq_H_eff_HOPS} is linearized by computing the non-linear term $\langle \hat{L}_j^{\dagger} \rangle_t$ with $\ket{\psi(t-\delta t)}$, which is a reasonable approximation as long as the timestep $\delta t$ is small.
For computing the electronic observables, at each timestep, the whole state needs to be projected onto the physical state:
\begin{equation}
\ket{\Psi(t)} \rightarrow \ket{\psi(t)} = \ket{\mathbf{0}}^{\text{bos}} \bra{\mathbf{0}}^{\text{bos}} \ket{\Psi(t)}, 
\end{equation}
where $\ket{\mathbf{0}}^{\text{bos}}\equiv \otimes_j \ket{0}^{\text{bos}}_j$ is the bosonic vacuum.
%
%
In practice, the Schr\"odinger equation \cref{eq:Schro_eq_H_eff_HOPS} is propagated in time by using the initial condition $\ket{\Psi(t=0)} = \ket{\psi^{(\mathbf{0})}(t=0)} \otimes \ket{\mathbf{0}}^{\text{bos}}$, where all the auxiliary states are set to zero and are then populated as time evolves. 
In principle, $k_{\text{max}}$ is infinite, but the populations of high-$k$ auxiliary states typically remains small, allowing for a truncation of the hierarchy.
In \cref{sec:PP,sec:Phys_dim_conv} we will discuss how the newly\hyp introduced \gls{PP} method allows for an optimal and automated selection of $k_{\text{max}}$.
\\
The restriction of the environment correlation function $\alpha(t)$ being a complex exponential can be lifted by noting that complex exponentials form a complete orthonormal set on $L^2$, and thus we can approximate 
\begin{equation}
\alpha_j(t-t') \approx \sum_{p=1} ^P g_p e^{-\kappa_p|t-t'| -i\omega_p(t-t')} \;,
\label{eq:env_corr_funct}
\end{equation}
for any square\hyp integrable function with arbitrary precision by increasing $P$.
The decomposition can be obtained, for instance, with the Laplace\hyp Pade method \cite{Laplace_Pade}, yielding a set of parameters $\omega_p$, $g_p$ and $\kappa_p$.
In this work, we will deal with the case $P=1$, corresponding to the case of a Lorentzian spectral density. 
For a presentation of the conceptually straightforward generalization to $P>1$ we refer to Ref. \cite{hops_hartmann_strunz,Gao_2022}.
\subsection{Improved stability for highly excited baths}\label{app:sec:aux_states_rescaling}
\begin{figure}[!t]
	\centering
	\ifthenelse{\boolean{buildtikzpics}}%
	{%
		\input{new_aux_states}%
	}%
	{%
		\includegraphics{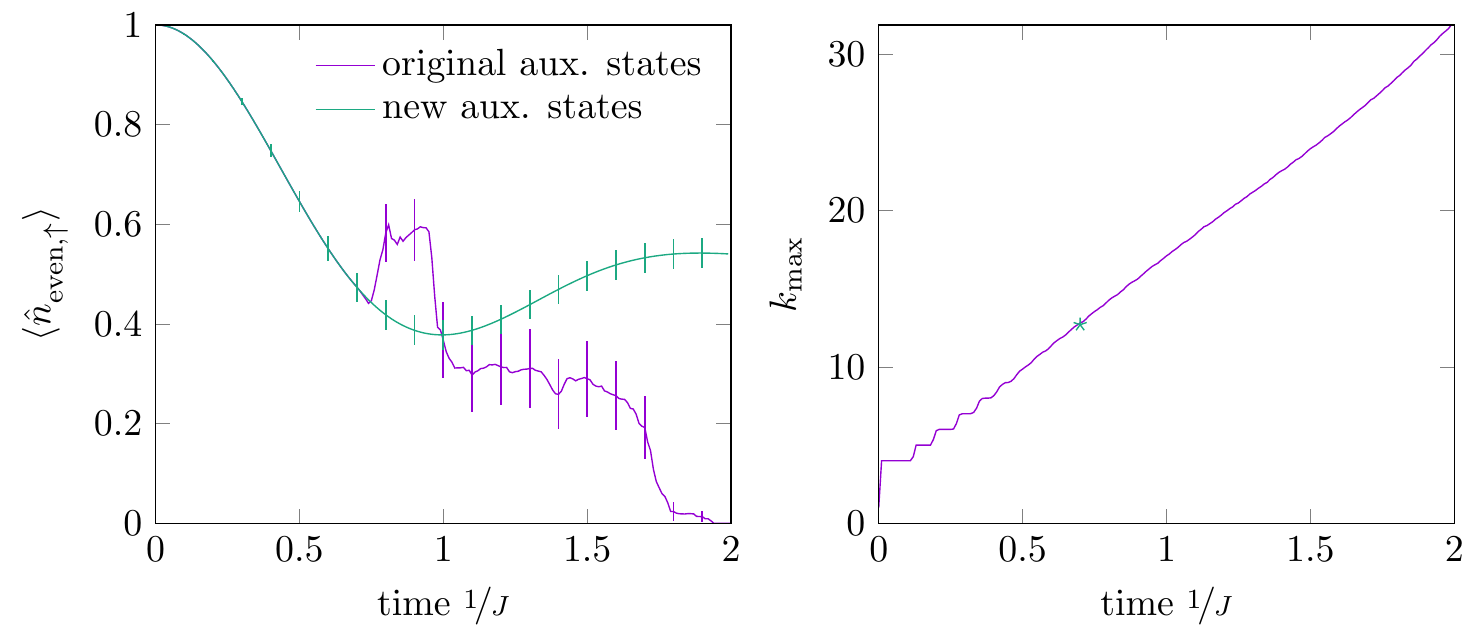}%
	}%
	\caption
	{
		\label{fig:new_aux_states} 
		Improved stability of \gls{HOPS} with auxiliary states transformed according to \cref{eq:new_auxiliary_states}.
		Many phononic modes get populated in the strongly non\hyp Markovian regime (right figure). 
		With the original \gls{HOPS} formulation, in such a case, the norm of the auxiliary states becomes very large and renders the method very unstable when computing observables with the normalized physical state (left figure).
		The calculations were performed for $20$ sites and averaged over $100$ trajectories with $g=J$ and $\kappa=0.1 \,J$. 
		All other parameters are analogue to \cref{fig:SzSz,fig:NfNf}.
	}
\end{figure}

For all the \gls{HOPS} calculations on the dissipative Hubbard\hyp Holstein model, we have rescaled the auxiliary states according to \cref{eq:new_auxiliary_states}.
This reduces the norm of the auxiliary states and prevents numerical errors arising from the normalization of the physical state that is performed at each time step when computing the observables. 
In \cref{fig:new_aux_states} we see that for a strong electron\hyp phonon coupling $g=1$, and a weak dissipation $\kappa=0.1$, the \gls{HOPS} method without a rescaling of the auxiliary states breaks down completely when $13$ bosonic modes are populated. 
In contrast, as shown in \cref{fig:k_max_surface}, with the new definition of the auxiliary states \gls{HOPS} can deal with up to $55$ occupied bosonic modes.
We want to point out that this is not an \gls{MPS}-related issue, as we encountered it also for \gls{ED} calculations.
\section{Method Benchmarks}\label{sec:benchmark}
Simulating the complicated interplay between electronic and dissipative, phononic degrees of freedom requires a careful understanding of the limitations of the used methods.
Even though \gls{HOPS}, as well as \gls{QJ}, are well\hyp established tools for the description of open quantum systems, here we combine these methods with a tensor\hyp network representation that comes along with its own approximations.
Additionally, we must consider the truncation in the enlarged phononic Hilbert space generated by the \gls{PP} mapping~\cref{sec:PP}.
It is therefore essential to understand the effect of the additional numerical approximations, particularly if we can control the numerical precision within each method by tuning typical control parameters such as the bond dimension or the discarded weight\cite{schollwoeck_mps_rev,time_ev_methods}.
A practical consequence of the method benchmark presented in the following is that even though both methods require similar numerical resources, their numerical accuracies complement each other with respect to the dissipation strength and electron\hyp phonon coupling.
Therefore, given a physical realization of some model parameters, our benchmark yields a comprehensive picture of which method is to be used for an optimal numerical outcome.
\subsection{\Acrlong{ED} and \acrlongpl{MPS}}\label{sec:ed_and_MPS}
\begin{figure}[!t]
	\centering
	\ifthenelse{\boolean{buildtikzpics}}%
	{%
		\input{hops_qj_ed_comparison_correlations}
	}%
	{%
		\includegraphics{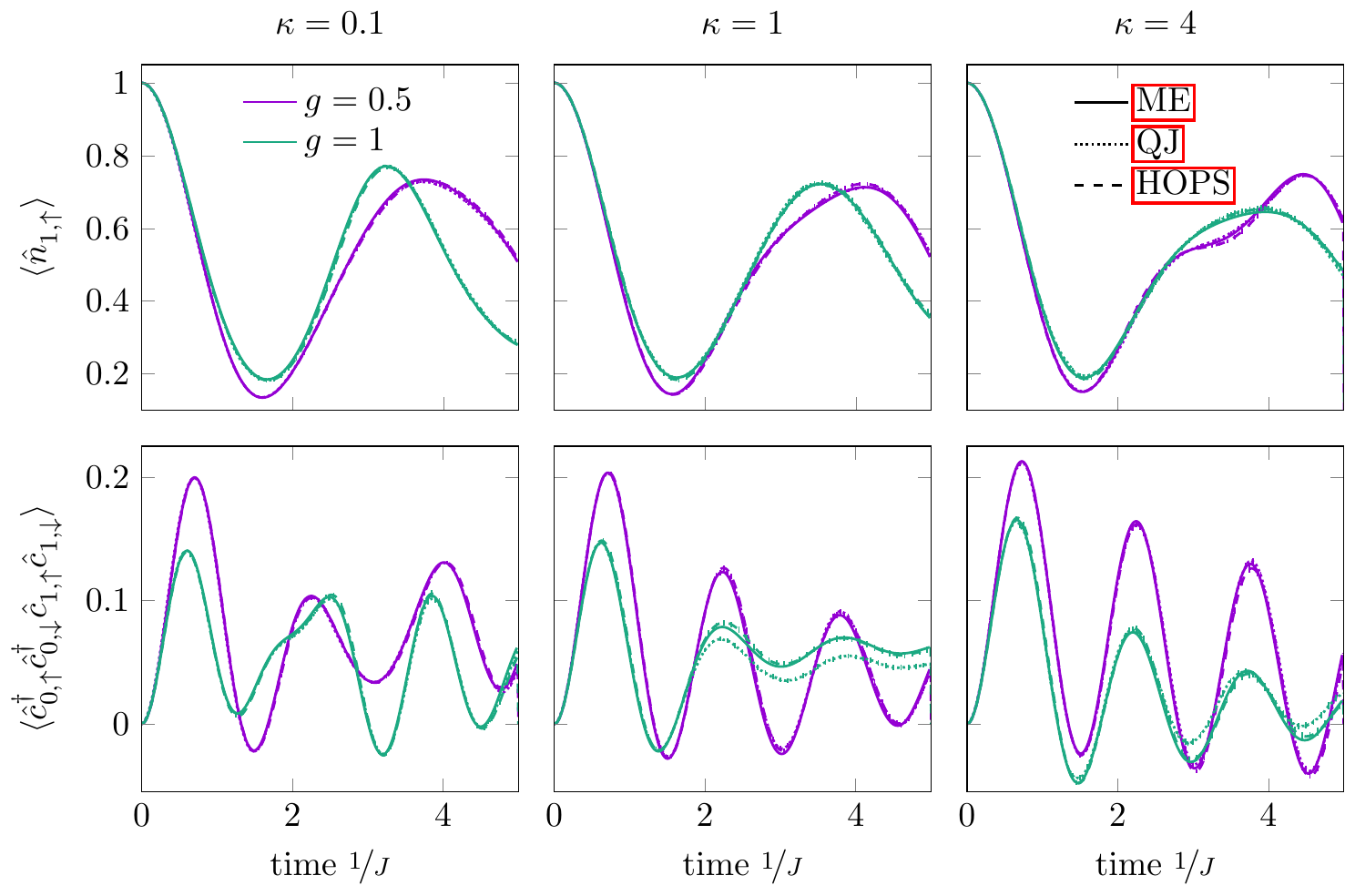}%
	}%
	\caption
	{
		\label{fig:comparison:hops-qj-ed:correlations}
		Comparing \gls{HOPS} and \gls{QJ} to the exact \gls{ME}. 
		All \gls{ME} results were computed with \gls{ED} for a system composed of two fermionic and two phononic sites. 
		For all plots, the Hamiltonian parameters were chosen to be $U=J$, $\omega = 2 \, J$, and $500$ trajectories and a time step $dt=0.005 \, J^{-1}$ were used. 
		The dissipation strength $\kappa$ was fixed to $0.1 \,J$ for the two upper plots and to $4 \,J$ for the two lower plots. 
		For all analyzed parameters, the linear and the non\hyp linear method present at most very small differences.
		Both \gls{HOPS} and \gls{QJ} agree very well with the exact \gls{ME} results, except for the case of intermediate and strong dissipation and large electron\hyp phonon coupling ($\kappa=1,4 \, J$, $g=J$) where \gls{QJ} exhibits deviations at later times $t>2 \, J^{-1}$.
	}
\end{figure}
Analyzing the ground state of~\cref{eq:hubbard-holstein-hamiltonian} already makes for a numerically involved problem.
Thus, faithfully simulating the dynamics following a global quantum quench in the presence of dissipation, we are equipped with a reasonable benchmark system.
Here, we prepare the system in a product state between the electronic and phononic system corresponding to a highly excited state of~\cref{eq:hubbard-holstein-hamiltonian}.
As a key feature, in the post\hyp quench dynamics, a potentially significant occupation of the bosonic, local degrees of freedom can occur, driven by the excess energy of the electronic system.
The latter competes with the effect of dissipation.
Considering large phonon frequencies $\omega~\sim \mathcal O(1)$, the relaxation separates into distinct time\hyp scales. Therefore, describing the dynamics of the overall system requires a large local Hilbert space dimension $\sim 10-60$ for the bosonic system.
Capturing these competing effects correctly is one of the most important points in practice, whereas any small, uncontrolled approximation already modifies the short\hyp time dynamics of correlation functions drastically.
Note that quenching from a product state, a large amount of energy is transferred into the system.
In that sense, our analysis refers to an extreme test case. In practice, for near\hyp equilibrium quenches, we expect both methods to perform reasonably also in the regime, which is complementary to the optimal one described in the following.
\paragraph{Comparison with \acrlong{ED}.}
The dynamics of the smallest meaningful Hubbard\hyp Holstein model, composed of two electrons and two phonons, can be described by the exact Lindblad master equation (\cref{eq:Lindblad}) via \gls{ED}.
\footnote%
{%
Note that due to the large local Hilbert space dimension required for the phonons, already the exact treatment of the two\hyp electron $+$ two phonon\hyp case is non\hyp trivial. 
For instance, the total Hilbert space dimension for the density matrix for two electrons and two phonons modelled by a $10$\hyp level harmonic oscillator is $\sim 2.6 \cdot 10^6$. For some \gls{MPS} calculation, we included up to $60$ phononic states, which would correspond to a total Hilbert space dimension of $\sim 3.3 \cdot 10^9$ for a two-site system.
}%
This is used as an exact reference to assess the precision and the computational complexity of the \gls{HOPS} and the \gls{QJ} methods before turning to large systems. 
We fix $U=J$ and $\omega=2\,J$, and study the performance of the \gls{HOPS} and the \gls{QJ} methods as a function of the electron\hyp phonon coupling $g$ and the dissipation strength $\kappa$. 
The dependence on the dissipation strength is particularly interesting because, in principle, the two methods are complementary:
for \gls{HOPS}, the environment becomes Markovian and thus trivial for $\kappa \to \infty$, 
whereas for \gls{QJ}, the non\hyp unitary part of the dynamics for the enlarged system becomes irrelevant in the limit $\kappa \to 0$.
We initialize the time\hyp evolution with the Neel state for the fermions and the vacuum for the phonons:
\begin{equation*}
\ket{\Psi}^{\text{\tiny{init}}} = \ket{\uparrow} ^{\text{\tiny{fer}}} _1 \ket{\downarrow} ^{\text{\tiny{fer}}} _2 \ket{0} ^{\text{\tiny{bos}}} _1 \ket{0} ^{\text{\tiny{bos}}} _2
\end{equation*}
and perform a global quench both in the electronic and in the phononic system.
We pick the number of spin\hyp up fermions on site one: $\langle \hat{n}_1^{\uparrow} \rangle$, and the pairing correlation between the two fermionic sites: $\langle \hat{c}^{\dagger}_{0, \uparrow} \: \hat{c}^{\dagger}_{0,\downarrow} \: \hat{c}^{\nodagger}_{1, \downarrow} \: \hat{c}^{\nodagger}_{1,\uparrow} \rangle$, as a single\hyp site and two\hyp site observable, respectively.
We choose to compare the two methods for very weak ($\kappa=0.1 \, J$), intermediate ($\kappa=J$) and very strong ($\kappa=4 \, J$) dissipation at the medium and strong electron\hyp phonon couplings $g=0.5 \, J$ and $g=J$. 
Our results are summarized in \cref{fig:comparison:hops-qj-ed:correlations}.
In general we observe excellent agreement for both \gls{HOPS} and \gls{QJ} with \gls{ME} at short times $t \leq 2 \, J^{-1}$.
The only notable deviation appears at larger simulation times in the \gls{QJ} results for the two-site observable, in the case of strong electron\hyp phonon coupling $g=1J$ and medium or strong dissipation $\kappa \geq 1 J$.
We believe that using a modified version of \gls{QJ}, or significantly decreasing the timestep and increasing the number of trajectories, would improve the agreement with the exact result.
However, with \gls{MPS} methods, using an excessively small timestep can lead to an accumulation of truncation errors and should be avoided.
Therefore, we suggest that, at least for a quench from a product state, \gls{HOPS} should be preferred over \gls{QJ} in the parameter regime mentioned above.
In \cref{app:sec:homodyne-detection}, we show that both the linear and the non-linear version of the homodyne detection unravelling do not yield accurate results for this model.
\paragraph{Comparison beyond exact diagonalization.}
\begin{figure}[!t]
	\centering
	\ifthenelse{\boolean{buildtikzpics}}%
	{%
		\input{hops_qj_comparison_time_dependent_szsz_correlations}
	}%
	{%
		\includegraphics{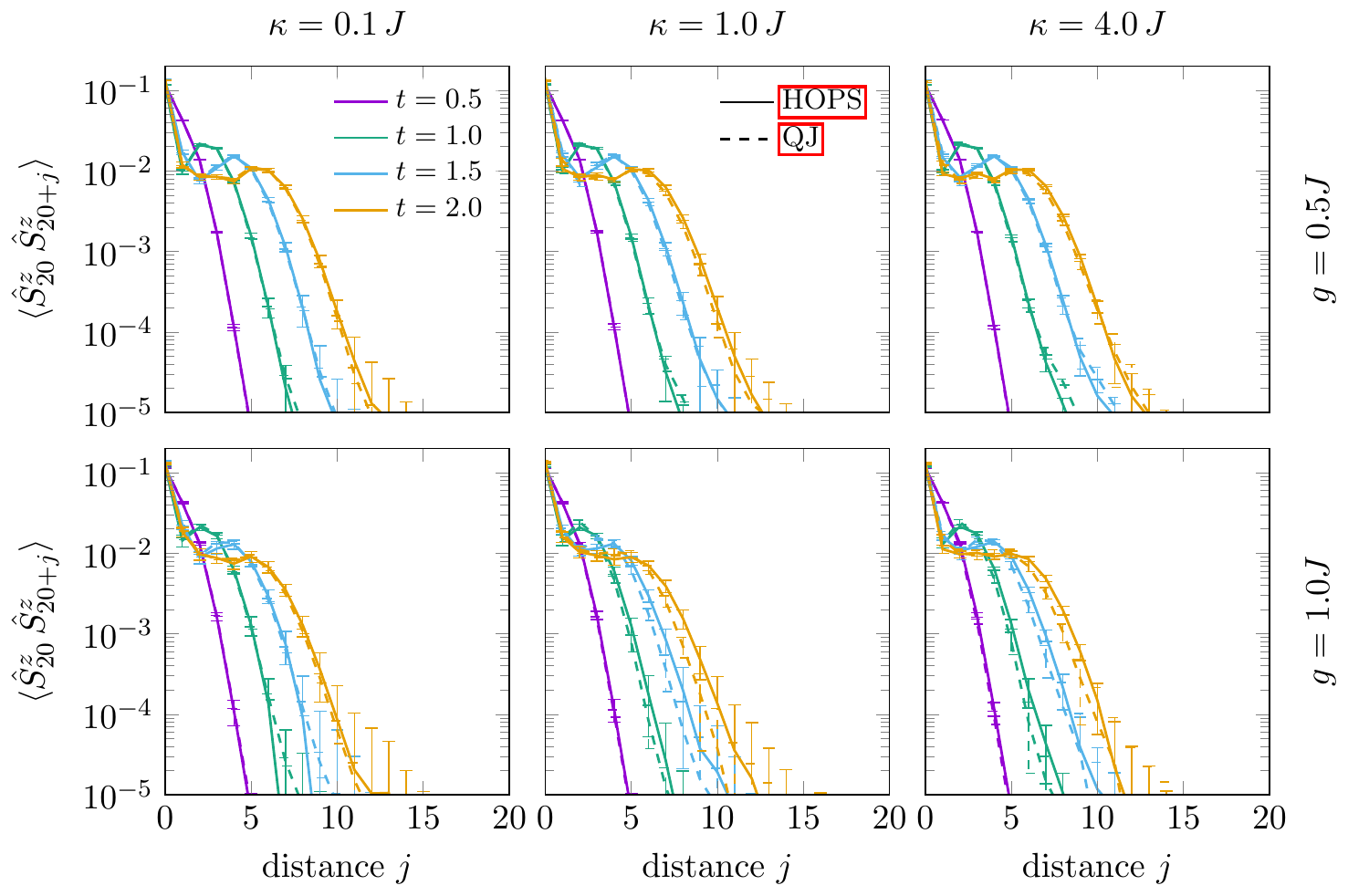}%
	}%
	\caption
	{
		\label{fig:SzSz}
		Spin-density correlations $\langle \hat{S}^z_{20} \, \hat{S}^z_{20+j} \rangle$ calculated using \gls{HOPS} and \gls{QJ} for a system with $L=40$ sites at half-filling, quenched from a Neel state with two different values of $g$ (columns) and three different values of $\kappa$ (rows).
		The two methods agree very well, even for long-range correlations.
		Only small deviations between \gls{HOPS} and \gls{QJ} are found for large electron\hyp phonon coupling $g=J$ and intermediate and strong dissipation $\kappa=1 \,J, \,4\,J$.
		We chose a timestep $\delta t = 0.005 \, J^{-1}$ for \gls{QJ} and $\delta t = 0.01 \, J^{-1}$ for \gls{HOPS}, since the latter method has shown to be less sensitive on the timestep.
	}
\end{figure}
\begin{figure}[h]
	\centering
	\ifthenelse{\boolean{buildtikzpics}}%
	{%
		\input{hops_qj_comparison_time_dependent_nfnf_correlations}
	}%
	{%
		\includegraphics{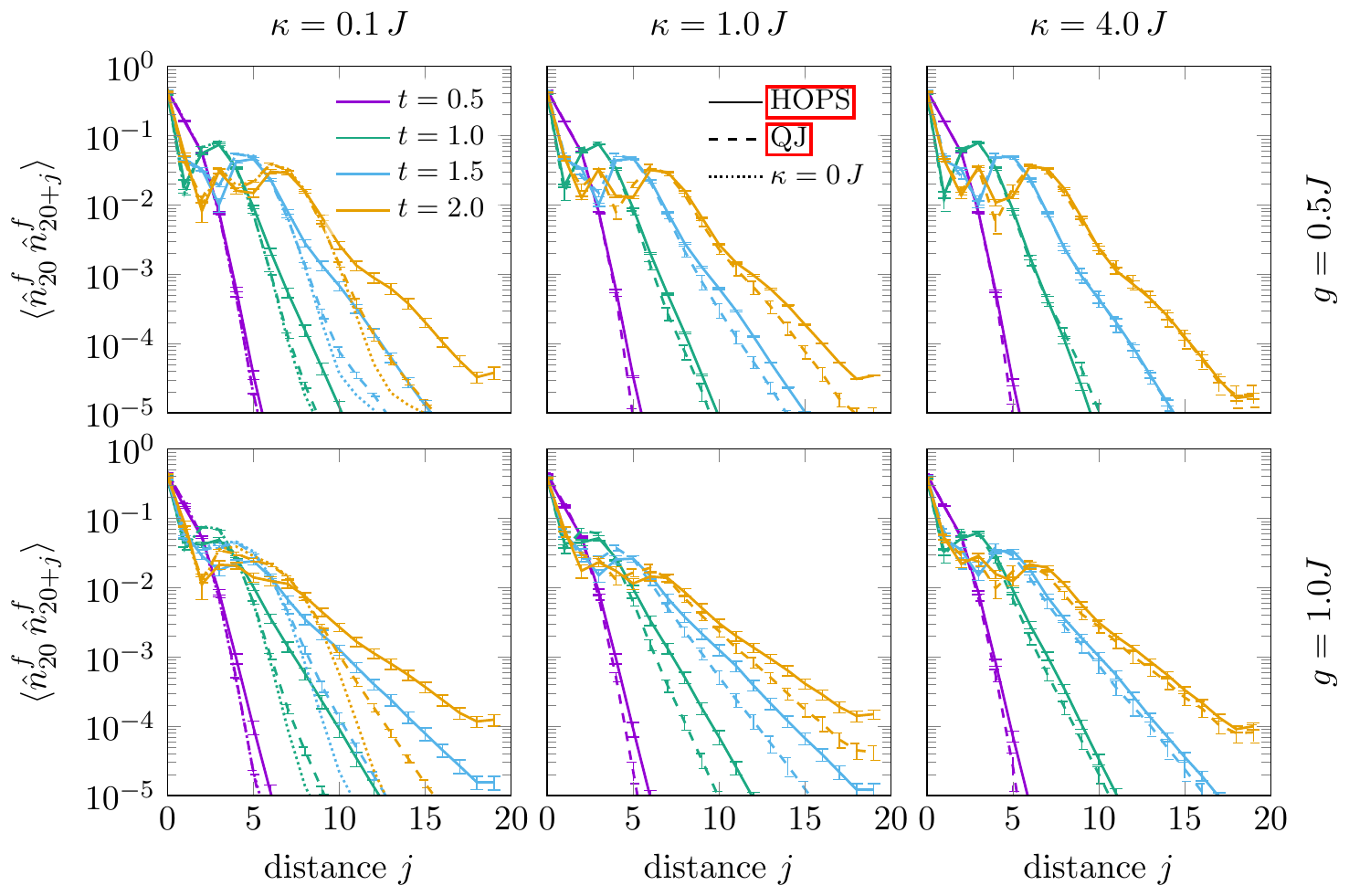}%
	}%
	\caption
	{
		\label{fig:NfNf}
		Charge-density correlations $\langle \hat{n}^f_{20} \, \hat{n}^f_{20+j} \rangle$ calculated with \gls{HOPS} and \gls{QJ} for the same quench as the one in \cref{fig:SzSz}.
		A disagreement is observed at later times and long distances.
		In the upper-left panel, at each time, we have added a dotted line that represents the case without dissipation, i.e. a simple Schr\"odinger evolution with the Hubbard-Holstein Hamiltonian.
		This indicates that the long-range correlations in case of very weak dissipation are better described by \gls{QJ} than by \gls{HOPS}.
	}
\end{figure}
We proceed with the comparison by considering the same parameters as in \cref{fig:comparison:hops-qj-ed:correlations} but increase the system size to $L=40$.
Such system sizes are far beyond reach for \gls{ED} methods, as well as density operator based time\hyp evolution schemes, in particular when considering a large number of phononic modes (here $\leq 40$) per site, too.
In order to ensure numerical convergence, throughout the benchmark calculations, we varied all relevant parameters.
\Cref{tab:simulation_parameters} displays the settings we found to produce faithful and converged results.
In particular, we fix the maximally allowed bond dimension to $m_\mathrm{max} = 6000$ and choose a time step $\delta t=0.01 \, J^{-1}$ for \gls{HOPS} and $\delta t=0.005 \, J^{-1}$ for \gls{QJ} and a discarded weight of $\delta=10^{-10}$. 
The maximally allowed hierarchy depth $k_{\text{\tiny{max}}}$ (for \gls{HOPS}) and local Hilbert space dimension of the phonons $\text{b}_{\text{\tiny{max}}}$ (for \gls{QJ}) are set to $k_{\text{\tiny{max}}} = b_{\text{\tiny{max}}} = 40$.
We find that these values are sufficient to describe the dynamics, and correspondingly, the actually exploited local dimensions never reach their respective upper limit.
Since the initial state is a product state, we start the time\hyp evolution with the global Krylov method and then switch to the \gls{2TDVP} method. 
Here, at least $\sim 50$ Krylov time\hyp evolution steps with otherwise identical numerical configuration are required in order to obtain converged results. 
Our comparisons aim to determine the model parameter regimes in which \gls{QJ} and \gls{HOPS} are capable of describing the many\hyp body post\hyp quench dynamics.
Since the dynamics are characterized by the spreading of correlations on different time scales, in the following, we concentrate on our results for the dynamics of spin-density and charge-density correlation functions w.r.t. the central site.
However, we note that during our investigations, both methods performed equally well when describing on\hyp site observables.
As shown in \cref{fig:SzSz}, the spin-density correlations agree very well for the two methods.
However, for the charge-density correlations displayed in \cref{fig:NfNf} we find deviations in the long\hyp distance behaviour for very weak dissipation.
An additional shoulder characterizes them in the tail of the correlation functions at times $t>1 \, J^{-1}$, occurring in the dynamics obtained from \gls{HOPS}.
This shoulder corresponds to an increased spreading of density correlations in the \gls{HOPS} result, compared to \gls{QJ} \footnote{We have checked that halving the timestep of \gls{HOPS} does not improve the results.}.
In order to clarify which method yields more reliable results in this regime, we performed a comparison to the quench dynamics in the absence of dissipation.
As shown in the upper-left panel of \cref{fig:NfNf} by the dotted curves, we find that \gls{QJ} smoothly connects to the non\hyp dissipative case.
We take this observation as an indicator that \gls{QJ} is more precise in the case of small dissipation strengths.
\begin{table}
	\centering
	\begin{tabular}{c c c c c}
		\toprule
			& \shortstack{ \textbf{d\hyp surface} \\ \cref{fig:k_max_surface,fig:4_points_dim_conv}} & \shortstack{ \textbf{\gls{QJ} and \gls{HOPS} } \\ \cref{fig:SzSz,fig:NfNf,fig:bond_dim} } & \shortstack{ \textbf{double } \\ \textbf{ occupations } \\ \cref{fig:hops:quench:hubbard-gs:double-occupations} } & \shortstack{ \textbf{ bipolaron } \\ \textbf{ metallicity } \\ \cref{fig:qj:quench:hubbard-holstein-gs:mobilities,fig:qj:quench:hubbard-holstein-gs:bipolarons:sp-ekin} } \\ [0.5ex] 
		\midrule
			$\delta$ & $10^{-10}$ & $10^{-10}$ & $10^{-10}$ &$10^{-10}$ \\
			$m_\mathrm{max}$ & 6000 & 6000 & 500 & 2000 \\
			$d_\mathrm{max}$ & 60 & 40 & 40 & 40 \\
			sites & 10 & 40 & 20 & 20 \\ 
			$\lvert \mathcal Q \rvert$ & 5 & 200 & 50 & 200\\ 
		\bottomrule
	\end{tabular}
	\caption
	{
		\label{tab:simulation_parameters} 
		Summary of the most relevant simulation parameters: 
		the max.~allowed discarded weight $\delta$, 
		the max.~allowed \gls{MPS} bond dimension $m_\mathrm{max}$, 
		the max.~allowed local dimension $d_\mathrm{max}$,
		and the overall number of trajectories $\lvert \mathcal Q \rvert$.
	}
\end{table} 
\subsection{Numerical complexity and stability}\label{sec:Phys_dim_conv}
\begin{figure}[!t]
	\centering
	\ifthenelse{\boolean{buildtikzpics}}%
	{%
		\input{hops_qj_kmax_comparison}
	}%
	{%
		\includegraphics{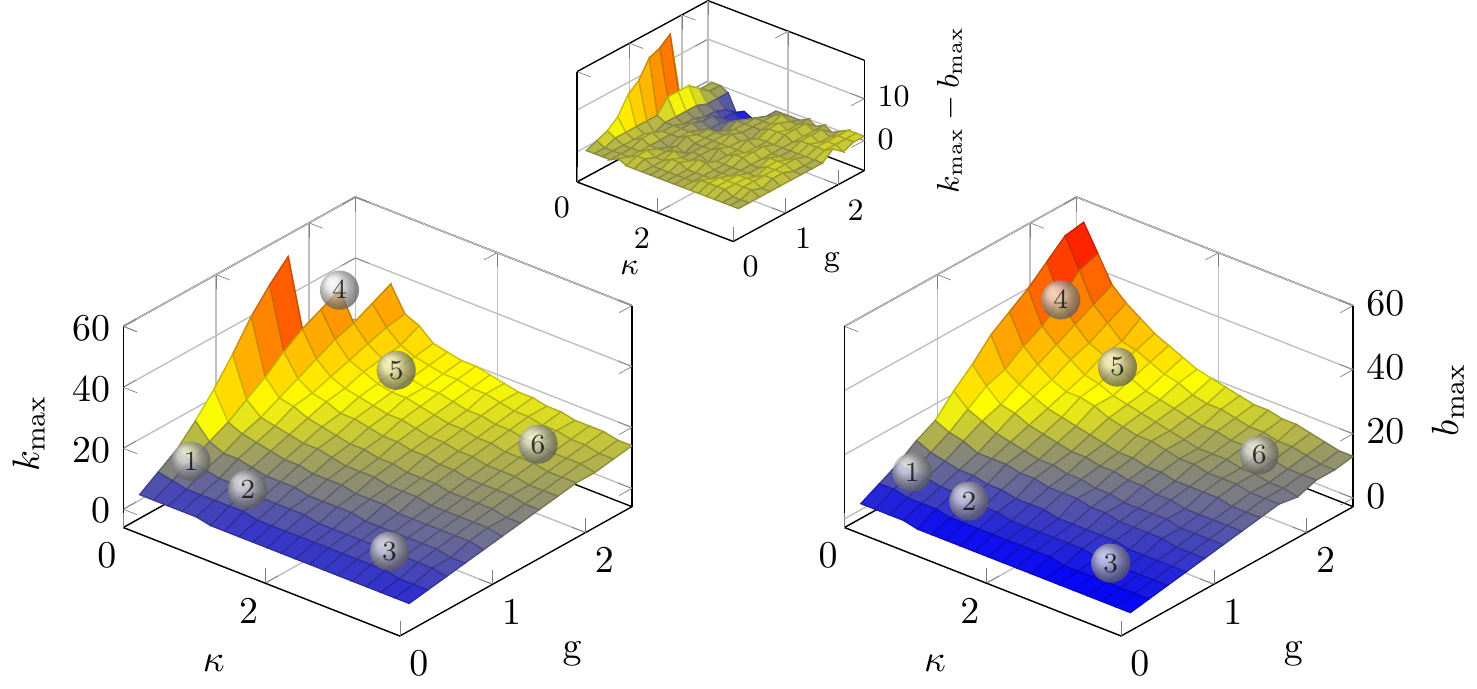}%
	}%
	\caption
	{
		\label{fig:k_max_surface} 
		Left: hierarchy dept $k_{\text{\tiny{max}}}$ for \gls{HOPS} as a function of $g$ and $\kappa$. 
		Right: local physical dimension of the phonons $\text{b}_{\text{\tiny{max}}}$ for \gls{QJ} $g$ and $\kappa$. 
		Center: difference between $k_{\text{\tiny{max}}}$ and $\text{b}_{\text{\tiny{max}}}$.
		The truncation is determined automatically via the \gls{PP} method by fixing a discarded weight $\delta=10^{-10}$. 
		For all calculations, the model parameters were $N_{\text{sites}}=10$, $U=J$, and $\omega = 2 \, J$ and time\hyp evolution has been performed until $T_{\text{max}}=2 \,J^{-1}$. We computed $10$ trajectories for each point.
		For the six points marked by a sphere, a convergence analysis of an observable is described in \cref{fig:4_points_dim_conv} and in the main text.
		Note that at very large $g$ and very small $\kappa$ (i.e., the scattered top left area in the left figure), \gls{HOPS} collapses, a finding which we discuss in the main text.
	}
\end{figure} 
From a practical point of view, it is important to clarify if the methods are numerically feasible in the identified optimal parameter regimes.
Here, we start by comparing the hierarchy depth $k_{\text{\tiny{max}}} \equiv d_\mathrm{max}$ for \gls{HOPS} with the local Hilbert space dimension of the phonons $\text{b}_{\text{\tiny{max}}} \equiv d_\mathrm{max}$ for \gls{QJ} for different values of $g$ and $\kappa$.
Using \gls{2TDVP} as time\hyp evolution method, the numerically most costly operations scale as $\mathcal O(m_\mathrm{max}^3 d_\mathrm{max}^2 \delta_\mathrm{max})$ and $\mathcal O(m_\mathrm{max}^2 d_\mathrm{max}^3 \delta_\mathrm{max}^2)$.
In case of considerably large local dimensions $d_\mathrm{max} > 10$, the latter operations become dominant and the applicability of \gls{QJ} and \gls{HOPS} depends on their required local Hilbert space dimensions.
In \cref{fig:k_max_surface} we show the evolution of $d_\mathrm{max} = k_{\text{\tiny{max}}}, \text{b}_{\text{\tiny{max}}}$ required to ensure an overall discarded weight $\delta = 10^{-10}$ throughout the time\hyp evolution.
Note that the \gls{PP}\hyp truncation scheme generically truncates the required local dimension so that the shown results already constitute the optimal number of local basis states that need to be kept.
Interestingly, we find that despite being conceptually very different, each method's required local Hilbert space dimensions $k_{\text{\tiny{max}}}$ (left plot) and $\text{b}_{\text{\tiny{max}}}$ (right plot), display a strikingly similar dependence on $g$ and $\kappa,$ throughout the whole analyzed parameter space. 
A broad connection between these two quantities is discussed for another model in Ref. \cite{Damanet2021}.
The shape of the surfaces drawn by $k_{\text{\tiny{max}}}$ and $\text{b}_{\text{\tiny{max}}}$ confirms our previous observation that in the case of strong electron\hyp phonon coupling and weak dissipation, many highly\hyp excited phononic modes are populated that can not escape due to dissipation, and thus large Hilbert space dimensions are required. 
Note that for \gls{HOPS} the top\hyp left corner of the $k_{\text{\tiny{max}}}$ surface is missing. 
This is due to the fact that for a few extreme cases of very strong electron\hyp phonon coupling and very weak dissipation, \gls{HOPS} becomes numerically unstable because the norm of the auxiliary states grows very large.
In \cref{app:sec:aux_states_rescaling} we show that, at least for the dissipative Hubbard\hyp Holstein model, this instability for the \gls{HOPS} method is much more severe when the original definition of the auxiliary states is adopted instead of the modified one of \cref{eq:new_auxiliary_states}.
We thus find that the numerical costs are equivalent for both methods when enforcing a certain discarded weight.
\begin{figure}[!t]
	\centering
	\ifthenelse{\boolean{buildtikzpics}}%
	{%
		\input{hops_qj_convergence}
	}%
	{%
		\includegraphics{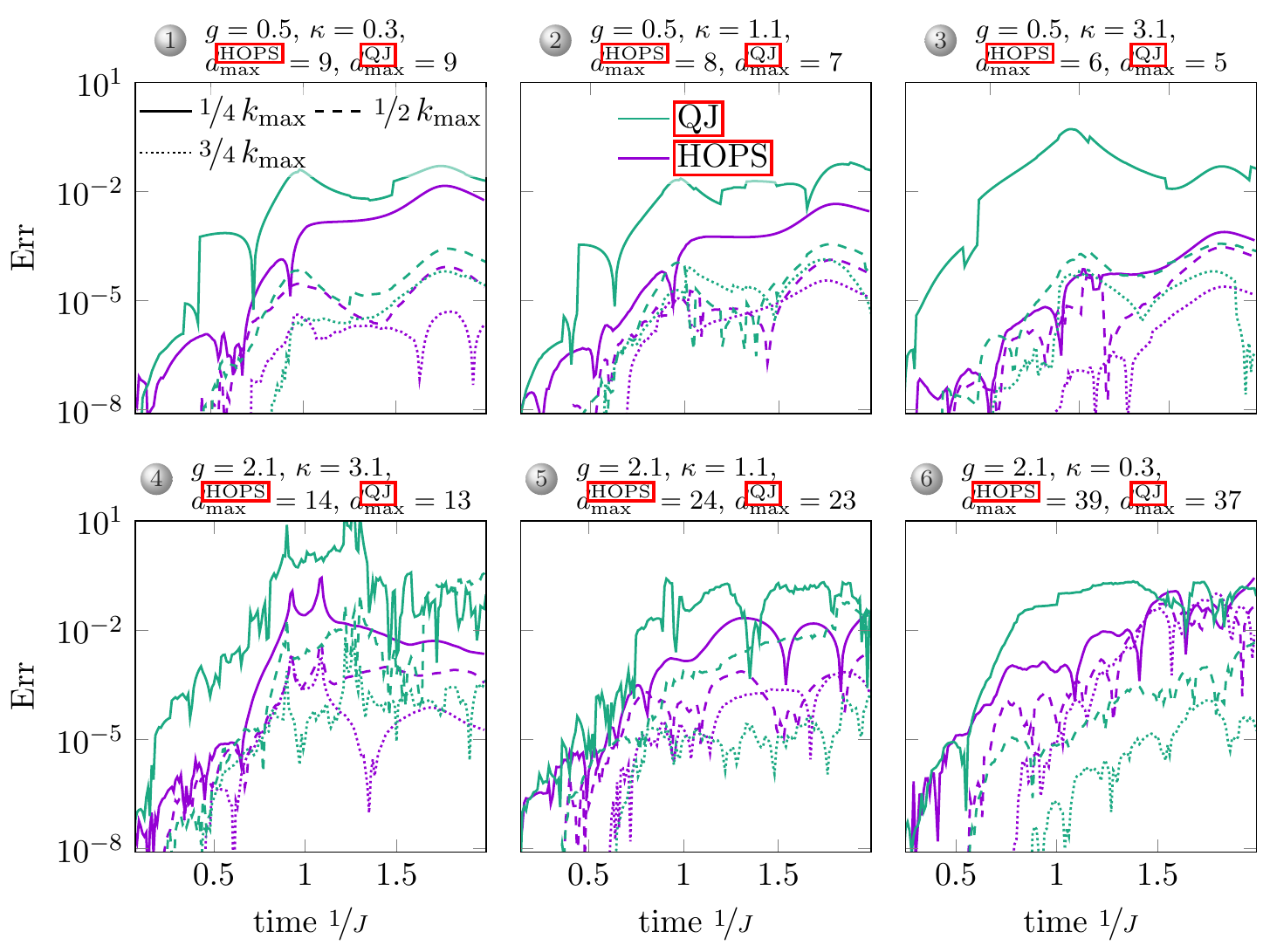}%
	}%
	\caption
	{
		\label{fig:4_points_dim_conv} 
		Convergence of nearest neighbour pairing correlation \cref{eq:nn-corr-pa} in the local Hilbert space dimension $j d_\mathrm{max}$.
		The chosen six parameter sets $(g,\kappa)$ are indicated in \cref{fig:k_max_surface}.
		At each time step, the relative error between a reference time evolution performed with the optimal local dimension $d_\mathrm{max}$ is evaluated.
		All the model and time evolution parameters are analogue to \cref{fig:k_max_surface}.
		Note that more than one trajectory is used only to avoid the risk of picking a particularly favourable or unfavourable combination of random numbers, but the error analysis here is not concerned with the statistical averaging performed for pure state methods.
	}
\end{figure}
When performing a time\hyp evolution, one is typically interested in the convergence of some specific observables and not in the approximation quality of the wave function controlled by the discarded weight.
Therefore, we pick six representative parameter points marked by circles in \cref{fig:k_max_surface} and studied the convergence of the nearest\hyp neighbor pairing correlation function:
\begin{equation}
\mathrm C^\mathrm{pa,d_\mathrm{max}}_\mathrm{nn} = \frac{1}{L-1} \sum_{j=1}^{L-1} \braket{\hat{c}^{\dagger}_{j, \uparrow} \: \hat{c}^{\dagger}_{j,\downarrow} \: \hat{c}^\nodagger_{j+1, \downarrow} \: \hat{c}^\nodagger_{j+1,\uparrow}} \;.
\label{eq:nn-corr-pa}
\end{equation}
We calculated its dependency on the maximally allowed local dimension, compared to a reference value $\hat{\text{C}}^{\text{pa,k}_{\text{max}}}_{\text{nn}}$ which was obtained fixing the discarded weight only and using the values of $k_\mathrm{max},\mathrm b_\mathrm{max}$ extracted from \cref{fig:k_max_surface}:
\begin{equation}
\text{Err}(j) = \left \lvert \left(\mathrm C^\mathrm{pa,d_\mathrm{max}}_\mathrm{nn} - \mathrm C^\mathrm{pa,d_\mathrm{max} j}_\mathrm{nn}\right)/\mathrm C^\mathrm{pa,d_\mathrm{max}}_\mathrm{nn} \right\rvert \; .
\end{equation}
Here, for both methods, we varied $j \in \{ 1/4, 1/2, 3/4 \}$, reducing the maximally allowed local dimension up to a quarter of the optimal value.
In \cref{fig:4_points_dim_conv}, we show the obtained convergence for the different fractions $j$ indicated by the different line styles. 
We observe that most of the time, the \gls{HOPS} curves lay below the \gls{QJ} curves, i.e., they exhibit less sensitivity on truncating the local Hilbert space dimension.
Noting that in \gls{HOPS}, the bosonic degrees of freedom represent auxiliary states with no direct physical meaning, it is reasonable to expect it to be somewhat less sensitive on truncations in the bosonic Hilbert space than \gls{QJ}.
\begin{figure}[!t]
	\centering
	\ifthenelse{\boolean{buildtikzpics}}%
	{%
		\input{hops_qj_bond_dimension_comparison}
	}%
	{%
		\includegraphics{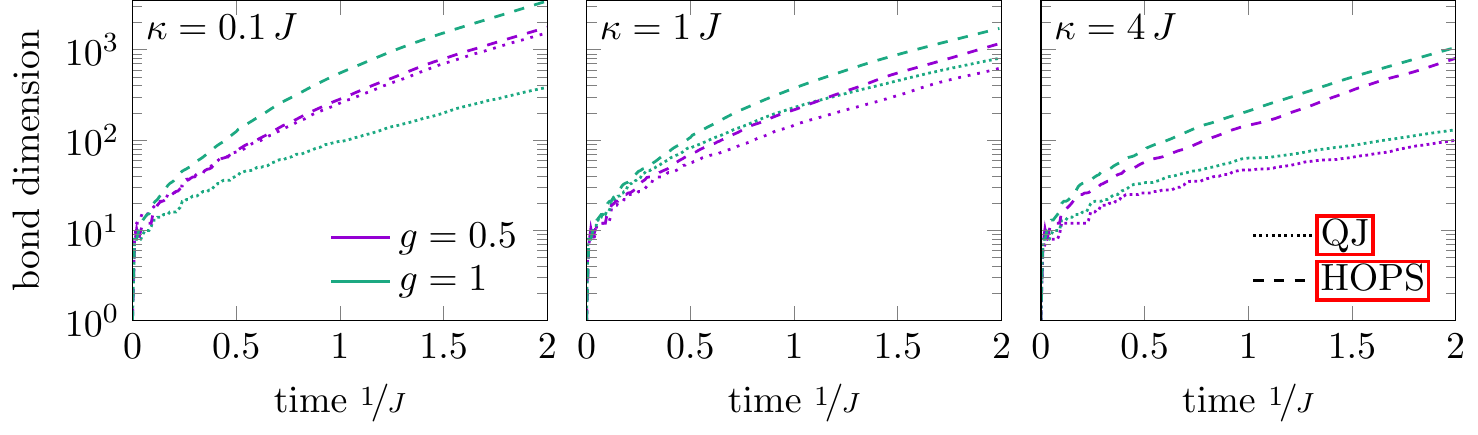}%
	}%
	\caption
	{
		\label{fig:bond_dim} 
		Bond dimension for \gls{QJ} and \gls{HOPS} during time evolution after the global quenches, specified in the caption of \cref{fig:SzSz}.
	}
\end{figure}
Aside from the local dimension, we also analyzed the bond dimension $m_\mathrm{max}$, which is of particular importance when using the \gls{PP} mapping, as it also controls the approximation quality of the phonon \glspl{1RDM} (c.f.,~\cref{sec:PP}).
The results are displayed in \cref{fig:bond_dim} for the same model parameters as for the benchmark calculation shown in \cref{fig:SzSz,fig:NfNf}.
Similarly to the local dimensions, the required bond dimensions decrease when the dissipation strength increases.
Notably, we find that for all six analyzed $(g,\kappa)$, \gls{QJ} features a smaller bond dimension than \gls{HOPS} when enforcing a constant discarded weight.
We investigated the possible origins of this surprising observation.
One possible reason may be buried in the fact that whenever a jump occurs, e.g., an annihilator is applied on a phononic site, the bond dimension drops significantly because a large portion of the local Hilbert space is projected out.
Moreover, it has been shown recently that repeated measurements reduce the support of lattice sites, on which correlations can spread significantly~\cite{Tang2020,Doggen2022} and thereby also reduce entanglement growth.
Since in \gls{QJ} the probability for a jump to happen is mainly controlled by the dissipation strength, we would expect considerably smaller bond dimensions to happen if $\kappa$ is large, as observed in the right panel of~\cref{fig:bond_dim}.
Furthermore, for small dissipation strengths and large electron\hyp phonon interactions, we also observed a significant increase in the required local dimension of \gls{HOPS}.
Since in the \gls{PP} mapping, the required local dimension is directly connected to the decay of the phonon \gls{1RDM} diagonal elements, \gls{HOPS} seems to have the tendency to create more substantial fluctuations in the phonon system in this parameter regime and thereby increases the overall bond dimension.
However, deciding whether the overall trend displayed in \cref{fig:bond_dim} is a peculiar feature of the analyzed systems or a general feature is beyond the scope of this work.
\section{Quantum State Diffusion}\label{app:sec:homodyne-detection}
\subsection{Linear and non\hyp linear homodyne detection}
\begin{figure}[!t]
	\centering
	\ifthenelse{\boolean{buildtikzpics}}%
	{%
		\input{homodyne_detection_me_comp}%
	}%
	{%
		\includegraphics{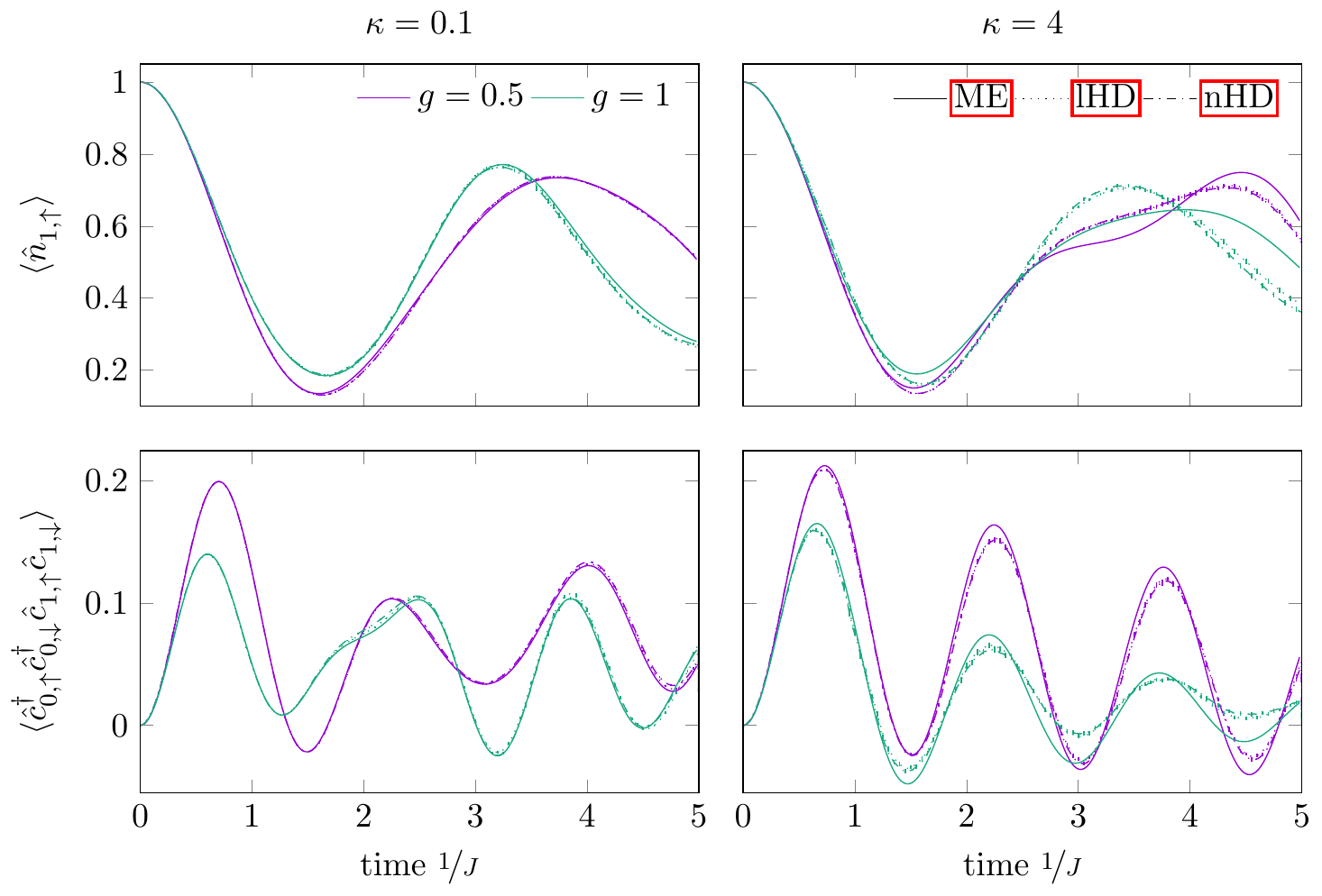}%
	}%
	\caption%
	{%
		\label{fig:Ed_Homodyne}%
		Comparing the linear and non\hyp linear homodyne detection to the exact \gls{ME}. 
		All results were computed with \gls{ED} for a system composed of two fermionic and two phononic sites.
		For all plots, the Hamiltonian parameters were chosen to be $U = \, J$, $\omega = 2\, J$, and 500 trajectories and a time step $dt = 0.005 \, J^{-1}$ were used. 
		The dissipation strength $\kappa$ was fixed to $ \kappa = 0.1 \, J$ for the two upper plots and to $\kappa=4 \, J$ for the two lower plots. 
		For all analyzed parameters, the linear and the non-linear method present at most very small differences.
		While both methods agree reasonably well with the exact results for a small dissipation strength, for strong dissipation, the homodyne detection results deviate strongly from the exact ones, and the problem becomes more severe for large electron-phonon coupling.
	}
\end{figure}
An alternative unravelling of the Lindblad master equation \cref{eq:Lindblad} is given by the so\hyp called \gls{lHD} \cite{DeVegaReview}.
Similarly to \gls{HOPS}, the stochastic part is represented by a random noise term contained in the effective Hamiltonian.
For each trajectory $\mathcal{Q}$, the time\hyp evolution is generated by the non\hyp hermitian Hamiltonian (\cite{wiseman_milburn_2009}) : 
\begin{equation}
\hat{H}_{\text{\tiny{eff}}}^{\text{\tiny{Q}}} = \hat{H}_{\text{\tiny{s}}} + i\sum_l\big[Z_l(t) \hat{D}_l -\frac{C_l}{2} \hat{D}_l^{\dagger} \hat{D}_l\big]\;,
\label{eq:Heff_lin_homodyne_det}
\end{equation}
where $\hat{H}_{\text{\tiny{s}}}$ is the system Hamiltonian, $\hat{D}_l$ are the Lindblad operators, and $Z_l(t)$ is a random number drawn from a real-valued Gaussian distribution with mean zero and standard deviation $\sigma$ given by the square root of the coupling parameter $C_l$ divided by the time step $\delta t$.
To show the equivalence between the Lindblad evolution and \gls{lHD} method, we time\hyp evolve a state $\ket{ \Psi(t) }$ to first order with the effective Hamiltonian of \cref{eq:Heff_lin_homodyne_det}, 
considering the case of only one Lindblad operator for clarity: 
\begin{equation}
\ket{ \Psi(t+\delta t) } = \big[ 1 +\delta t \big( -i \hat{H}_s + \hat{D} Z(t) -\frac{C}{2} \hat{D}^{\dagger}\hat{D} \big) \big] \ket{ \Psi(t) } \;.
\label{eq:homodyne_det_diff_form}
\end{equation}
To first order in $\delta t$ (recalling that $Z^2$ is $\mathcal{O}(\delta t^{-1})$ ), the outer product of \cref{eq:homodyne_det_diff_form} with its hermitian conjugate reads:
\begin{equation*}
\begin{split}
\ket{ \Psi(t+\delta t) }\bra{ \Psi(t+\delta t) } &= \ket{ \Psi(t) }\bra{ \Psi(t) } + \delta t \big ( -i \hat{H}_s + \hat{D} Z(t) -\frac{C}{2} \hat{D}^{\dagger} \hat{D} \big) \ket{ \Psi(t) }\bra{ \Psi(t) } \\
& +\ket{ \Psi(t) }\bra{ \Psi(t) } \delta t \big( +i \hat{H}_s + \hat{D}^{\dagger} Z(t) -\frac{C}{2} \hat{D}^{\dagger}\hat{D} \big)\\
&+ \delta t^2 Z^2(t) \hat{D} \ket{ \Psi(t) }\bra{ \Psi(t) } \hat{D}^{\dagger} \;.
\label{eq:homodyne_projector}
\end{split}
\end{equation*}
Now, by making use of the mean and the variance of $Z$, namely $\mathcal{E} [Z(t)] = 0$ and $\mathcal{E} [Z^2(t)] = C/\delta t$ we compute the ensemble average over the projectors:
\begin{equation*}
\begin{split}
\mathcal{E} [ \ket{ \Psi(t+\delta t) }\bra{ \Psi(t+\delta t) } ] &\equiv \hat{\rho}(t+\delta t) = \hat{\rho}(t) + \delta t( -i \hat{H}_s -\frac{C}{2} \hat{D}^{\dagger} \hat{D} ) \hat{\rho}(t) \\
&+ \hat{\rho}(t) \delta t ( +i \hat{H}_s -\frac{C}{2} \hat{D}^{\dagger}\hat{D} ) + \delta t^2 \frac{C}{\delta t} \hat{D} \hat{\rho}(t) \hat{D}^{\dagger} \ \\
& = \hat{\rho}(t) +\delta t \big( -i[ \hat{H}_s, \hat{\rho}(t) ] -\frac{C}{2} \{ \hat{D}^{\dagger} \hat{D},\hat{\rho}(t) \} + C \hat{D} \hat{\rho}(t) \hat{D}^{\dagger} \big) \;,
\label{eq:homodyne_projector_averaging}
\end{split}
\end{equation*}
which, in the limit $\delta t \to 0$, is the Lindblad equation.
For the case of the dissipative Hubbard\hyp Holstein model considered in \cref{sec:benchmark,sec:results}, the effective Hamiltonian reads:
\begin{equation}
\hat{H}_{\text{\tiny{eff}}}^{\text{\tiny{Q}}} = \hat{H}_{\text{\tiny{HH}}} + i\sum_{j=1}^L\big[Z_j(t) \hat{a}_j -\frac{\kappa}{2} \hat{a}_j^{\dagger} \hat{a}_j\big] \; ,
\label{eq:Heff_lin_homodyne_det_Hu_Ho}
\end{equation}
with the constant coupling being the dissipation strength $\kappa$.
To try to lower the number of trajectories needed to converge the observables for this pure\hyp state method, a modification of \cref{eq:Heff_lin_homodyne_det} called \gls{nlHD} can be used (\cite{wiseman_milburn_2009}): 
\begin{equation}
\hat{H}_{\text{\tiny{eff}}}^{\text{\tiny{Q}}} = \hat{H}_{\text{\tiny{s}}} + i\sum_{j=1}^L \big[Z_j(t) \hat{D}_j -\frac{C_j}{2} \hat{D}_j^{\dagger} \hat{D}_j +C_j\bra{\Psi(t)}( \hat{D}_j^{\dagger}+ \hat{D}_j)\ket{\Psi(t)}\hat{D}_j \big]\;.
\label{eq:Heff_nonlin_homodyne_det}
\end{equation}
Analogously to what is done for \gls{HOPS}, the non\hyp linear dynamics generated by Hamiltonian \cref{eq:Heff_nonlin_homodyne_det} are linearized by computing the expectation value with the state $\ket{\Psi(t-\delta t)}$, which is a reasonable approximation as long as the time step $\delta$ is small.
We show the \gls{ED} comparison of both the linear and the non\hyp linear homodyne detection methods to the \gls{ME} methods for the same parameters used in \cref{fig:comparison:hops-qj-ed:correlations}.
\Cref{fig:Ed_Homodyne} shows that lHD and nlHD work well for small dissipation but fail to yield correct results both for single\hyp site and for two\hyp site observables in the case of large dissipation. 
We thus conclude that the \gls{QJ} method is more suitable to be used as a comparison to \gls{HOPS}. 

\subsection{Exact factorization of the time\hyp evolution operator}
The matrix elements of the non\hyp hermitian part of the effective Hamiltonian can be computed exactly, both for the linear and the non\hyp linear case. 
Also, the \gls{MPO}\hyp representation of the phononic displacement operator used for the computation in \cref{sec:results} is obtained in a completely analogous way.
We first consider the linear case \cref{eq:Heff_lin_homodyne_det_Hu_Ho}, define $\hat{B} \equiv \sum_{j=1}^L\big[Z_l(t) \hat{a}_l -\frac{\kappa}{2} \hat{a}_l^{\dagger} \hat{a}_l\big]$ and start by factorizing the exponential of the effective Hamiltonian via a second\hyp order Trotter decomposition:
\begin{equation}
e^{-i( \hat{H}_{\text{\tiny{HH}}} +i\hat{B} )\delta t} \approx e^{\hat{B} \delta t/2}\cdot e^{-i \hat{H}_{\text{\tiny{HH}}} \delta t}\cdot e^{\hat{B}\delta t/2} +\mathcal{O}(\delta t^3). 
\end{equation} 
We then focus on calculating the exponential $e^{B\delta t}$. 
Since the terms acting on each site commute, the expression
\begin{equation*}
e^{\sum_{j=1}^{L}\big[Z_j (t) \hat{a}_j -\frac{\kappa}{2} \hat{a}_j^{\dagger} \hat{a}_j\big]\delta t}= e^{\big[Z_1(t) \hat{a}_1 -\frac{\kappa}{2} \hat{a}_1^{\dagger} \hat{a}_1\big]\delta t}\cdot e^{\big[Z_2(t) \hat{a}_2 -\frac{\kappa}{2} \hat{a}_2^{\dagger} \hat{a}_2\big]\delta t}\dots e^{\big[Z_L(t) \hat{a}_L -\frac{\kappa}{2} \hat{a}_L^{\dagger} \hat{a}_L\big]\delta t}
\end{equation*}
is exact.
We consider the expression for one site and drop the site subscript and the explicit time dependency of $Z$:
\begin{equation}
e^{\big[Z \hat{a}-\frac{\kappa}{2} \hat{a}^{\dagger} \hat{a} \big]\delta t} \;.
\label{eq:original_operator}
\end{equation}
We now want to write this exponential as a product of two exponentials.
We use the following theorem from Ref. \cite{Hall_book}: 
Given two operators $\hat{X}$ and $\hat{Y}$, if $[\hat{X}, \hat{Y}] = s\hat{Y} $ with $s \in \mathbb{C}, s \neq 2\pi in, n\in \mathbb{N}$, then $e^{\hat{X}} \cdot e^{\hat{Y}} = \exp{( \hat{X} +\frac{s}{1-e^{-s}} \hat{Y} )}$.
Applied to \cref{eq:original_operator}, this theorem implies that:
\begin{equation}
e^{\big[Z \hat{a} -\frac{\kappa}{2} \hat{a}^{\dagger} \hat{a} \big]\delta t} = e^{\big[-\frac{\kappa}{2} \hat{a}^{\dagger} \hat{a} +\frac{s}{1-e^{-s}}\Tilde{Z} \hat{a} \big]\delta t} = e^{-\frac{\kappa}{2} \hat{a}^{\dagger} \hat{a} \delta t}\cdot e^{\Tilde{Z} \hat{a}\delta t} \;,
\label{eq:splitting}
\end{equation} 
with $\Tilde{Z} = Z\frac{1-e^{-s}}{s}$, $s = \frac{\kappa}{2}\delta t\;$.
Finally, the factorized operator reads:
\begin{equation}
e^{\big[Z \hat{a}-\frac{\kappa}{2} \hat{a}^{\dagger} \hat{a} \big]\delta t}= e^{-\frac{\kappa}{2} \hat{a}^{\dagger} \hat{a} \delta t}\cdot e^{Z\frac{1-e^{-\kappa\delta t/2}}{\kappa\delta t/2} \hat{a} \delta t}= e^{-\frac{\kappa}{2} \hat{a} ^{\dagger} \hat{a} \delta t} \cdot e^{Z\frac{1-e^{-\kappa\delta t/2}}{\kappa/2} \hat{a} }.
\label{eq:split_exponential}
\end{equation}
The operator $e^{Z\frac{1-e^{-K\delta t/2}}{\kappa/2} \hat{a}}$ does not conserve the bosonic particle number.
The $U(1)$ symmetry is restored in the \gls{PP} mapping, by replacing the annihilator $\hat{a}$ with $\hat{a} \otimes \hat{b} ^{\dagger}$, where $\hat{b}^{\dagger}$ is the balancing operator acting on the bath site. 
By defining the prefactor as $\gamma(Z)$ we get:
\begin{equation*}
e^{-\frac{\kappa}{2} \hat{a}^{\dagger} \hat{a} \delta t}\cdot e^{\gamma(Z) \hat{a} \otimes \hat{b}^{\dagger}}\;.
\end{equation*}
We now want to calculate the \gls{MPO} representation of the dissipative operator:
We thus compute the matrix elements:
\begin{align}
&\bra{n,n'}e^{-\frac{\kappa}{2} \hat{a}^{\dagger} \hat{a} \delta t}\cdot e^{\gamma(Z) \hat{a}\otimes \hat{b}^{\dagger}}\ket{m,m'} = e^{-\frac{\kappa}{2}n\delta t}\sum_{l=0}^{\infty} \frac{\gamma(z)^l}{l!}\bra{n} \hat{a}^l\ket{m}\bra{n'}( \hat{b}^{\dagger})^l\ket{m'} = \\
& e^{-\frac{\kappa}{2}n\delta t}\sum_{l=0} \frac{\gamma(z)^l}{l!}\sqrt{\frac{(l+n)!}{n!}}\delta t_{n+l,m}\delta t_{n',m'+l} =
\begin{cases}
& 0, \hspace{0.5cm}n > m \\
& \frac{e^{-\frac{\kappa}{2}n\delta t}}{(m-n)!}\gamma(Z)^{m-n} \sqrt{\frac{m!}{n!}}\delta t_{n'-m',m-n}, \hspace{0.5cm} \text{otherwise.}
\end{cases}
\end{align}
We can rewrite the rank 4-tensor $\delta t_{n'-m',m-n}$ as 
\begin{equation*}
\delta t_{n'-m',m-n} = \sum_{a = 0}^{d-1}\delta t_{n'-m',a}\delta t_{m-n,a}\;.
\end{equation*}
Thus we get the expression:
\begin{equation}
e^{-\frac{\kappa}{2} \hat{a}^{\dagger} \hat{a} \delta t}\cdot e^{\gamma(Z) \hat{a}\otimes \hat{b}^{\dagger}} = \sum_{n,m,n',m',a} \frac{e^{-\frac{\kappa}{2}n\delta t}}{(m-n)!}\gamma(Z)^{m-n} \sqrt{\frac{m!}{n!}}\hspace{0.1cm}W_{1,a}^{(p)n,m}W_{a,1}^{(pp)n',m'}\ket{n}\bra{m}\otimes\ket{n'}\bra{m'}\;,
\label{eq:factorized_MPO}
\end{equation}
with
\begin{equation}
\begin{cases}
W_{1,a}^{(p)n,m} = \delta t_{m-n,a} \\
W_{a,1}^{(pp)n',m'} = \Tilde{W}_{a,1}^{(pp)n',m'} = \delta t_{n'-m',a}\;.
\end{cases}
\label{eq:W_tensors}
\end{equation}
At this point, obtaining the exact factorization of the effective Hamiltonian for the non\hyp linear homodyne detection is straightforward.
We start by defining $\kappa\bra{\Psi(t)}( \hat{a}_j^{\dagger}+\hat{a}_j)\ket{\Psi(t)}\equiv f$, considering a single site, dropping the $j$ subscript and writing
\begin{equation}
e^{ (Z+f)\delta t \hat{a} -\frac{\kappa}{2}\delta t \hat{a}^{\dagger} \hat{a} }.
\end{equation}
We see that the operator has the same form as \ref{eq:original_operator} with $Z+f$ instead of $f$. 
Thus the factorized operator has the form:
\begin{equation}
e^{ (Z+f)\delta t \hat{a} -\frac{\kappa}{2}\delta t \hat{a}^{\dagger} \hat{a}}=e^{-\frac{\kappa}{2} \hat{a}^{\dagger} \hat{a}\delta t}\cdot e^{(Z+f)\frac{1-e^{-\kappa\delta t/2}}{\kappa/2} \hat{a} }\;.
\end{equation}
The \gls{MPO} form of this operators is given by \cref{eq:factorized_MPO,eq:W_tensors} with $\gamma(Z)=(Z+f)\frac{1-e^{-\kappa \delta t/2}}{\kappa/2}$.
\section{Physical Motivation for the System-Environment Model}\label{app:sec:phys_motiv}
Typical physical systems are immersed in a single global environment. 
For example, electrons in a real material are coupled to the atoms in the crystal structure, which vibrate collectively through excited phonon modes. 
In this section, we sketch out the justification and physical approximations required for mapping a system coupled globally to an environment with a continuum of energy modes to the toy models that we have considered in this paper, where we have an effective (independent) mode coupled locally to each site of the lattice, with an effective correlation function that decays in time.
We begin with a system-environment interaction in the linear form,
\begin{equation}
\hat{H}_{\rm Int} = \sum_{j,k} g_{j,k} \hat{L}_j \hat{a}^{\dagger}_k + g^*_{j,k} \hat{L}_j^{\dagger} \hat{a}_k,
\end{equation}
where $\hat{L}_j$ act on system site $j$, $\hat{a}_k$ annihilates an excitation in mode $k$ of the environment and the $g_{j,k}$ are some complex coefficients describing the coupling strength which in general are $k$ dependent and may also be spatially inhomogeneous. 
We can then define \textit{effective} environment modes,
\begin{equation}
\begin{split}
\hat{\tilde{B}}\nodagger_j &= \sum_{k} g^*_{j,k} \hat{a}_k \\
\hat{\tilde{B}}_j^{\dagger} &= \sum_{k} g_{j,k} \hat{a}_k^{\dagger},
\end{split}
\end{equation}
allowing us to write the interaction Hamiltonian as,
\begin{equation}
\hat{H}_{\rm Int} = \sum_{j} \hat{L}_j \hat{\tilde{B}}^{\dagger}_j + \hat{L}_j^{\dagger} \hat{\tilde{B}}\nodagger_j,
\end{equation}
which is now in the form of the electron-phonon coupling in the Hubbard\hyp Holstein model considered in the main text. 
However, we also need to consider the correlations between different effective environment modes, which in general will be non\hyp zero and so not independent,
\begin{equation}
\begin{split}
\langle \hat{\tilde{B}}_{j'}(t') \hat{\tilde{B}}^{\dagger}_j(t) \rangle & = \sum_{k,k'} g_{j,k} g^*_{j',k'} \langle \hat{a}_{k'}(t') \hat{a}^{\dagger}_k(t) \rangle \\
& = \sum_{k,k'} g_{j,k} g^*_{j',k'} e^{-i\omega_kt + i\omega_{k'} t'} \langle \hat{a}_{k'} \hat{a}^{\dagger}_k \rangle \\
& = \sum_{k,k'} g_{j,k} g^*_{j',k'} e^{-i\omega_kt + i\omega_{k'} t'} \delta_{k,k'} \\
& = \sum_{k} g_{j,k} g^*_{j',k} e^{-i\omega(t-t')},
\end{split}
\end{equation}
where in the second to last line, we have used the (zero\hyp temperature) relation, $ \langle \hat{a}_{k'} \hat{a}^{\dagger}_k \rangle = \delta_{k,k'} $, valid if the operators $\hat{a}^{\dagger}_k$ are the eigenmodes of the environment Hamiltonian, i.e., the environment is a collection of non\hyp interacting bosons $\hat{H}_E = \sum_k \omega_k \hat{a}^{\dagger}_k \hat{a}_k $.
Next, we assume that the magnitudes of the coupling coefficients are homogeneous, but there can be a relative phase factor,
\begin{equation}
g_{j,k} = g_k e^{-ikja},
\end{equation}
where $a$ is the spacing between lattice sites. We then arrive at the expression for the correlation functions,
\begin{equation}
\langle \tilde{B}_{j'}(t') \tilde{B}^{\dagger}_j(t) \rangle = \sum_{k} |g_k|^2 e^{-ika(j - j')} e^{-i\omega(t-t')}.
\label{eq:bath_corr_funct_stuart}
\end{equation}
\\
Following \cite{stuart_phd_thesis}, we consider strong lattice confinement so that the eigenstates of the harmonic oscillator can approximate the localized basis for the fermions. 
Then, the coupling coefficients between such fermionic states and a continuous bosonic excitation in the environment described by a plane wave can be written as:
\begin{equation}
g_k^{\alpha,\beta} \propto \int \mathrm{d} z \, \Phi^{\alpha*}(z)\Phi^{\beta}(z) e^{-ikz},
\end{equation}
where $\Phi^{\alpha}(z)$ is the $\alpha$\hyp th eigenstate of the harmonic oscillator:
\begin{equation}
\Phi^{\alpha}(z) = \frac{1}{ \sqrt{2^{\alpha}n! } } (\pi a^2)^{-1/4} e^{\frac {z^2}{2 a_z} } H^{\alpha} \left( \frac{z}{a_z} \right) \, ,
\end{equation}
where $a_z=\sqrt{1/m\omega_z}$ and $H^{\alpha}$ are the Hermite polynomials. 
Assuming only the ground states $\Phi^{0}(z)$ to be occupied, we can compute the coupling coefficients exactly:
\begin{equation}
g_k = g e^{-k^2a^2/2},
\label{eq:gk_analytic_integral}
\end{equation}
where we have assumed a momentum\hyp independent prefactor $g$.
We now consider a linear dispersion relation $\omega = c k$ and insert the expression for $g_k$ into \cref{eq:bath_corr_funct_stuart}. 
If $ka \gg 1$, then for $j \neq j'$, we get a large oscillating component in the sum, which leads to a vanishingly small correlation. 
This corresponds to the so-called large wave-vector limit, which is valid if the characteristic wavelength of excitations in the environment $\lambda_{\rm eff}$ is much smaller than the spacing between system lattice sites. 
Approximating the sum with an integral for $j=j'$ we obtain:
\begin{equation}
\begin{split}
\langle \hat{\tilde{B}}^\nodagger_{j'}(t') \hat{\tilde{B}}^{\dagger}_j(t) \rangle & = \sum_{k} |g_k|^2 e^{-i\omega(t-t')} \approx \int_0^{\infty} \mathrm{d} k \, |g_k|^2 e^{-ick(t-t')} \approx \int_0^{\infty} \mathrm{d} k \, g ^2 e^{-k^2a^2-ick(t-t')} 
\\ & =\sqrt{ \frac{\pi}{a^2}} g ^2 e ^{-c^2/4a^2 (t-t')^2 } \equiv{\alpha(t-t')}.
\end{split}
\label{eq:bath_correlation_function_gaussian}
\end{equation}
\begin{figure}[!t]
	\centering
	\ifthenelse{\boolean{buildtikzpics}}%
	{%
		\input{fit_correlation_function}%
	}%
	{%
		\includegraphics{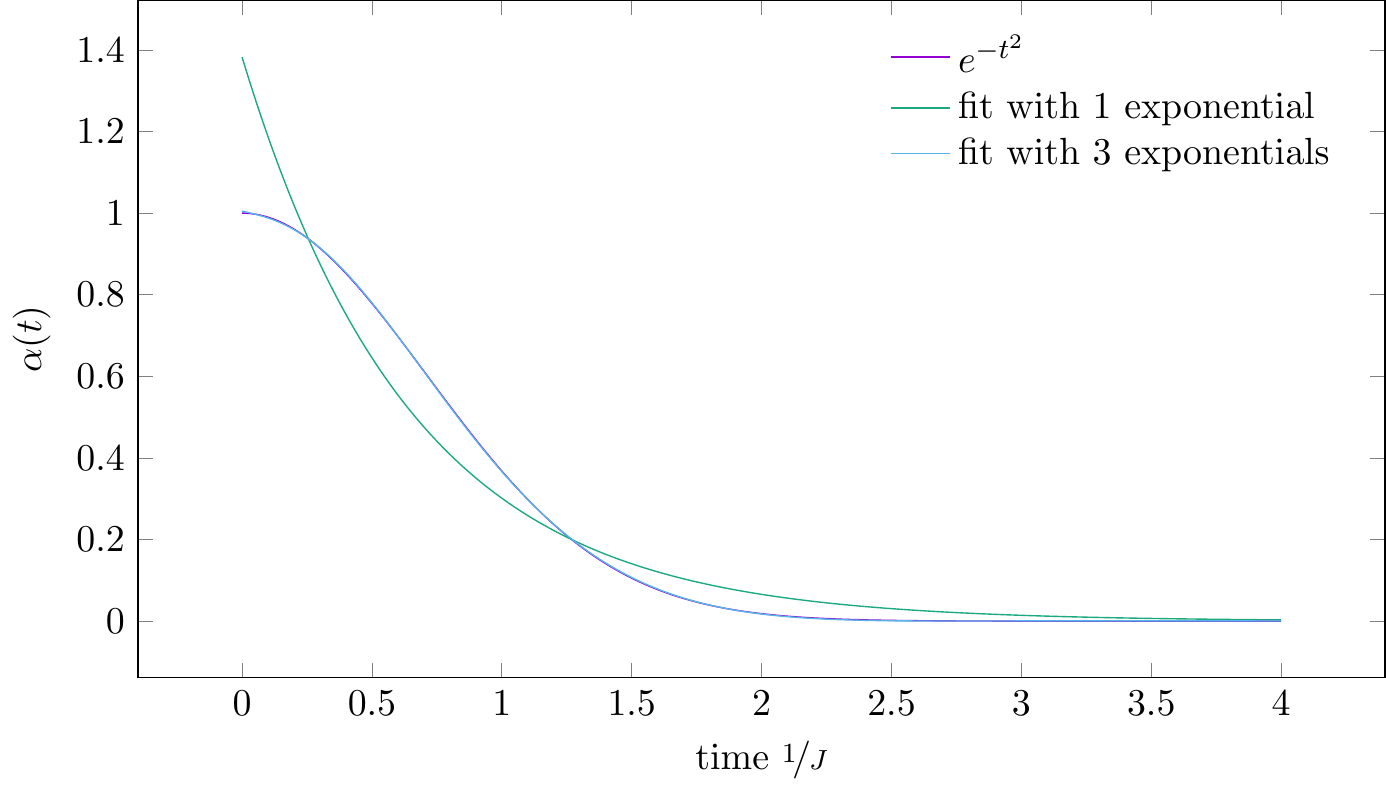}%
	}%
	\caption
	{
		\label{fig:fit_correlation_function}
		Bath correlation function \cref{eq:bath_correlation_function_gaussian} approximated with 1 and with 3 complex exponentials via the Laplace-Pade method for $c=g=a$.
	}
\end{figure}
In \cref{fig:fit_correlation_function} we approximate the correlation function \cref{eq:bath_correlation_function_gaussian} via the Laplace\hyp Pade method \cite{Laplace_Pade}.
It can be seen that already three complex exponentials suffice to reproduce the correlation function fairly well. 
\\
This then allows us to connect our work presented here to a wider variety of more realistic physical systems. 
An interesting future research direction would be analyzing what happens when this small wavelength limit is not satisfied, giving rise to strong correlations between the different environment modes.

\section{Failure of the Markovian Description of the Dissipative Hubbard\hyp Holstein Model}\label{app:sec:markovian_me}
The non\hyp Markovian method outlined in \cref{sec:hops} and the Markovian one for the enlarged physical system (electrons $+$ phonons) discussed in \cref{sec:q-jumps} are numerically challenging.
Thus, one could wonder whether a much simpler Markovian master equation for the electronic system only would suffice to describe the dynamics correctly. 
Such an equation was derived in \cite{hops_stuart_francois} and reads:
\begin{equation}
\partial_t \hat{\rho} = -i [\hat{H}_{\text{\tiny{f}}}, \hat{\rho}] +g^2 \big( \sum_{j=1}^L \hat{n}_j\hat{\rho} \hat{n}_j -\frac{1}{2} \{ (\hat{n}_j)^2, \hat{\rho} \} \big ) \;,
\label{eq:markovian_master_eq_for_electrons}
\end{equation}
where $\hat{H}_{\text{\tiny{f}}}$ is the Hubbard Hamiltonian, $g$ the electron-phonon coupling and $\hat{n}_j$ the number operator acting on the $j-$th fermionic site.
Note that the Lindblad equation \cref{eq:markovian_master_eq_for_electrons} has been derived via the Markovian and the Born (i.e., weak coupling) approximation and is thus not expected to provide a valid description for large values of the electron\hyp phonon coupling $g$.
\begin{figure}[!t]
	\centering
	\ifthenelse{\boolean{buildtikzpics}}%
	{%
		\input{ed_markovian_non_markovian}%
	}%
	{%
		\includegraphics{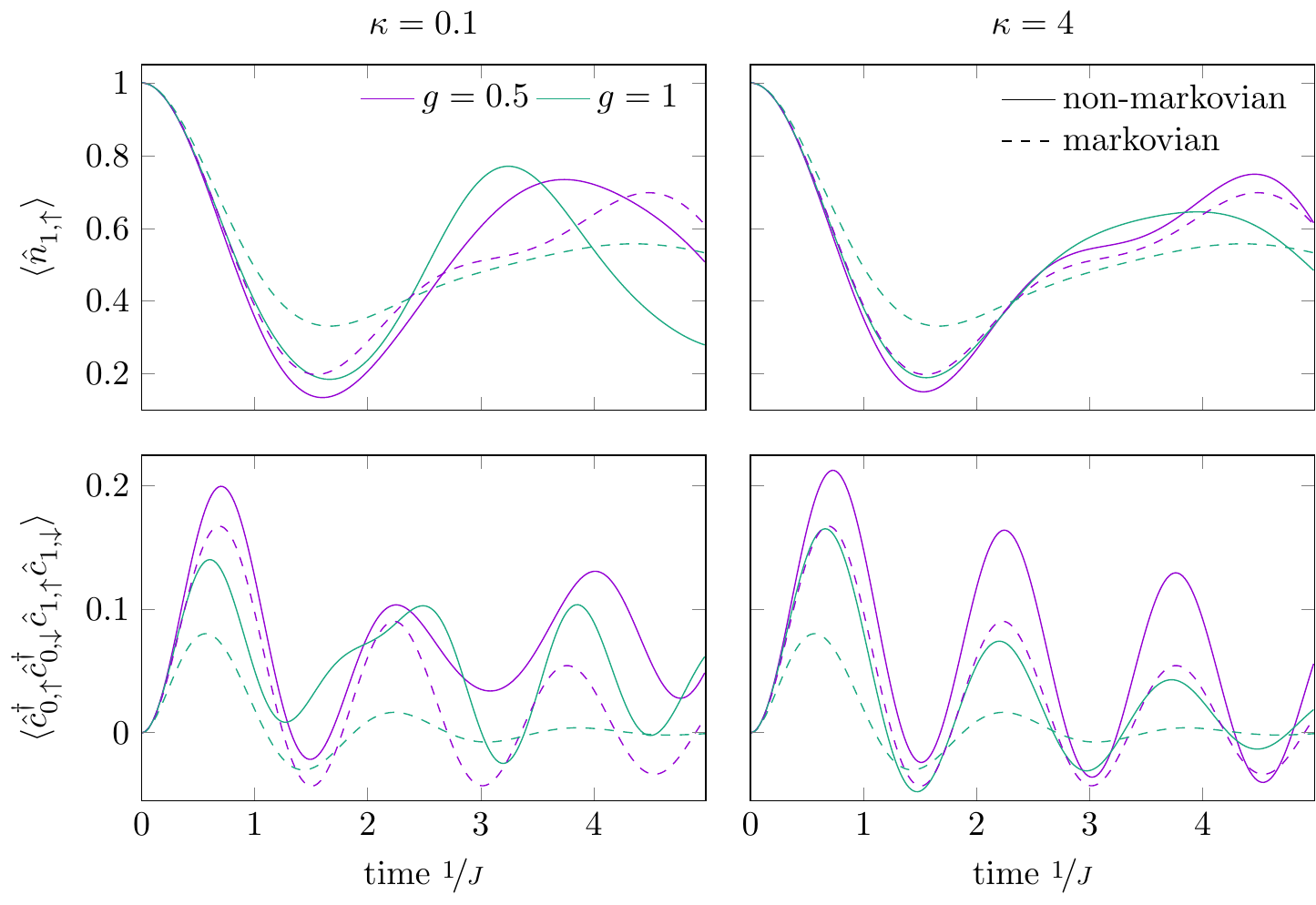}%
	}%
	\caption
	{
		\label{fig:ed_markovian_non_markovian} 
		Comparison between a master equation for the "electron $+$ phonon" system and a master equation for the electron system only. 
		For all plots, the Hamiltonian parameters were chosen to be $U=J$ and $\omega = 2J$. 
		The dissipation strength $\kappa$ was fixed to $0.1 \, J$ for the left plots and $4 \, J$ for the right plots. 
		The results obtained with the two methods strongly deviate from one another, showing that the electron dynamics of the systems considered here cannot be captured by the naive Lindblad master equation of the form \cref{eq:markovian_master_eq_for_electrons}.
}
\end{figure}
The exact\hyp diagonalization comparison between the master equation for the enlarged system \cref{eq:Lindblad} and the master equation for the electronic system only \cref{eq:markovian_master_eq_for_electrons}, shows that the latter is not suited for describing the non-Markovian bath that arises when the phonons are traced out.

\end{document}

%% file: header.tex
\usepackage{lineno}
\usepackage{graphicx}  
\usepackage{dcolumn}   
\usepackage{bm}        
\usepackage{amssymb}   
\usepackage{amsmath,stmaryrd}
\usepackage{blkarray, multirow, graphicx, diagbox, color, xcolor, colortbl}
\usepackage{bbm, bbold}
\usepackage{glossaries}
\usepackage{hyphenat}
\usepackage{ifthen}
\usepackage{xkeyval}
\usepackage{moreverb}
\usepackage{rotating}
\usepackage{wrapfig}
\usepackage{slashbox}
\usepackage{xspace}
\usepackage{nicefrac}
\usepackage[]{units}
\usepackage{physics}
\usepackage{booktabs}
\usepackage{braket}
\usepackage[inline]{enumitem}
\usepackage{tabto}
\usepackage{listings}
\usepackage{xstring}
\def\ReplaceStr#1{%
	\IfSubStr{#1}{p}{%
		\StrSubstitute{#1}{p}{.}}{#1}}

\usepackage[caption=false]{subfig}
\usepackage[customcolors]{hf-tikz} 
\usepackage{tikz}
\usepackage{calc}
\usetikzlibrary{external}
\usepackage{pgffor}
\usepackage{pgfplots}
\pgfplotsset{compat=1.13}
\usepackage{pgfplotstable}
\usepgfplotslibrary{groupplots}
\usepgfplotslibrary{fillbetween}

\tikzstyle{n} = [draw,shape=ellipse,minimum size=1.5em,inner sep=0pt,fill=white!20, minimum width=2.5em]
\tikzstyle{Init} = [n,color=green,fill=green!20,text=black]
\tikzstyle{Fin} = [n,color=red,fill=red!20,text=black]
\tikzstyle{Ghost} = [minimum size=1.5em,inner sep=0pt,color=white,text=black]
\tikzstyle{Multiple} = [draw,shape=rect,minimum size=2em,inner sep=0pt]

\tikzstyle{ghostA} = [text=red!70,thick, minimum size=2*(5pt-\pgflinewidth), inner sep=0pt, outer sep=0pt]
\tikzstyle{ghostB} = [text=blue!70,thick, minimum size=2*(5pt-\pgflinewidth), inner sep=0pt, outer sep=0pt]
\tikzstyle{siteA} = [regular polygon, regular polygon sides=3, shape border rotate= 30, draw=red!50,fill=red!20,thick,inner sep=0pt,minimum width=1.5em,font=\footnotesize]
\tikzstyle{siteB} = [regular polygon, regular polygon sides=3, shape border rotate= -30, draw=green!50,fill=green!20,thick,inner sep=0pt,minimum width=1.5em,font=\footnotesize]
\tikzstyle{op} = [regular polygon, regular polygon sides=4, draw=orange!50, fill=orange!20, thick, inner sep=0.2pt, minimum width=1.25em, minimum height=1.5em,font=\footnotesize]
\tikzstyle{opghost} = [regular polygon, regular polygon sides=4, thick, inner sep=0.2pt, minimum width=1.25em, minimum height=1.5em,font=\footnotesize]
\tikzstyle{site} = [circle,draw=blue!50,fill=blue!20,thick,inner sep=0.2pt,minimum width=1.25em,font=\footnotesize]
\tikzstyle{hiddensite} = [circle,draw=white!50,fill=white!20,thick,inner sep=0.2pt,minimum width=1.25em,font=\footnotesize]
\tikzstyle{nosite} = [circle,draw=white,fill=white,thick,inner sep=0.1pt,minimum width=1.5em]
\tikzstyle{ghost} = [font=\footnotesize]
\tikzstyle{intersite} = [regular polygon, regular polygon sides=4, shape border rotate= 45, draw=black!50,fill=black!20,thick,inner sep=0pt,minimum width=1.5em]
\tikzstyle{ld} = [inner sep=1pt, font=\small]
\tikzstyle{unsite} = [circle, outer sep=0pt,inner sep=0.2pt,minimum width=1.25em]

\definecolor{colorA}{rgb} {0.58,0,0.8275}
\definecolor{colorB}{rgb} {0.11,0.663,0.51}
\definecolor{colorC}{rgb} {0.3373,0.7059,0.9137}
\definecolor{colorD}{rgb} {0.902,0.6235,0}
\definecolor{colorE}{rgb} {0.9451,0.902,0.3255}
\definecolor{colorF}{rgb} {0.3373,0.3255,0.902}
\definecolor{colorG}{rgb} {0.9451,0.3255,0.3373}
\definecolor{cbColorA}{HTML} {810F7C}
\definecolor{cbColorB}{HTML} {006D2C}
\definecolor{cbColorC}{HTML} {7BCCC4}
\definecolor{cbColorD}{HTML} {C51B8A}
\definecolor{cbColorE}{HTML} {0868AC}

\usetikzlibrary
{
	calc,
	decorations,
	plotmarks,
	patterns,
	positioning,
	petri,
	arrows,
	decorations.markings,
	backgrounds,
	fit,
	graphs,
	shapes.geometric,
	decorations.pathmorphing,
	shapes.misc,
	shapes,
	tikzmark,
	pgfplots.colorbrewer,
	fpu,
}

\usepackage{ocgx}

\usetikzlibrary{ocgx}

\pgfplotsset{
        cycle from colormap manual style/.style={
            x=3cm,y=10pt,ytick=\empty,
            stack plots=y,
            every axis plot/.style={line width=2pt},
        },
}

\tikzset{>=stealth}

\tikzset{->-/.style={decoration={
			markings,
			mark=at position .5 with {\arrow{>}}},postaction={decorate}}}

\tikzset{-<-/.style={decoration={
			markings,
			mark=at position .5 with {\arrow{<}}},postaction={decorate}}}

\tikzstyle{orientedsnake} = [
decorate, 
decoration={snake},
->
]  
\tikzstyle{orientedshortarrow} = [
decoration={markings,
	mark=at position .33 with {\arrow{>}}},
postaction={decorate}
]  
\tikzstyle{orientedlongarrow} = [
decoration={markings,
	mark=at position .67 with {\arrow{>}}},
postaction={decorate}
]
\tikzset{dbl/.style={double,
		double equal sign distance,
		-implies,
		shorten >=10pt,
		shorten <=10pt}}
\tikzset{
	between/.style args={#1 and #2}{
		at = ($(#1)!0.5!(#2)$)
	}
}

\tikzstyle{process} = [rectangle, minimum width=3cm, minimum height=1cm, text centered, text width=5cm, draw=black]
\tikzstyle{io} = [trapezium, trapezium left angle=70, trapezium right angle=110, minimum width=3cm, minimum height=1cm, text centered, text width=7cm, draw=black]
\tikzstyle{choose} = [diamond, inner sep=1pt, minimum width=2cm, minimum height=2cm, text centered, text width=1.5cm, draw=black]
\tikzstyle{arrow} =[thick,->, >=stealth]

\newcommand{\nodagger}[0]{{\vphantom{\dagger}}}
\newcommand{\noprime}[0]{{\vphantom{\prime}}}

\newcommand{\discussive}[1]{\textcolor{black}{#1}}

\pgfmathdeclarefunction{peierlspotential}{6}%
{
	\pgfmathparse
	{
		#2 / (#3 * #5) 
		* sin(deg(#1 * #3 - #3 * #5 * (x)))
		* exp(-( ( #1 - #5 * ((x) - #4) ) )^2.0 / ( #6^2.0 ) )
	}%
}

\pgfmathdeclarefunction{linearFct}{2}%
{%
	\pgfmathparse{#1*x+#2}%
}
\pgfmathdeclarefunction{quadFct}{3}%
{%
	\pgfmathparse{#1*x*x+#2*x+#3}%
}
\pgfmathdeclarefunction{logFct}{2}%
{%
	\pgfmathparse{#1*log10(x)+#2}%
}
\pgfmathdeclarefunction{expFct}{2}%
{%
	\pgfmathparse{#1*exp(-x/#2)}%
}

\newboolean{buildCurrentBlockInline}
\setboolean{buildCurrentBlockInline}{false}
\newboolean{rebuildCurrentBlock}
\setboolean{rebuildCurrentBlock}{false}

\newboolean{rebuildData}
\setboolean{rebuildData}{false}

\newboolean{removeData}
\setboolean{removeData}{false}

\graphicspath{{figures/}}

\newboolean{buildtikzpics}
\setboolean{buildtikzpics}{false}
\newif\ifrebuildtikz
\newif\ifChangeMode
\ChangeModetrue
\ChangeModefalse
\ifthenelse{\boolean{buildtikzpics}}
{
	\rebuildtikztrue
	\usetikzlibrary{external}
	\tikzexternalize[optimize=false,prefix=figures/autogen/]%
}
{
	\rebuildtikzfalse
}

\usepackage{ulem}
\usepackage[colorlinks,bookmarks=false,citecolor=blue,linkcolor=black,urlcolor=blue]{hyperref}
\usepackage{cleveref}
\Crefname{appendix}{Appendix}{Appendices}
\Crefname{equation}{Equation}{Equations}
\Crefname{figure}{Figure}{Figures}
\Crefname{section}{Section}{Sections}
\Crefname{tabular}{Tabular}{Tabulars}
\crefname{appendix}{App.}{Apps.}
\crefname{equation}{Eq.}{Eqs.}
\crefname{figure}{Fig.}{Figs.}
\crefname{section}{Sec.}{Secs.}
\crefname{tabular}{Tab.}{Tabs.}

\newcommand{\ingoing}[1]{#1}

\ExplSyntaxOn
\DeclareExpandableDocumentCommand \eval { m } { \fp_eval:n { #1 } }
\ExplSyntaxOff

\makeatletter
\def\pgfplotsutil@decstringcounter#1{%
 \begingroup
  \c@pgf@counta=#1\relax
  \advance\c@pgf@counta by -1
  \edef#1{\the\c@pgf@counta}%
  \pgfmath@smuggleone#1%
 \endgroup
}%

\pgfplotsset{
/pgfplots/each nth point*/.style 2 args={%
/pgfplots/x filter/.append code={%
 \ifnum\coordindex=0
  \def\c@pgfplots@eachnthpoint@xfilter{0}%
  \edef\c@pgfplots@eachnthpoint@xfilter@cmp{#1}%
 \else
  \ifnum\coordindex>#2\relax
   \pgfplotsutil@advancestringcounter\c@pgfplots@eachnthpoint@xfilter
   \ifx\c@pgfplots@eachnthpoint@xfilter@cmp\c@pgfplots@eachnthpoint@xfilter
    \def\c@pgfplots@eachnthpoint@xfilter{0}%
   \else
    \let\pgfmathresult\pgfutil@empty
   \fi
  \fi
 \fi
}%
},
}
\makeatother

\newcommand{\printpgfnumberorder}[1]%
{%
	\pgfmathfloatparsenumber{#1}%
	\pgfmathfloattomacro{\pgfmathresult}{\Fn}{\Mn}{\En}%
	\pgfmathparse{\Fn==2 ? "-" : ""}%
	\edef\Sn{\pgfmathresult}%
	\Sn 10^{\En}%
}

\newcounter{marknumber}
\pgfplotsset{
	error bars/every nth mark/.style={
		/pgfplots/error bars/draw error bar/.prefix code={
			\pgfmathtruncatemacro\marknumbercheck{mod(floor(\themarknumber/2),#1)}
			\ifnum\marknumbercheck=0
			\else
			\begin{scope}[opacity=0]
				\fi
			},
			/pgfplots/error bars/draw error bar/.append code={
				\ifnum\marknumbercheck=0
				\else
			\end{scope}
			\fi
			\stepcounter{marknumber}    
		}
	}
}

%% file: acronyms.tex
\newacronym{PCMO}{PCMO}{praseodymium\hyp calcium\hyp manganite}
\newacronym{1D}{1D}{one\hyp dimensional}
\newacronym{MPS}{MPS}{matrix\hyp product state}
\newacronym{MPO}{MPO}{matrix\hyp product operator}
\newacronym{SVD}{SVD}{singular\hyp value decomposition}
\newacronym{QCS}{QCS}{quantum\hyp computer simulator}
\newacronym{QC}{QC}{quantum computer}
\newacronym{FSM}{FSM}{finite\hyp state machine}
\newacronym{ACA}{ACA}{adaptive cross\hyp approximation}
\newacronym{CDW}{CDW}{charge\hyp density wave}
\newacronym{SDW}{SDW}{spin\hyp density wave}
\newacronym{ARPES}{ARPES}{angle-resolved photoemission spectroscopy}
\newacronym{OBC}{OBC}{open-boundary conditions}
\newacronym{PBC}{PBC}{periodic-boundary conditions}
\newacronym{TEBD}{TEBD}{time-evolving block decimation}
\newacronym{TDVP}{TDVP}{time-dependent variational principle}
\newacronym{2TDVP}{2TDVP}{two-site time-dependent variational principle}
\newacronym{iff}{iff}{if and only if}
\newacronym{DFT}{DFT}{density\hyp functional theory}
\newacronym{DMFT}{DMFT}{dynamical mean\hyp field theory}
\newacronym{DMRG}{DMRG}{density\hyp matrix renormalization group}
\newacronym{QMC}{QMC}{quantum Monte Carlo}
\newacronym{ED}{ED}{exact diagonalization}
\newacronym{AIM}{AIM}{Anderson impurity model}
\newacronym{SIAM}{SIAM}{single impurity Anderson model}
\newacronym{LDA}{LDA}{local\hyp density approximation}
\newacronym{LBNL}{LBNL}{Lawrence Berkeley National Laboratory}
\newacronym{VQE}{VQE}{variational\hyp quantum eigensolver}
\newacronym{QPT}{QPT}{quantum phase transition}
\newacronym{QCP}{QCP}{quantum critical point}
\newacronym{ETH}{ETH}{eigenstate thermalization hypothesis}
\newacronym{AKLT}{AKLT}{Affleck\hyp Lieb\hyp Kennedy\hyp Tasaki}
\newglossaryentry{QR}{name={QR},description={QR decomposition}}
\newacronym{TNS}{TNS}{tensor\hyp network state}
\newacronym{PS}{PS}{pseudo site}
\newacronym{PP}{PP}{projected purification}
\newacronym{PP-DMRG}{PP-DMRG}{projected purified \gls{DMRG}}
\newacronym{PP-DMRG_long}{PP-DMRG}{projected purified \acrlong{DMRG}}
\newacronym{LBO}{LBO}{local\hyp basis optimization}
\newacronym{SC}{SC}{superconducting}
\newacronym{1RDM}{1RDM}{single\hyp site reduced density\hyp matrix}
\newacronym{OQS}{OQS}{open quantum system}
\newacronym{QJ}{QJ}{quantum jumps}
\newacronym{HOPS}{HOPS}{hierarchy of pure states}
\newacronym{HEOM}{HEOM}{hierarchy of equations of motion}
\newacronym{MBL}{MBL}{many\hyp body localization}
\newacronym{ME}{ME}{master equation}
\newacronym{lHD}{lHD}{linear homodyne detection}
\newacronym{nlHD}{nlHD}{non-linear homodyne detection}